\documentclass[showpacs,twocolumn,aps,prx,superscriptaddress,10pt]{revtex4-1}
\usepackage{amsmath,amssymb,amsfonts,bm,graphicx}
\usepackage{braket}

\usepackage{graphicx}
\usepackage{amsmath}
\usepackage{bm}
\usepackage{hyperref}
\hypersetup{%
	pdftitle={Competing order dynamics}, %
	pdfauthor={Zhiyuan Sun},%
	pdfpagemode={UseNone},%
	pdfstartview={FitH},%
	breaklinks=true,%
	citecolor=blue,%
	colorlinks=true,%
	linkcolor=blue,%
	urlcolor=blue
}

\usepackage[T1]{fontenc}
\usepackage{fourier}
\usepackage{baskervald}

\newcommand{\unit}[1]{\,\mathrm{#1}} % for specifying units in math mode
\newcommand{\equa}[1]{Eq.~\eqref{#1}} % 

\newcommand{\RomanNumeralCaps}[1]
{\textit{\MakeUppercase{\romannumeral #1}}}

\graphicspath{{./}}

\begin{document}

\title{Transient trapping into metastable states in systems with competing orders}

\author{Zhiyuan Sun}
\affiliation{Department of Physics, Columbia University,
	538 West 120th Street, New York, New York 10027, USA}

\author{Andrew J. Millis}
\affiliation{Department of Physics, Columbia University,
	538 West 120th Street, New York, New York 10027, USA}
\affiliation{Center for Computational Quantum Physics, The Flatiron Institute, 162 5th Avenue, New York, New York 10010, USA}
\date{\today}

\begin{abstract}
The quench dynamics of a system involving two competing orders is investigated using a Ginzburg-Landau theory with relaxational dynamics. We consider the scenario where a pump rapidly heats the system to a high temperature, after which the system cools down to its equilibrium temperature. We study the evolution of the order parameter amplitude and fluctuations in the resulting time dependent free energy landscape.
Exponentially growing thermal fluctuations dominate the dynamics. The system typically evolves into the phase associated with the faster-relaxing order parameter, even if it is not the global free energy  minimum.  This theory offers a natural explanation for the widespread experimental observation that metastable states may be induced by laser induced collapse of a dominant equilibrium order parameter.
\end{abstract}

\maketitle

%%%%%%%%%%%%%%%%%%%%%%%%%%%%%%%%%%%%%%%%%%%%%%%%%%%%%%%%%%%%%%%%%%%%%%%%%%%%%%%%%%%%%%%%%%%%%%%%%
\section{Introduction}
Dynamical phase transitions \cite{Hohenberg1977,Polkovnikov.2011,Bray.1994}, in which  systems are tuned through a phase transition by time variation of system parameters, are a fundamental topic of longstanding interest in many areas of science. For example, it is believed that cosmological expansion tuned the universe through the electroweak symmetry breaking transition \cite{Kibble1976}. Supercooled liquids are a widely studied terrestrial example.  Spinodal decomposition  \cite{Langer1975,Binder1987,Carter1998,Tierno2016} and  Kibble-Zurek (KZ)  \cite{Zurek.1996,Zurek1985,Biroli.2010,Chandran.2012} theories have addressed important aspects of dynamical phase transition physics for systems characterized by an order parameter which is  tuned through a first or second order transition respectively.  

Systems with multiple competing or intertwined orders are of great current interest in condensed matter physics \cite{Gruner.1988,Fradkin2015}. Examples include high T$_c$ cuprates  and transition metal dicalcogenides in which  superconductivity and spin and/or charge density wave order compete and coexist  as well as `colossal' magnetoresistance manganites where  ferromagnetic metal and charge ordered antiferromagnetic insulating states compete at low temperatures \cite{Tokura2006,Zhang2016}. Recent developments in ``ultrafast'' experimental technique \cite{Basov2011,Fausti2011,Nicoletti2014,Zhang2018,Zhang2018a,Nicoletti2018,Cremin2019,Niwa.2019,Zong2019,Suzuki2019,Rini2007} have made it possible to  dynamically suppress one or more order parameters and study the subsequent evolution, raising the possibility of  ``steering'' the order parameters into a desired metastable state.

The purpose of this paper is to provide theoretical insight into  dynamical phase transitions in systems with multiple order parameters  and in particular to draw attention to the crucial importance of the relative magnitudes of order parameter relaxation rates. We consider  systems in which the relevant degrees of freedom are space-time dependent order parameter fields $\psi_i(\mathbf{r},t)$ defined from a fundamental theory by integrating out microscopic degrees of freedom such as electrons. For notational simplicity we deal here with a system with two real order parameters. Adding more real order parameters or making the order parameters complex does not alter our conclusions.   

%%%%%%%%%%%%%%%%%%%%%%%%%%%%%%%%%%%%%%%%%
\begin{figure}[b]
	\includegraphics[width=\linewidth]{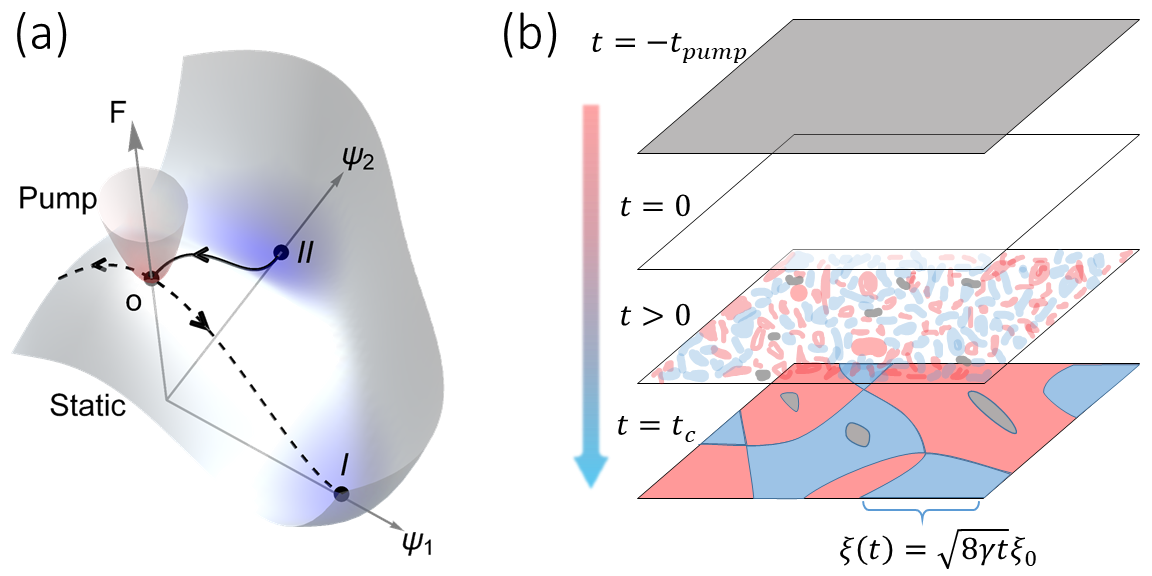}
	\caption{(a) Equilibrium free energy landscape for two competing order parameters $\psi_1$ and $\psi_2$. The energy is represented both as a height and with color, with  lower energy appearing bluer.  The point labeled $\RomanNumeralCaps{2}$ is a global free energy minimum and the point $\RomanNumeralCaps{1}$ is a locally stable minimum. The surface labeled `Pump' is a free energy landscape with only one minimum, at the origin, corresponding to a high temperature state established by a pump pulse. The pump destroys order \RomanNumeralCaps{2} as shown by the black solid trajectory. In the subsequent cooling process, exponential growth of thermal fluctuations leads the system into order \RomanNumeralCaps{1}, as shown by the dashed trajectory.
    (b) The real space illustration of order parameter evolution in the fast cooling limit. Gray means order $\RomanNumeralCaps{2}$ and blue/red means positive/negative order  $\RomanNumeralCaps{1}$. Large domains are formed at time $t_c$ with only a small volume fraction being the original order \RomanNumeralCaps{2}. 
	}
	\label{fig:quartic_free_energy}
\end{figure}
%%%%%%%%%%%%%%%%%%%%%%%%%%%%%%%%%%%%%%%%%%%%%%%%%%%%%%

We  consider two broad classes of behavior described by the equilibrium free energy landscapes sketched in Fig.~\ref{fig:quartic_free_energy} (a) strictly {\em competing orders}, where the free energy has two local minima, such that in each minimum only one of the two order parameters is nonzero, and Fig.~\ref{fig:path}  {\em intertwined order }parameters, where at the global minimum   both order parameters are nonzero, but a metastable minimum exists in which only one of the order parameters is nonzero. When such  systems are exposed to an experimentally relevant  pump pulse they may be driven to a point in phase space near the origin ($o$)  as indicated by  the solid line trajectory  in Fig.~\ref{fig:quartic_free_energy}(a). We show that after the pump is turned off, the time evolution is dominated by the exponential amplification of very long wavelength fluctuations of the order parameters, and that even a modest difference in relaxation rates will drive the system to the minimum related to the faster evolving order, even if it is not the global free energy minimum. The probability of trapping into this metastable state is close to unity. The probability of going back to the equilibrium order scales as $p_2 \sim \zeta^\delta$ where $\zeta \ll 1$ is the same Ginzburg parameter that controls the validity of static mean field theory and $\delta$ is a positive number defined latter. Nucleation dynamics operating on much longer time scales will lead eventually to relaxation to the global minimum \cite{Binder1976,Hohenberg1977}, but this physics is not explicitly considered here.  %While on a formal level the treatment is general, we have in mind the case that $\psi_1$ corresponds to  superconducting (in the case of cuprates, TMDC) or ferromagnetic metallic (manganites) order while  $\psi_2$ corresponds to charge/stripe/orbital order. Because charge/stripe/orbital order couples to the lattice it is natural to assume that the dynamics of $\psi_2$ is slower than the dynamics of $\psi_1$. 

Transient dynamics in systems with a single order parameter have been extensively discussed; the literature is too large to review here but we note that a recent study of a quench to the superconducting state has derived from microscopics the model A dynamics used here \cite{Lemonik.2017}.  Transient dynamics in systems with competing orders has been previously discussed in terms of  deterministic dynamics of spatially uniform order parameters \cite{Kung2013,RossTagaras2019,Dolgirev2019}. Our paper goes beyond the previous work by studying the formation and growth of spatial fluctuations and focusing on the difference in order parameter time constants. 

Section~\ref{sec:physical_picture} describes the physical picture and defines the formalism. Section~\ref{sec:dynamics} has the general solution to the dynamical problem. Section~\ref{sec:fast_cooling} discusses the fast cooling limit which illustrates the essential physics. Section~\ref{sec:finite_cooling} analyzes the effect of slower cooling rate. Section~\ref{sec:intertwined} discusses systems with intertwined orders. Section~\ref{sec:experiments} contains detailed comparison to recent experiments. Section~\ref{sec:discussion} is a summary and conclusion, containing a discussion of the assumptions made and consequences of relaxing them. Appendices give detailed derivations of some of the formulas in the main text.

\section{Physical picture and formalism}
\label{sec:physical_picture}
We consider a system in which the important degrees of freedom are space-time dependent order parameter fields $\psi_i(\mathbf{r},t)$ obtained from a fundamental theory by integrating out quasiparticles. We assume that  the order parameter fields evolve according to dissipative [relaxational Time-Dependent-Ginzburg-Landau (TDGL) or ``Model A''] dynamics \cite{Gorkov1968, Cyrot1973, Hohenberg1977,Lemonik.2017}  defined by a free energy functional $F$ which is time dependent because of the applied pump field: 
\begin{align}
\frac{1}{\gamma_i }\partial_t \psi_i(\mathbf{r},t)
=
-\frac{1}{E_c}\frac{\delta F(t)}{\delta \psi_i (\mathbf{r},t)} + \eta_i(\mathbf{r},t)
\,.
\label{eqn:TDGL}
\end{align} 
Here $\gamma_i$ are the corresponding relaxation rates, $E_c$ is the condensation energy density and $\eta$ is a noise field determined by the microscopic degrees of freedom that were integrated out to obtain the order parameter theory. The free energy functionals are assumed to be of the general form
\begin{align}
& F[\psi_1,\psi_2] = E_c\int d^D\mathbf{r} \, \left(f_1 + f_2 +f_c \right) \,,\notag\\
& f_i= -\alpha_i \psi_i^2 + (\xi_{i0} \nabla \psi_i)^2 + \psi_i^4 \,
\,
\label{eqn:free_energy}
\end{align}
as sketched in Fig.~\ref{fig:quartic_free_energy}(a). 
Here the $\xi_{i0}$ are the bare coherence lengths, $D$ is the spatial dimension and
\begin{equation}
f_c =c \psi_1^2 \psi_2^2
\,
\label{eqn:fc}
\end{equation}
describes the interaction between the two order parameters in the competing order case.  

In our convention, $\psi_i$, $\alpha_i$, $c$ and $f_i$ are dimensionless, intensive and defined such that the quartic term in the free energy has coefficient $1$ and the $\alpha_i$ are of the order of unity at zero temperature.
For cooperation ($c<0$) or weak competition ($0<c<2$), $F$ has a single minimum. For $c>2$, $F$ has two locally stable minima if $\alpha_1$ and $\alpha_2$ $>0$ and $\frac{2}{c}<\frac{\alpha_1}{\alpha_2}<\frac{c}{2}$ (see Appendix \ref{appendix:phase_diagram}). We study the $c>2$ (multiple minima) case in this paper.
Following  usual practice we assume that all parameters are temperature independent except the $\alpha_i = \kappa_i (T_{ci}-T)/T_{ci}$, which are positive at low temperatures, negative at high temperatures, vary smoothly with temperature and vanish at the respective critical temperatures  $T = T_{ci}$ (we assume linear temperature dependence for simplicity). 
The nonequilibrium enters the formalism as a  time dependence of the $\alpha_i$, determined by the time dependence of the effective temperature $T(t)$ of the microscopic degrees of freedom that were integrated out.
We focus on the case $\alpha_2>\alpha_1$ but $\gamma_2\alpha_2<\gamma_1\alpha_1$ so minimum \RomanNumeralCaps{2} is the equilibrium free energy minimum but the dynamics associated with minimum \RomanNumeralCaps{1} is faster.

The condensation energy density $E_c$, in combination with $\xi_{i0}$, sets the relevant microscopic scales. An important  dimensionless measure of the thermal fluctuations is 
\begin{equation}
G_{i}(T)=\frac{T}{E_c\xi_{0i}^D}
\,.
\label{Gdef}
\end{equation}
The Ginzburg parameter defined in the conventional theory of critical phenomena is $G(T_c)\alpha^\frac{D-4}{2}$ with mean field theory applying when the parameter is much less than unity.  For example, in the weak coupling case of conventional superconductors, $G \sim \left( \mathrm{gap}/\mathrm{fermi \,\, energy}\right)^{D-1}$. The treatment that follows is formally valid in the $G\ll1$ limit.

The stochastic Eq.~\eqref{eqn:TDGL} may be recast as a Fokker-Planck equation for  a  probability functional $\rho[\{\psi_k\}]$  that gives the distribution of fluctuations around the mean field value (see, e.g., \cite{Kramers1940,Hohenberg1977,Risken.1996}  and Appendix~\ref{FP_equation}). In the linearized approximation used below the probability functional  is a direct product $\rho[\psi] = \prod_{k} \rho_{k}(\psi_{k})$ where
\begin{align}
\rho_k =  \frac{1}{\sqrt{2\pi  D_k(t) }} 
e^{-\frac{\psi_{k}^2}{2 D_k(t)}}
\label{eqn:gausssian_solution_diffusion}
\end{align}
is a Gaussian distribution for each Fourier mode $\psi_{k}$ of the field with time-dependent variance $D_k(t)=\langle \psi_{k}(t)\psi_{-k}(t) \rangle$ which we calculate below.

We are primarily interested in understanding experiments in which a system is highly excited by a pump pulse  and the subsequent evolution  is probed. We  assume that the pump does not couple directly to the order parameters; rather, it excites microscopic degrees of freedom (e.g. electron quasiparticles or phonons) which thermalize very quickly (relative to the order parameter   timescales) to a quasiequilibrium state  described by an effective temperature $T(t)$ \cite{He.2016}. The effective temperature  is maintained by the pump at a high value $T_H$ for some time. After the pump is turned off, $T(t)$ evolves over a timescale $t_m$ to the true thermal equilibrium temperature $T_L$ determined by the bath. In most experiments, the bath is the lattice and $t_m$ is just the electron-phonon thermalization time scale which is typically of the order of picoseconds.
This assumption is in essence the two temperature model of Rothwarf and Taylor \cite{Rothwarf.1967} in which one set of degrees of freedom (e.g., electrons or a particular set of phonon modes) is excited to a high temperature and then relaxes back to the equilibrium temperature set by the rest of the system. 

Within these assumptions, the instantaneous value of $T(t)$ determines the parameters $\alpha_i$ of the free energy and the noise. We take the noise correlators 
\begin{equation}
\left<\eta_i(\mathbf{r},t)\eta_i(\mathbf{r}^\prime, t^\prime)\right>=\frac{2T(t)}{\gamma_i E_c} \delta(\mathbf{r}-\mathbf{r^\prime},t-t^\prime) 
\,
\label{etacorrelator}
\end{equation}
to be local in space and time and consistent with the fluctuation-dissipation theorem.
Here  the Boltzmann constant $k_B$ is set to unity and the average is over a probability distribution of the noise field. For the slow dynamics we consider here, the relevant frequency/momentum scale is well below those of the microscopic degrees of freedom that are integrated out, justifying the locality assumption of \equa{etacorrelator}.

%%%%%%%%%%%%%%%%%%%%%%%%%%%%%%%%%%%%%%%%%
\begin{figure}
	\includegraphics[width= 1.0 \linewidth]{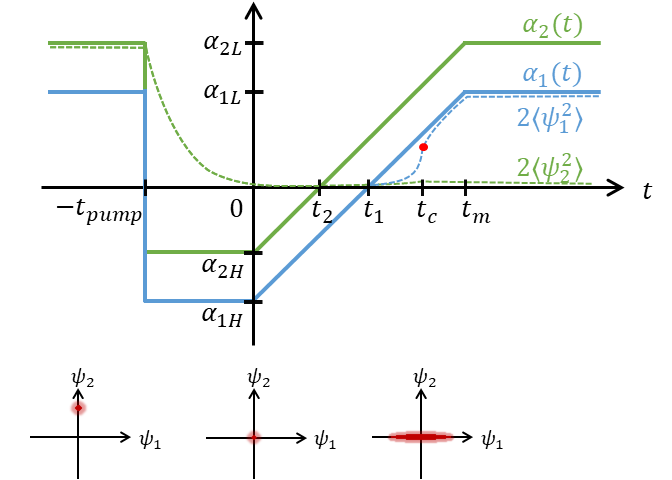}
	\caption{Upper panel: time evolution of quadratic free energy coefficients $\alpha_i(t)$ (solid lines) and mean square values of order parameters (dashed lines). Different colors denote different orders. 
	The pump maintains the system at a high temperature $T_H$ (negative $\alpha=\alpha_{iH}$) for the time $-t_{pump}<t<0$, after which the temperature relaxes to the equilibrium one $T_L$ (positive $\alpha=\alpha_{iL}$) over a time $t_m$. Shown is the linear cooling profile used to derive exact formulas in the slow cooling case. The mean field order parameter is suppressed during the high temperature stage. Order parameter fluctuation $\langle \psi_i(r)^2 \rangle$ starts to grow exponentially after $t_i$, as shown by the dashed curves. The red dot denotes the point of crossover to nonlinear dynamics ($\langle \psi_1^2 \rangle \sim \alpha_1(t)$) at time $t_c$. Lower panel: the local order parameter probability distributions $\rho\left(\psi_1,\psi_2\right)$ corresponding to the time intervals vertically above.} 
	\label{fig:quench_profile}
\end{figure}
%%%%%%%%%%%%%%%%%%%%%%%%%%%%%%%%%%%%%%%%%%%%%%%%%%%%%%

Representative time histories of $\alpha(t)$ are shown in Fig.~\ref{fig:quench_profile}. There are three time regimes: (a){\em  pump on}, covering the  time interval $-t_{pump}<t<0$ in which the  temperature $T=T_H$ and corresondingly the quadratic coefficient $\alpha=\alpha_{H} <0$; (b){\em relaxation},  time $0<t<t_m$, during which  $T$ evolves from $T_H$ through $T_c$ to $T_L$ while $\alpha$ evolves from $\alpha_H<0$ through $\alpha=0$ to $
\alpha_L>0$; and (c) {\em evolution}, time $t>t_m$,  $T=T_L$ and $\alpha=\alpha_L$.   It is convenient also to introduce the time $t_0=\frac{T_H-T_c}{T_H-T_L}t_m$, at which $T=T_c$ and $\alpha(t)=0$ (the $t_0$ times for order parameters $\psi_1$ and $\psi_2$ are labeled as $t_1$ and $t_2$ in Fig.~\ref{fig:quench_profile}). 

The physical picture is that for $t\in (-t_{pump},\, t_0)$, the free energy landscape has its only minimum at $\psi=0$ (point $O$ in Fig.~\ref{fig:quartic_free_energy}(a)).
Thus the high temperature produced by the pump suppresses the mean field order parameter to nearly zero and the fluctuations remain bounded. 
After time $t_0$, point $O$ becomes unstable,  the long wavelength fluctuations start to grow exponentially with time and the correlation length grows as $\xi\sim \xi_0 \sqrt{8\gamma t}$. This process leads to the creation of large domains where most domains are in phase \RomanNumeralCaps{1} as shown in Fig.~\ref{fig:quartic_free_energy}(b).

\section{Dynamics}  
\label{sec:dynamics}
In this section we present a general solution of Eq.~\eqref{eqn:TDGL} with time dependent $\alpha$ and noise correlators as described above. The initial condition is
\begin{equation}
\psi(\mathbf{r})^{init}=\bar{\psi}+\sum_ke^{i\mathbf{k}\cdot \mathbf{r}}\delta\psi_k\equiv \bar{\psi}+V\int \frac{d^D k}{(2\pi)^D}e^{i\mathbf{k}\cdot \mathbf{r}}\delta\psi_k
\label{initial}
\end{equation}
where $\bar{\psi}$ is the initial order parameter, $\delta \psi_k$ represents  thermal fluctuations about the initial ordered state and $V$ is the system volume. For simplicity of notation, here and henceforth we suppress the index $i$ labelling the different order parameters wherever possible. In  the initial state, fluctuations are assumed small: $\bar{\psi}^2\gg \left<\delta \psi(r=0)^2\right>\equiv\sum_k\left<\delta\psi_k\delta\psi_{-k}\right>\sim G$. The pump acts to decrease $\bar{\psi}^2$ and increase the fluctuations. If $\bar{\psi}^2$ remains large compared to the mean square  fluctuation amplitude, the state of the system is determined by a straightforward deterministic dynamics. This case is discussed briefly below, but our main interest is in situations in which the pump drives the initial order parameter to a value smaller than the root mean square fluctuation amplitude and the physics is determined by the  evolution of the fluctuations.  

Because \equa{eqn:TDGL} is first order in time it has no ``memory'', so the evolution over one time regime fixes initial conditions for the next one. We will first consider the evolution over the decaying order parameter regime $t<t_0$; the resulting state of the system at $t_0$ is then the initial condition for the subsequent evolution. 
 
In the pump-on regime the system is hot (temperature $T=T_H$) so the free energy is dominated by a large quadratic term which justifies the use of linearized dynamics even when the order parameter is not small. This means that in the pump-on regime we may study
\begin{equation}
\frac{1}{\gamma} \partial_t \psi_{k} =  2 \alpha_{k}(t)  \psi_{k}  + \eta_{k}
\label{eqn:fourier_mode_linear_evolution}
\end{equation}
where
\begin{equation}
\alpha_k(t)=\alpha(t)-\xi_0^2k^2<0
\,.
\label{alphadef}
\end{equation} 
We assume that the dynamics in the pump-on regime drives the order parameter to a small enough value that we may continue to use the linearized approximation throughout the $t<t_0$ regime and for some time into the growing fluctuations ($t>t_0$) regime. Conditions for the validity of this approximation will be presented below. The solution of Eq.~\eqref{eqn:fourier_mode_linear_evolution} may be written as
\begin{equation}
\psi_k(t)=\psi_k^{init} e^{S_k\left(t,-t_{pump}\right)} +\gamma\int_{-t_{pump}}^{t}dt^\prime \eta_k(t^\prime) e^{S_k\left(t,t^\prime\right)}
\label{linearsolution}
\end{equation}
where  the first term gives the propagation forward in time of the initial order parameter  $\psi_k^{init}=\bar{\psi}\delta_{k,0}+\delta\psi_k$ with mean field part $\bar{\psi}$ and small fluctuations $\delta \psi$. The second term represents the propagation forward of fluctuations created by the noise after $-t_{pump}$. The accumulated phase $S$ is defined as
\begin{equation}
S_k(t_a,t_b)= 2\gamma \int_{t_b}^{t_a} dt \alpha_k(t)
\,
\label{eqn:time_dependent_solution}
\end{equation}
and $S_0(t_a,t_b)$ has the interpretation as the signed area enclose by the solid lines and the time axis in Fig.~\ref{fig:quench_profile}.
With \equa{etacorrelator}, the square of \equa{linearsolution} yields
the fluctuation amplitude 
\begin{equation}
D_k(t)=
D_k^{init} e^{2S_k(t,-t_{pump})} +
\frac{2\gamma}{E_c V} \int_{-t_{pump}}^{t}dt^\prime T(t^\prime) e^{2S_k(t,t^\prime)} 
\label{variance11}
\end{equation}
defined in Eq.~\eqref{eqn:gausssian_solution_diffusion} where $D_k^{init}$ is the initial fluctuation amplitude.

In the linear cooling profile approximation the initial mean field order parameter amplitude evolves as
\begin{equation}
\bar{\psi}(t_0)=e^{-|\alpha_H|\gamma t_0}e^{-2|\alpha_H|\gamma t_{pump}}\bar{\psi}
\label{meanfieldorderparameter}
\end{equation}
where the factor $e^{-2|\alpha_H|\gamma t_{pump}}$ gives the exponential suppression during the pump-on stage and the $e^{-|\alpha_H|\gamma t_0}$ factor gives the additional suppression during $t \in (0,\, t_0)$. We assume that $|\alpha_H|\gamma(2t_{pump}+t_0)$ is large enough that the mean field order parameter is reduced to a very small value at $t=t_0$, less than the mean square fluctuations. As shown in detail in section \ref{sec:finite_cooling}, this assumption plus a small Ginzburg parameter implies that at time $t_0$, the system is well prepared in a disordered state with negligible mean field order parameter and small fluctuations.

We now consider  the evolution of the distribution at times after $t_0$, where the system has cooled below the transition temperature  ($\alpha(t)>0$) so long wavelength modes with $k< \sqrt{\alpha}\xi_0^{-1}$ grow exponentially with time. As long as the fluctuation does not become too large the linearized equation may be used for the dynamics so
\begin{align}
D_k(t)=&e^{2S_0(t,t_0) -4k^2\xi_0^2\gamma\left(t-t_0\right)}
\notag\\
&\left(D_k(t_0)+\frac{2\gamma}{E_cV}\int_{t_0}^{t}dt^\prime e^{-2S_k(t^\prime,t_0)}T(t^\prime)\right)
\label{variance3}
\end{align}
where as before the first term represents the propagation forward in time of the fluctuations existing at $t_0$ while the second term represents the additional contributions generated by the noise thereafter. A detailed analysis given in Appendix~\ref{sec:dynamics} shows that the second term in \equa{variance3} is of the same order as the first in the situations of interest here. 

The key observation is that long wavelength modes with $4\xi_0^2k^2\gamma(t-t_0)<2S_0(t,t_0)$ are exponentially amplified and we are interested in long times $\gamma(t-t_0)\gg 1$ for which the growth is substantial ($e^{2S_0(t,t_0)}\sim G^{-1}$). The exponential growth continues until  the local mean square fluctuation amplitude of one of the order parameters becomes large enough  that the  nonlinearity becomes important to the dynamics, i.e., until $t$ reaches the crossover time $t_c$ defined by  
\begin{equation}
\langle \psi_i(r=0)^2 \rangle_{t=t_c} \sim \alpha_i/2
\,.
\label{linearregime}
\end{equation}
To compute  $\langle \psi_i(0)^2 \rangle $ we observe that at long times the important momentum dependence is controlled by the $e^{-4k^2\xi_0^2\gamma\left(t-t_0\right)}$ factor.  Defining the time-dependent correlation length by
\begin{align}
\xi^2(t)=\xi^2_0 \left(8 \gamma(t-t_0) +1/a \right)
\label{eqn:correlation_length}
\end{align}
we perform the momentum integral to obtain (up to an unimportant overall factor) the correlation function in real space 
\begin{align}
\left<\psi(0)\psi(r)\right>_t
&=
V \int \frac{d^Dk}{(2\pi)^D}D_k(t) e^{i \mathbf{k} \cdot \mathbf{r}}
\notag\\
&= \frac{G/a}{\left(16\pi\gamma\left(t-t_0\right)\right)^\frac{D}{2}}
e^{2S_0(t,t_0)} 
e^{-\frac{r^2}{2\xi(t)^2}}
\,.
\label{localoft}
\end{align}
The details of the pump and initial cooling enter Eq.~\eqref{eqn:correlation_length} and \eqref{localoft} via the $a$ which varies from $\sim \alpha_L$ in the fast cooling limit to $\alpha_{KZ}$ in the slow cooling case (to be discussed in Sec. \ref{sec:finite_cooling}). This variation leads to  corrections that are subleading relative to the terms we consider and we will not explicity notate this dependence henceforth. 

Eqs.~\eqref{localoft} is the first key result. The exponential factors show that the fluctuations grow exponentially in time, and the spatial correlation is governed by the universal correlation length growth law $\xi \sim \xi_0 \sqrt{8\gamma t}$ which does not depend on the equilibrium value of $\alpha$ or temperature-time profile.

 %The physical picture after $t_0$ is, although the initial fluctuation is tiny due to the small Ginzburg parameter $G \ll 1$, the exponentially growing factor amplifies it to be \begin{align}\left<\psi(0)^2\right>_{t_c} \sim \alpha(t_c)/2\label{eqn:t_c}\end{align} at the crossover time $t_c$ to induce the phase transition. If at this time for order \RomanNumeralCaps{1}, we have $\langle \psi_2^2 \rangle/\langle \psi_1^2 \rangle \ll 1$ satisfied, the system is trapped into the metastable phase \RomanNumeralCaps{1}.

\section{Fast cooling limit}
\label{sec:fast_cooling}
\begin{figure}[b]
	\includegraphics[width=1 \linewidth]{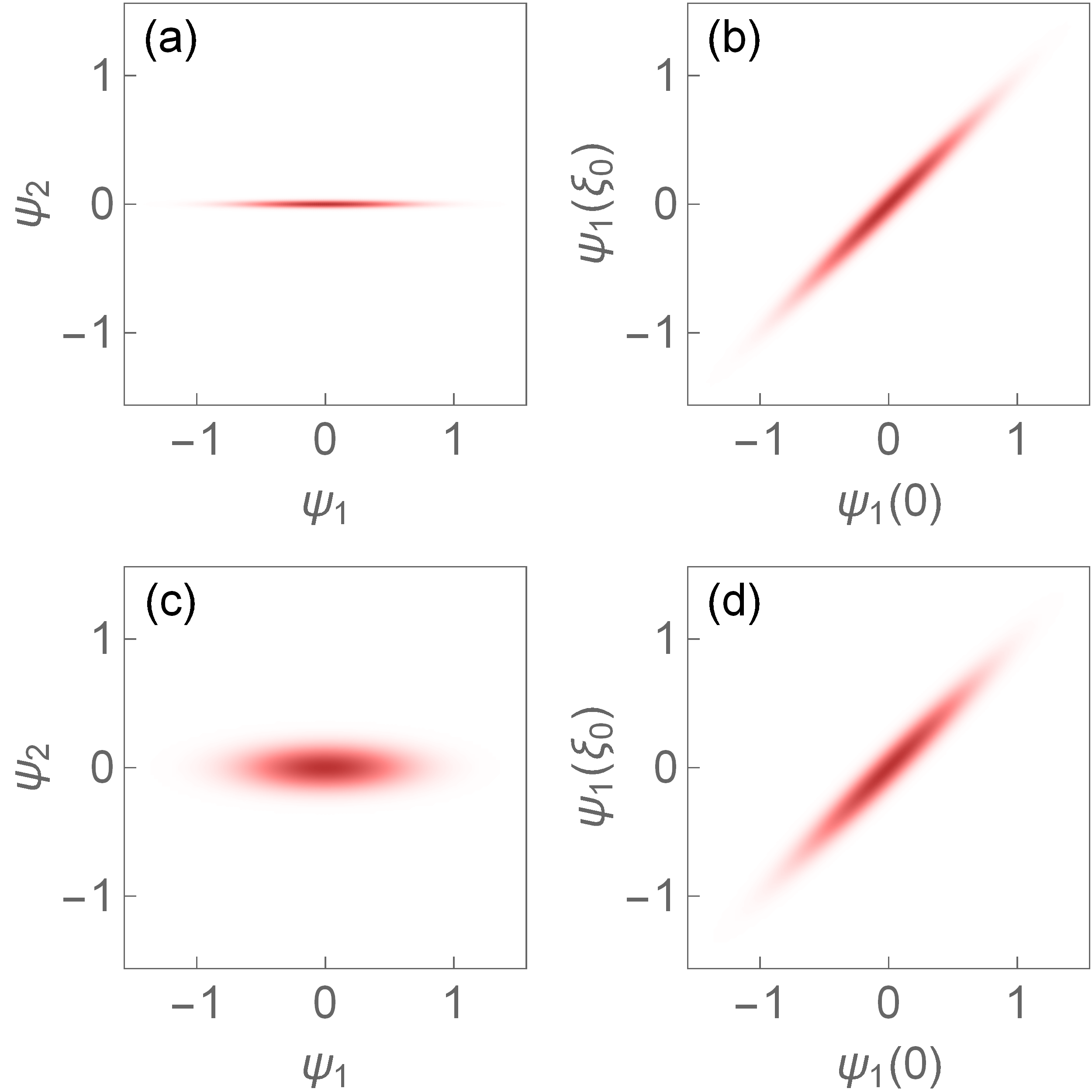}
	\caption{Panel (a)(c) are the density plots of the Gaussian probability distribution $\rho\left(\psi_1(r),\psi_2(r)\right)$ computed at time $t=2.0 \unit{ps}$ for $G_1 = G_2 =10^{-5}$ (panel (a)) and $t=1.0 \unit{ps}$ for $G_1 = G_2 =10^{-2}$ (panel (c)) in the fast cooling limit $t_m=0$. Regions of higher value of $\rho$ appear redder.  Panel (b)(d) show the two point probability $\rho\left(\psi_1(0),\psi_1 (\xi_0)\right)$ distributions for the same two cases.
	The common parameters used are $(\alpha_{1L},\,\alpha_{2L}) =(1.0,\, 1.1)$,  $(\gamma_1,\,\gamma_2) =(2,\, 1) \unit{ps}^{-1}$, $\xi_{0i}=\xi_{0}$ and $D=3$.}
	\label{fig:gaussian}
\end{figure}

In this section we consider the fast cooling limit $t_m\rightarrow 0$ which illustrates the essential physics with minimal complexity.
In the fast cooling limit  the phase for the exponential amplification term in \equa{localoft} is simply
\begin{equation}
2 S_0(t,t_0) 
= 4 \alpha_L \gamma t
\,
\label{phase_fast_cooling_limit}
\end{equation} 
and Eq.~\eqref{linearregime} for the crossover time $t_c$ is
\begin{align}
4\alpha_L \gamma t_c
= \ln \frac{1}{\zeta}  + \frac{D}{2} \ln \left(4 \alpha_L \gamma t_c\right)
\label{eqn:time}
\end{align}
where we have set $a$ in \equa{localoft} to $\alpha_L$ without altering the leading behavior, and have defined
\begin{equation}
\zeta= 2(4\pi)^{-\frac{D}{2}} \alpha_L^{D/2-2} G
\label{zetadef}
\end{equation} 
which is in effect the usual Ginzburg parameter of the theory of critical phenomena.  Similar results were previously presented by Lemonik and Mitra~\cite{Lemonik.2018}  who noted the importance of the Ginzburg parameter in setting post-quench timescales. We see that $ t_c$ is logarithmically large if $\zeta$ is small, i.e. if mean field theory works well for equilibrium. At time $t_c$ of the fast order \RomanNumeralCaps{1}, the fluctuation of the slow order \RomanNumeralCaps{2} is smaller by the ratio
\begin{align}
\frac{\langle \psi_2^2 \rangle}{\langle \psi_1^2 \rangle}=
\frac{\alpha_{1L}}{\alpha_{2L}}
\left( \frac{\gamma_1}{\gamma_2} \right)^{D/2}
\frac{G_2}{G_1} 
\left(
\frac{1}{\zeta_1} \left(\ln \frac{1}{\zeta_1} \right)^{D/2}
\right)^{\frac{\alpha_{2L}\gamma_2}{\alpha_{1L}\gamma_1}-1}
\,
\label{eqn:psi_ratio_fast}
\end{align}
which is much less than unity if $\gamma_2\alpha_{2L}<\gamma_1\alpha_{1L}$ and $\zeta \ll 1$.
At this stage of the evolution the two order parameters are independent and the joint distribution of local  amplitudes at a position $r$  is the product of Gaussians:
\begin{equation}
\rho\left(\psi_1(r),\psi_2(r)\right)
=\frac
{\mathrm{Exp}
	\left[-\frac{\psi_1^2}{2\langle \psi_1^2\rangle}-\frac{\psi_2^2}{2\langle \psi_2^2\rangle}\right]}
{2\pi \sqrt{ \langle\psi_1^2\rangle\langle \psi_2^2\rangle}}
\,.
\label{Ppsi1psi2}
\end{equation}
For $\gamma_2\alpha_{L2}<\gamma_1\alpha_{L1}$ the mean square values are very different, leading to the highly anisotropic joint distribution function shown in Fig.~\ref{fig:gaussian}(a)(c). The probability distribution describing the space dependence of the fluctuations is derived in Appendix \ref{appendix:joint_probability} and is plotted in Fig.~\ref{fig:gaussian}(b)(d) for order \RomanNumeralCaps{1}. We see that the fluctuations of order parameter \RomanNumeralCaps{1}  are highly correlated over scales out to $\xi(t)\gg \xi_0$ (the fluctuations of the slower order parameter \RomanNumeralCaps{2} are correlated over slightly shorter distances).

%%%%%%%%%%%%%%%%%%%%%%%%%%%%%%%%%%%%%%%%%%%%%%%%%%%%%%
%%%%%%%%%%%%%%%%%%%%%%%%%%%%%%%%%%%%%%%%%%%%%%%%%%%%%%

Thus the physical picture at $t=t_c$ is of order parameter domains of typical size $\xi(t_c)\approx\sqrt{\ln\frac{1}{\zeta}} \sqrt{\frac{2}{\alpha_L}} \xi_0\gg\xi_0$ within which the order parameters are nearly uniform and normally distributed and with the typical value of $\psi_1$ much larger than $\psi_2$, as illustrated in Fig.~\ref{fig:quartic_free_energy}(b). Moreover, the typical value of local $\psi$ is now much larger than the new fluctuation scale induced by the noise, so that in the subsequent evolution the mean field dynamics dominates over the effect of the noise. Therefore, to study the subsequent evolution it suffices to consider the evolution within a domain, which is described by the uniform, deterministic TDGL equations, written here for the competing orders case:
\begin{eqnarray}
\gamma_1^{-1}\partial_t\psi_1&=&2\alpha_1\psi_1-4\psi_1^3-2c\psi_2^2\psi_1
\,,
\nonumber
\\
\gamma_2^{-1}\partial_t\psi_2&=&2\alpha_2\psi_2-4\psi_2^3-2c\psi_1^2\psi_2
\label{LocalTDGL}
\end{eqnarray}
with initial conditions chosen from the joint probability distribution $\rho\left(\psi_1 (0),\psi_2 (0) \right)$. The issues associated with matching the solutions at the domain walls are a coarsening problem discussed briefly below.

%Nonlinearity is onset in the subsequent stage of dynamics. However, the correlation length $\xi_{it}$ is much larger than the equilibrium correlation length $\xi_i$ and the dynamics inside each domain can be approximately described by the uniform TDGL equation, which says the $\psi_i$ in each domain evolves towards its optimized value $\sqrt{\alpha_i/2}$ exponentially.

\begin{figure}[t]
	\includegraphics[width=0.9 \linewidth]{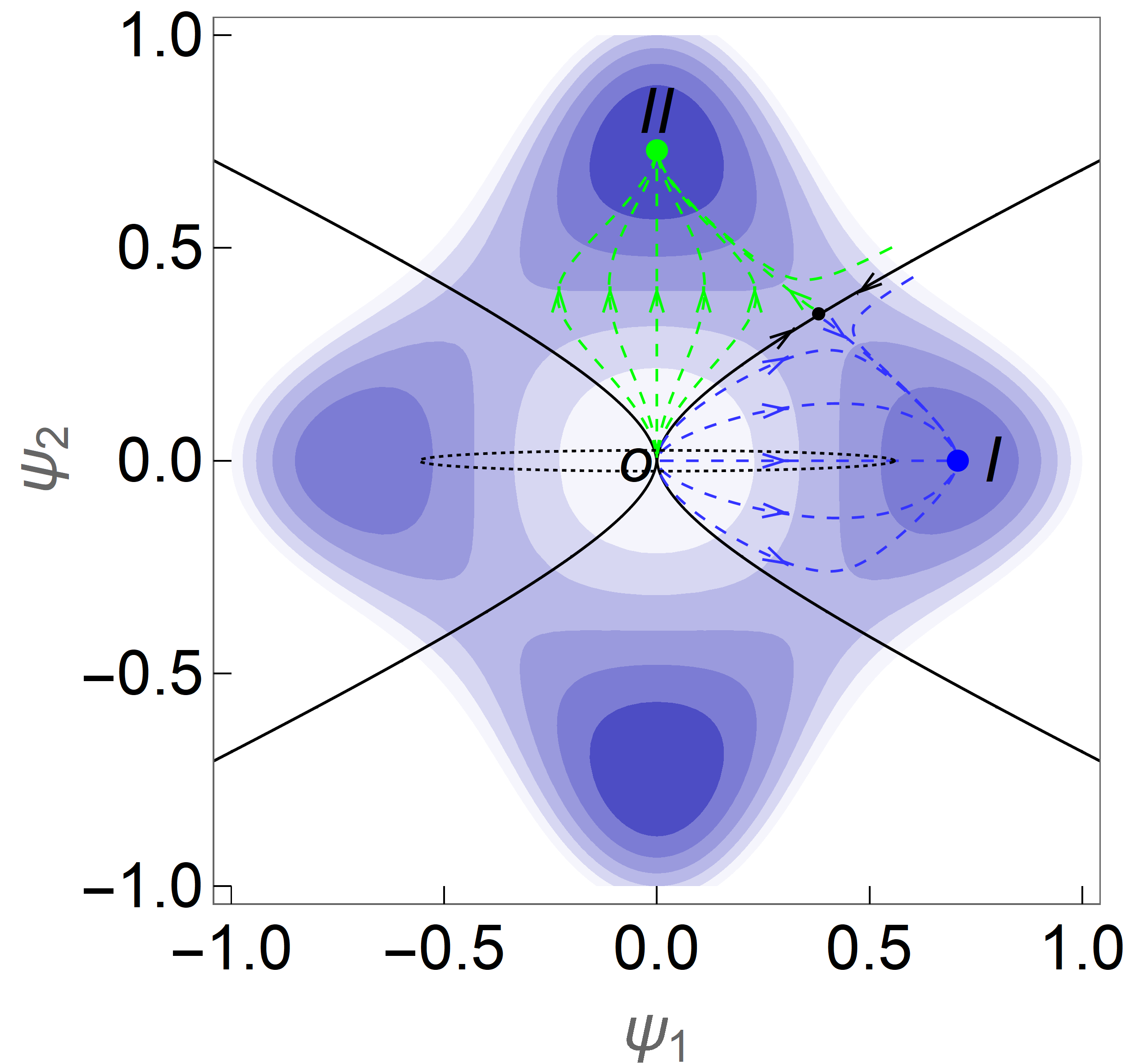}
	\caption{Basins of attraction of the two orders on the contour plot of the free energy landscape. Lower energy appears bluer. The arrows show the direction of order parameter dynamics. Black solid line separates the basins. The parameters used are $(\alpha_{1L},\,\alpha_{2L}) =(1.0,\, 1.1)$,  $(\gamma_1,\,\gamma_2) =(2,\, 1) \unit{ps}^{-1}$ and $c=6$. Black dotted line illustrates the magnitudes of $\psi_i$ fluctuations at time $t=2.0 \unit{ps}$ for $G_1 = G_2 =10^{-4}$ and spatial dimension $D=3$. Most of the probability lies in the basin of order \RomanNumeralCaps{1} meaning most volume of the system will be trapped into \RomanNumeralCaps{1} afterwards. }
	\label{fig:basin}
\end{figure}

The flow defined by \equa{LocalTDGL} has a simple phase space structure with  stable fixed points defined by the minima of $F$, as shown in Fig.~\ref{fig:basin}. Each initial condition defines a trajectory that flows into one of the minima. For the physically relevant case $\alpha_{1L},\,\alpha_{2L}>0$ with $\frac{2}{c}<\frac{\alpha_{1L}}{\alpha_{2L}}<\frac{c}{2}$ there are four fixed points, with basins of attraction separated by a four-branched separatrix curve. We may estimate the position of the separatrix by matching the small $\psi$ regime, where the exponential growth requires
\begin{equation}
\psi_1=\lambda \psi_2^{1/\Delta}
\label{eq:separatrixsmall}
\end{equation}
to the requirement that the separatrix goes through the saddle point $(\psi_1^2, \psi_2^2)=(c \alpha_{2L}-2\alpha_{1L},\, c\alpha_{1L} -2\alpha_{2L})/(c^2-4)$.  Here $\Delta=\gamma_2\alpha_{2L}/\gamma_1\alpha_{1L}<1$ and the coefficient is fixed by the matching condition. %%%%%%%%%%%%%%%%%%%%%%%%%%%%%%%%%%%%%%%%%%%%%%%%%%%%%%
%%%%%%%%%%%%%%%%%%%%%%%%%%%%%%%%%%%%%%%%%%%%%%%%%%%%%%

By finding the relative weights of the probability distribution in the different basins of attraction we can estimate the relative volume fractions of the different order parameter domains. The volume fraction $p_2$ of order \RomanNumeralCaps{2} domains is just the probability of $(\psi_1,\, \psi_2)$ lying in the basin of attraction of minimum \RomanNumeralCaps{2} in Fig.~\ref{fig:basin}, which at time $t=t_c$ can be estimated  as
\begin{align}
p_2 & \approx 4\int_{0}^{\infty} d\psi_2 \rho(\psi_1=0,\psi_2) \lambda \psi_2^{1/\Delta} 
\notag\\
&= \frac{1}{\pi} 2^{\frac{1}{2\Delta} +\frac{1}{2}} \Gamma\left(\frac{1}{2}\left(1+\frac{1}{\Delta} \right) \right) 
\lambda \langle \psi_1^2 \rangle^{\frac{1/\Delta -1}{2}}
\left(\frac{\langle \psi_2^2 \rangle}{\langle \psi_1^2 \rangle }\right)^{\frac{1}{2\Delta}}
\notag\\
&= \vartheta \left( \ln \frac{1}{\zeta_1} \right)^{-\frac{D}{2} \delta} \zeta_1 ^{\delta} 
\,
\label{eqn:p2}
\end{align}
where $\delta = (1/\Delta-1)/2>0$ and $\vartheta \sim 1$ can be found in Appendix~\ref{appendix:coefficients}. Thus the proportion of $\psi_2$ domains is suppressed by a power law of the Ginzburg parameter $\zeta \ll 1$ and is negligibly small even if the time scales are just slightly different. \equa{eqn:p2} is the second key result of this paper.
 
%where we have defined the trapping exponent $\delta = \frac{1}{2} \left( \Delta-1 \right) >0 $. The coefficient $\vartheta$ can be found in supplemental material and is order one if none of the $\alpha_i$ are too small. Therefore, the criterion of trapping into phase \RomanNumeralCaps{1} is just $\delta >0$, i.e., $\alpha_1 \gamma_1 > \alpha_2 \gamma_2$. If it is satisfied, $p_2$ approaches $0$ as a power law of the critical Ginzburg parameter $\zeta$.

\emph{Life time of the metastable state---}Each domain then evolves to the appropriate minimum; the evolution takes a time of the order of $\frac{1}{\alpha_{L} \gamma} \ln \frac{1}{\zeta}$, after which the physical picture is of a set of domains, most of which have   $\psi_1=\pm \sqrt{\alpha_{1L}/2}$ and $\psi_2\approx 0$ (i.e. are in phase  \RomanNumeralCaps{1}) while a small volume fraction of the sample are  phase \RomanNumeralCaps{2} domains where   $\psi_2=\pm \sqrt{\alpha_{2L}/2}$ and $\psi_1\approx 0$, as illustrated in Fig.~\ref{fig:quartic_free_energy}(b). However, long range order is not established.
The subsequent evolution is determined by spontaneous nucleation of phase \RomanNumeralCaps{2} regions in the dominant phase \RomanNumeralCaps{1} domains, and by growth of the existing $\psi_2$ domains.  The timescale for ultimate equilibration thus depends both on nucleation rates and on domain wall dynamics, both of which are beyond the scope of this paper. We do, however, provide a likely lower limit on the equilibration time by considering the free growth of $\psi_2$ domains, assuming no domain wall pinning and  nucleation which is an exponentially slow process. The speed of domain wall motion is at the order of $v \sim \gamma \xi_0$ as long as the free energy difference $\delta f$ between the two minima is order one. Assuming the phase \RomanNumeralCaps{2} domains are evenly distributed among the phase \RomanNumeralCaps{1} domains we  can estimate the equilibration time as
\begin{align}
t_{life} \sim \left( \frac{1}{p_2} \right)^\frac{1}{D} \xi(t_c)/v \sim \frac{1}{\gamma} \left( \ln \frac{1}{\zeta} \right)^{\frac{1}{2}(1+\delta)}  \left(\frac{1}{\zeta}\right)^{\frac{\delta}{D}}
\,.
\label{eqn:td}
\end{align}
%For $\gamma=1 \unit{ps}^{-1}$, $\delta=1$ and $\zeta=10^{-4}$, this yields the life time $t_{life} \sim 200 \unit{ps}$ for three dimension, consistent with many experiments \cite{Fausti2011,Zhang2018,Zhang2018a,Cremin2019,Niwa.2019} but these estimates depend very sensitively on parameters.

If the order parameters are complex ones corresponding to $U(1)$ symmetry breaking, as in the case of superconductivity (SC) being order \RomanNumeralCaps{1} and charge density wave (CDW) being order \RomanNumeralCaps{2}, one just needs to double the degrees of freedom for each order before nonlinear dynamics is onset. As a result, \equa{eqn:p2} gives the offset of trapping probability as 
\begin{equation}
p_2 \sim \left( \ln \frac{1}{\zeta} \right)^{-D \delta} \zeta^{2\delta}	
\label{eqn:p2_2}
\end{equation}
and \equa{eqn:td} is modified to
\begin{align}
t_{life} \sim \frac{1}{\gamma} \left( \ln \frac{1}{\zeta} \right)^{\frac{1}{2} +\delta}  \left(\frac{1}{\zeta}\right)^{\frac{2\delta}{D}}
\,.
\label{eqn:td_2}
\end{align}
After $t_c$, the different SC regions are characterized by different order parameter phases which can continuously synchronize to leave behind topological vortices. The number density of the vortices scales as $n\sim 1/\xi(t_c)^2 \sim \left( \xi_0^2 \frac{2}{\alpha_L} \ln \frac{1}{\zeta} \right)^{-1} $, different from the Kibble-Zurek scaling \cite{Zurek1985} since the latter applies to the slow cooling limit.

\section{Finite cooling rate}
\label{sec:finite_cooling}
We now ask how the physics is modified as the cooling time $t_m$ is increased from zero. The essential picture derived in the previous section of long length-scale domains of one or the other order parameter still applies, but because the time  $t_0$ of transition from exponential decay to exponential growth is earlier for order \RomanNumeralCaps{2}  than that for order \RomanNumeralCaps{1}, the $\psi_2$ fluctuations will have a longer period of growth than the $\psi_1$ fluctuations. The longer period of growth will compensate for the faster dynamics of $\psi_1$, meaning that the condition on the difference in relaxation rates required for the system to evolve to minimum \RomanNumeralCaps{1} becomes more stringent.  A second issue is that the cross over to nonlinear dynamics may occur at a time $t<t_m$ before the relaxation of the $\alpha$ to its equilibrium value is complete, meaning that the free energy landscape at the point of crossing  to nonlinearity differs from the equilibrium one.  For these reasons  the behavior for given $t_m$ depends on the ratio of relaxation rates $\gamma_2/\gamma_1$ in a somewhat complicated manner. The various regimes are shown in Fig.~\ref{fig:criterions}. 

\subsection{The state at $t_0$}
We first characterize the state at $t=t_0$. If the mean field order parameter is reduced to a small value, then the thermal fluctuations existing at time $t=-t_{pump}$ (the first term in \equa{variance11}) will be reduced to a completely negligible level so the order parameter at $t=t_0$ and subsequent times is determined entirely by the random noise. We distinguish {\em fast cooling} ($|\alpha_H|\gamma t_0 \le1$) and {\em slow cooling} ($|\alpha_H|\gamma t_0 \ge1$) regimes according to whether the cooling to $t_0$ after the pump is turned off has a significant effect on the order parameter.
The linearized dynamics means that the corresponding distribution function is the product of Gaussians given in Eq.~\eqref{eqn:gausssian_solution_diffusion}. Averaging the solution for $\psi_k(t_0)$  over the noise using Eq.~\eqref{etacorrelator} shows that the fluctuation distribution half-width defined in Eq.~\eqref{eqn:gausssian_solution_diffusion} is 
\begin{equation}
D_k(t_0)=\frac{2\gamma}{E_cV}\int_{-t_{pump}}^{t_0}dt^\prime e^{2S_k(t_0,t^\prime)}T(t^\prime)
\,.
\label{variance1}
\end{equation}
The integral in Eq.~\eqref{variance1}  may easily be evaluated numerically, and in the linear quench approximation may be expressed exactly in terms of error functions (see Appendix \ref{appendix:exact_solution}). Here we present results in important limits which explicate the basic physics. In the fast cooling limit  the portion of the integral from $t=0\rightarrow t_0$ makes a negligible contribution and we find
\begin{equation}
D_k(t_0)
\approx
\frac{T_H}{2E_c V}\frac{1}{|\alpha_H|+\xi_0^2k^2}
\label{PhiH}
\end{equation}
indicating that in the fast cooling limit  the fluctuations at $t=t_0$ are those of the hot thermal state created by the pump, with  distance $|\alpha_H|$ from criticality and correlation length $\xi_H=\xi_0/\sqrt{|\alpha_H|}$. 
In the slow cooling regime only times near $t_0$ are important and we find
\begin{align}
D_k(t_0) &\approx
\frac{T_c}{E_c V}
\frac{1}{\alpha_{KZ}}\int_0^\infty dv Exp\left[-v^2-2\xi_{kz}^2k^2v\right]
\notag\\
&=
\left\{
\begin{array}{lr}
\frac{\sqrt{\pi}}{2\alpha_{KZ}} \frac{T_c}{E_c V} \,,\quad &  k \ll 1/\xi_{KZ}   \\
1/(2\xi_0^2 k^2) \,,\quad & 
k \gg 1/\xi_{KZ}
\end{array}
\right.
\label{PhiKZ}
\end{align} 
where the  effective distance from criticality $\alpha_{KZ}$ and corresponding correlation length $\xi_{KZ}$ are given by
\begin{equation}
\alpha_{KZ}=\sqrt{|\alpha_H|/(2\gamma t_0)},
\hspace{0.2in}\xi_{KZ}=\xi_0/\sqrt{\alpha_{KZ}}
\label{akz}
\end{equation}
which depend on the square root of the cooling rate, consistent with Kibble-Zurek scaling  \cite{Zurek.1996,Zurek1985} and the mean field exponents of the problem at hand. 

The correlation function in real space is given by
\begin{equation}
\left<\psi(0) \psi(r) \right>_t=
V \int^{k_c} \frac{d^Dk}{(2\pi)^D}D_k(t) e^{i \mathbf{k} \cdot \mathbf{r}}
\label{psilocal}
\end{equation}
where  the upper cutoff $k_c$ which we expect to be of the order of a few times $\xi_0^{-1}$ is required to make the integral finite as $r\rightarrow 0$. The momentum integral is dominated by large momenta for which $D_k\sim (k\xi_0)^{-2}$, so the local fluctuation amplitude is
\begin{equation}
\left<\psi^2(0)\right>_{t_0} =
\left\{
\begin{array}{lr}
\frac{G_{D=2}}{8\pi} \ln\left(1+k_c^2\xi^2\right) \,, & (D=2)  \\
\frac{G_{D=3} k_c\xi_0}{2\pi^2}\left(1-\frac{1}{k_c\xi} \tan^{-1}k_c\xi\right)
\,,& (D=3)
\end{array}
\right .
\label{fluctrapid}
\end{equation}
where G (Eq.~\eqref{Gdef}) is to be evaluated at temperature $T=T_H, T_c$ in the fast and slow quench limits respectively and $\xi$ takes the value appropriate for the relevant limit. %The second term means the correlation length of the fluctuations prepared at $t_0$ and $a$ interpolates from $|\alpha_H|$ to $\alpha_{KZ}$ as the cooling rate varies from the fast to the slow limit. As the cooling rate is varied from the fast to the slow limit, $G/a$ interpolates from $G(T_H)/(2|\alpha_H|)+G(T_L)/(2\alpha_L)$ to $G(T_c)\sqrt{\pi}/\alpha_{KZ}$. As long as $T_H/T_L$ is not very large, the different choice of temperature only causes subleading effects and we choose $G=G(T_c)$ in the following. 

To summarize, if the initial pump and subsequent cooling are strong enough to drive the initial mean field order parameter to a level smaller than the local root mean square fluctuation then at time $t=t_0$ the order parameter fluctuations  in both the rapid  and  slow quench cases are described by a Gaussian  (mean field-like) probability characterized by a temperature ($T_H$ or $T_c$), a distance from criticality ($|\alpha_H|$ or $\alpha_{KZ}$) and the associated correlation length $\xi^2_{H,KZ}=\xi_0^2/\alpha_{H,KZ}$. The local mean square local fluctuation order parameters are of the order of $G$, justifying \equa{localoft} even in the slow cooling case. Using Eqs.~\eqref{meanfieldorderparameter} and \eqref{fluctrapid} we see the criterion for suppression of the mean field order parameter  is roughly $|\alpha_H|\gamma\left(2t_{pump}+t_0\right)\gg \ln (1/G)$.  The linearized analysis used here requires that $\left<\psi^2(r=0)\right> \ll \alpha$. A renormalization group-improved treatment will break down when $G \alpha(t)^\frac{D-4}{2} \sim 1$ (the Ginzburg criterion) which sets an upper limit on the cooling time $t_m \sim \zeta^{-1/(1-D/4)} /\gamma \equiv t_{mc}$ in the slow quench regime. 

It is important to remember that the $T_c$ (and thus $t_0$) of the two different order parameters are different as are the bare correlation lengths and relaxation constants, so especially in the slow cooling limit  the probability distribution functions of the two order parameters will differ, especially at long wavelengths, although Eq.~\eqref{fluctrapid} shows that the local fluctuation amplitudes, which are determined by short wavelength fluctuations, are not too different for the two order parameters.

\subsection{The trapping condition}
\label{sec:parameter_diagram}
Now we analyze the effect of finite cooling rate on the condition for trapping into phase \RomanNumeralCaps{1}.
To simply the formulas, in the main text we focus on the exponential growth and neglect power-law prefactors, thus approximating $\langle \psi^2 \rangle_t \sim Ge^{2S(t,t_0)}$. Our main focus will be on establishing how small $\gamma_2$ must be relative to $\gamma_1$ for  the system to evolve with high probability into the metastable minimum \RomanNumeralCaps{1}. We will find dependence of the critical ratio $\Delta=\frac{\gamma_2\alpha_{L2}}{\gamma_1\alpha_{L1}}$ is a scaling function of the variable $t_m/t_{mu}$, where $t_{mu}$ is the cooling time at which the onset of nonlinearity $t_c$ coincides with the equilibration time $t_m$. 

\begin{figure}
	\includegraphics[width=0.9 \linewidth]{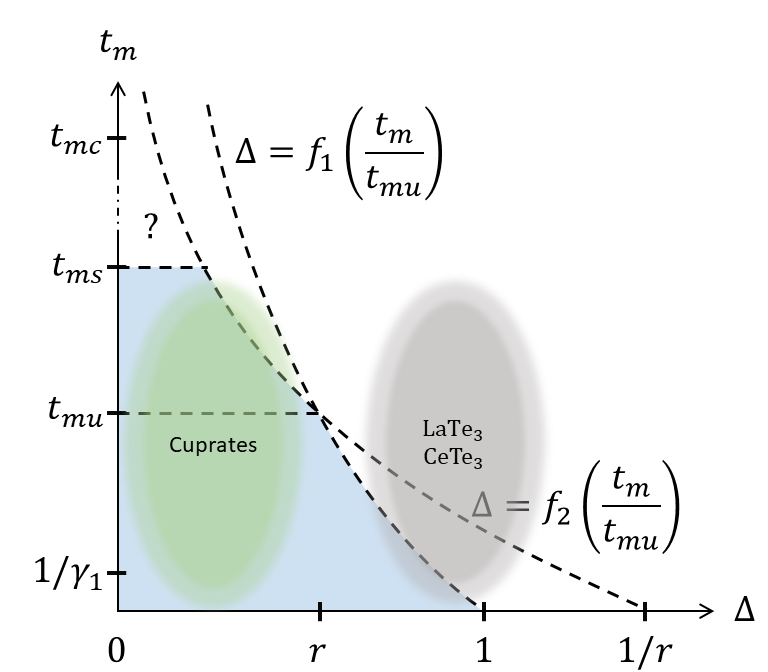}
	\caption{A schematic `phase' diagram delineating the behavior on the cooling time $t_m$ versus $\Delta=\gamma_2\alpha_{2L}/(\gamma_1\alpha_{1L})$ plane. The horizontal axis can be viewed as $\gamma_2$ while all other parameters are fixed. In the blue region, trapping into phase \RomanNumeralCaps{1} happens because $\langle \psi_2^2 \rangle/\langle \psi_1^2 \rangle<1$ at $t_c$.  The region $t_m > t_{ms}$ is unexplored in this paper and discussed in Sec.~\ref{sec:parameter_diagram}. The green/gray ellipses are our tentative guess of where the cuprates/rare earth tri-tellurides lie on this diagram.}
	\label{fig:criterions}
\end{figure}

To begin the analysis we note that for $t_m>0$ and $t>t_m$  the accumulated phase becomes (after eliminating $t_0$ in favor of $\alpha_L$)
\begin{equation}
2S_0(t,t_0)= 4 \alpha_L \gamma\left(t- \frac{t_m}{2}\frac{2|\alpha_H|+\alpha_L}{|\alpha_H|+\alpha_L}\right)
\label{phase_slow_cooling}
\end{equation}
and Eq.~\eqref{eqn:time} for the crossover time becomes 
\begin{equation}
t_c = \frac{t_m}{2}\frac{2|\alpha_H|+\alpha_L}{|\alpha_H|+\alpha_L}+ \frac{1}{4\gamma \alpha_{L}} \ln \frac{1}{\zeta}
\label{tc_slow_cooling}
\end{equation}
while  Eq.~\eqref{eqn:psi_ratio_fast} becomes
\begin{align}
\frac{\langle \psi_2^2 \rangle}{\langle \psi_1^2 \rangle}=e^{2\alpha_{2L}\gamma_2t_m\frac{|\alpha_{1H}|\alpha_{2L}-|\alpha_{2H}|\alpha_{1L}}{\left(|\alpha_{2H}|+\alpha_{2L}\right)\left(|\alpha_{1H}|+\alpha_{1L}\right)}}
\left(
\frac{1}{\zeta_1} 
\right)^{\frac{\alpha_{2L}\gamma_2}{\alpha_{1L}\gamma_1}-1}
\,.
\label{eqn:psi_ratio_slow}
\end{align}
The factor $\left(
1/\zeta_1 
\right)^{\frac{\alpha_{2L}\gamma_2}{\alpha_{1L}\gamma_1}-1}$ is Eq.~\eqref{eqn:psi_ratio_fast} with only the leading term in $t_c$  retained and the exponential factor expresses the additional growth of  $\psi_2$ due $\alpha_2$ crossing zero earlier than $\alpha_1$.
When 
\begin{equation}
t_m=\frac{|\alpha_{1H}|+\alpha_{1L}}{2 \gamma_1 \alpha_{1L}^2} \ln\frac{1}{\zeta_1}
=
\frac{1}{2\gamma_1\alpha_{1L}}
\frac{T_H-T_L}{T_{c1}-T_L}
\ln\frac{1}{\zeta_1}
\equiv t_{mu}
\label{tc=tm}
\end{equation}
we have $t_c=t_m$, i.e., the onset of nonlinearity occurs at $t=t_m$.
Thus the onset of nonlinearity occurs before equilibration only for cooling rates very slow relative to the basic order parameter timescales by a factor of the order of the log of the Ginzburg parameter.
Using
\begin{equation}
\frac{|\alpha_{1H}|\alpha_{2L}-|\alpha_{2H}|\alpha_{1L}}{\left(|\alpha_{2H}|+\alpha_{2L}\right)\left(|\alpha_{1H}|+\alpha_{1L}\right)}=\frac{T_{c2}-T_{c1}}{T_H-T_L}
\,
\end{equation}
we see that  $\langle \psi_2^2 \rangle/\langle \psi_1^2 \rangle<1$ provided that $\Delta= \alpha_{2L}\gamma_2/(\alpha_{1L}\gamma_1)$ is less than a critical value defined by
\begin{equation}
\frac{1}{
1+
\frac{T_{c2}-T_{c1}}{T_{c1}-T_L}\frac{t_m}{t_{mu}}}\equiv f_1\left(\frac{t_m}{t_{mu}}\right)
\,
\label{condition1}
\end{equation}
as shown in Fig.~\ref{fig:criterions}.
In the $t_m\rightarrow0$ limit Eq.~\eqref{condition1} reverts to the previous result $\alpha_2\gamma_2<\alpha_1\gamma_1$ (up to logarithmic corrections), but as $t_m$ increases the constraint on $\gamma_2$ becomes more stringent and when $t_m=t_{mu}$ Eq.~\eqref{condition1} becomes
\begin{equation}
\Delta=\frac{T_{c1} - T_L}{
T_{c2} - T_L
}
\equiv r<1
\,.
\label{condition12}
\end{equation}

For $t_m>t_{mu}$, $\psi_1$  reaches nonlinearity before $t_m$ which is the regime considered by Kibble-Zurek theory \cite{Zurek1985}. Note that relaxational dynamics predicts a logarithmic correction to Kibble-Zurek scaling, as described in Appendix~\ref{appendix:slow_cooling}. In this time regime the accumulated phase  may be written
\begin{equation}
2S_0(t,t_0)= 2 \gamma \frac{\alpha_{L}}{t_m-t_0}(t-t_0)^2
\label{phase_ultra_slow_cooling}
\end{equation}
and after some algebra Eq.~\eqref{linearregime} for phase \RomanNumeralCaps{1} can be written as
\begin{equation}
t_c = t_{1}+\left(t_m-t_{1}\right)\sqrt{\frac{t_{mu}}{{t_m}}
 } 
\,.
\label{tc_ultra_slow_cooling1}
\end{equation}
which is obviously before $t_{m}$ if $t_m>t_{mu}$.
The condition $S(t_c,t_{2})<S(t_c,t_{1})$ becomes
\begin{equation}
\Delta<\frac{t_{mu}}{t_m}\frac{r}
{
\left(1-r\left(1-\sqrt{\frac{t_{mu}}{t_m}}\right)\right)^2
}
\equiv f_2\left(\frac{t_m}{t_{mu}}\right)
\label{ultra_slow_condition}
\end{equation}
which reduces to our previous Eq.~\eqref{condition12} when $t_m=t_{mu}$ and drops as $1/t_m$ for large $t_m$ so that as the equilibration time becomes extremely long, the system would evolve to the equilibrium minimum unless the relxation rate $\gamma_2$ becomes exceptionally small. 

Even if Eq.~\eqref{ultra_slow_condition} is satisfied, the system will only evolve to the metastable minimum if $\alpha_{i}$ are such that the metastable minimum exists at the time order \RomanNumeralCaps{1} crosses to nonlinearity, i.e., if $\alpha_{2}(t_c) < \frac{c}{2} \alpha_1(t_c)$
which yields 
\begin{align}
t_{m}< t_{mu}
\frac{(T_{c1}-T_{L})^2}{ (T_{c2}-T_{c1})^2} 
\left(
\frac{c}{2} \frac{\kappa_1 T_{c2}}{\kappa_2 T_{c1}}
-1
\right)^2
\equiv t_{ms} 		
\,.
\label{conditionminimum}
\end{align}
If $t_m>t_{ms}$, as denoted by `?' in Fig.~\ref{fig:criterions}, the following would happen in the cooling process: at $t_c$, the time for $\psi_1$ crossing over to nonlinear dynamics, the free energy landscape has not recovered enough such that order I is not yet a local minimum. Thus trapping into the order I state won’t necessarily happen even if $\langle \psi_2^2 \rangle \ll \langle \psi_1^2 \rangle$  at this time. 
We see that typically $t_{ms}$ cannot be too much larger than $t_{mu}$, unless either $T_{c2}-T_{c1}$ is very small or $\kappa_1 \gg\kappa_2$ or $c\gg 2$. The various time scales are collected in Table~\ref{tbl:time_scales}.

Below the upper critical dimension $D=4$, our `time dependent fluctuation around mean field' approach fails if at the predicted cross over time $t_c$, the system is still inside the critical regime $G \alpha(t)^\frac{D-4}{2} \sim 1$ (the Ginzburg criterion) where even renormalization group improved treatment breaks down. This imposes an ultimate upper limit 
\begin{align}
t_{mc} \equiv \frac{1}{\gamma_1}  \zeta^{-\frac{1}{1-D/4}}  \,
\end{align}
for the cooling time, see Appendix~\ref{appendix:slow_cooling}.
This much larger time scale is deep inside the `?' region and is also labeled in Fig.~\ref{fig:criterions}.

\begin{table}
	\begin{ruledtabular}
		\begin{tabular}{ll}
			Symbol  & Physical meaning
			\\
			\hline
			$-t_{pump}$ & When pump pulse arrives
			\\[3pt]
			$t_0$ & When the temperature crosses $T_c$
			\\[3pt]
			$t_1\,,\, t_2$ & $t_0$ for order \RomanNumeralCaps{1}, \RomanNumeralCaps{2} 
			\\[3pt]
			$t_m$ & Cooling (electron-phonon thermalization) time scale 
			\\[3pt]
			$t_c$ & The time that fluctuation becomes comparable to $1$
			\\[3pt]
			$t_{mu}$ & The cooling time scale that makes $t_c=t_{m}$ (\equa{tc=tm})
			\\[3pt]
			$t_{ms}$ & See discussion around \equa{conditionminimum}
			\\[3pt]
			$t_{mc}$ & The cooling time scale where mean field theory fails
			\\[3pt]
			$\Delta$ & The critical ratio $\gamma_2\alpha_{L2}/\left(\gamma_1\alpha_{L1}\right)$
			\\[3pt]
			$\delta$ & $(1/\Delta-1)/2$
		\end{tabular}
	\end{ruledtabular}
	\caption{Physical meaning of various time scales and $\Delta$.}
	\label{tbl:time_scales}
\end{table}

%sets a stringent limit lower limit on the cooling rates for For even slower cooling rate, since the correlation length at $t_c$ is much smaller than that of equilibrium, the dynamics after $t_c$ can be analyzed by solving the uniform TDGL \equa{LocalTDGL} inside each domain and we leave it for future work.}

%In principle, if we further slow down the cooling process such that $t_{m} > t_{mc} = \frac{\alpha_L + |\alpha_H|}{2\alpha_L^2 \gamma } \left( 1/\zeta \right)^{1/(1-D/4)}$, the predicted crossover happens inside the critical regime $\zeta(\alpha(t_c)) \sim 1$ of the equilibrium theory. Even if one replaces $\alpha_i$ by the physical ones, the renormalized perturbation theory has infrared divergence for $D<4$ and thus fails here according to the Ginzburg criterion. Therefore, the probability function is not a direct product of Gaussians and our formalism does not apply to this case. However, $t_{mc}$ is a rather large time since it is amplified by the power law of the Ginzburg parameter and is unrealistic in the cooling process led by electron phonon thermalization. 
%%%%%%%%%%%%%%%%%%%%%%%%%%%%%%%%%%%%%%%%%%%%%%%%%%%%%%
%%%%%%%%%%%%%%%%%%%%%%%%%%%%%%%%%%%%%%%%%%%%%%%%%%%%%%

\section{Intertwined order}
\label{sec:intertwined}
%%%%%%%%%%%%%%%%%%%%%%%%%%%%%%%%%%%%%%%%%%%%%%%%%%%%%%
\begin{figure}
	\includegraphics[width=0.9 \linewidth]{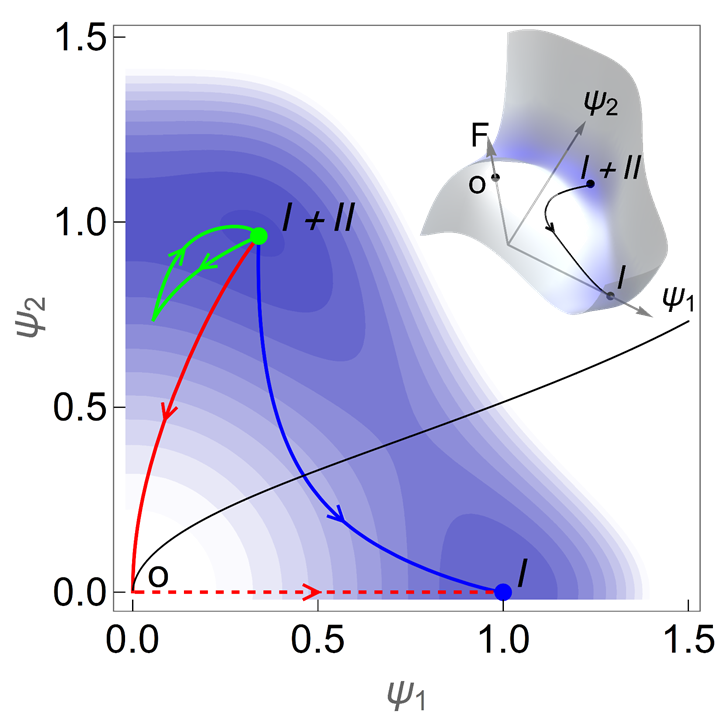}
	\caption{Contour plot of the free energy landscape for intertwined orders. Lower energy appears bluer. The parameters used are $(\alpha_1,\,\alpha_2) =(2.0,\, 2.05)$,  $(\gamma_1,\,\gamma_2) =(2,\, 1) \unit{ps}^{-1}$, $c=1$ and $d_1=4$. The lines are different trajectories the system undergoes in the pump-cooling process. Red dashed line means the process is led by exponentially growing fluctuations. Thin black line is the boundary of the basin of attraction of minimum \RomanNumeralCaps{1}. Inset is the free energy landscape plotted in 3D. }
	\label{fig:path}
\end{figure}
%%%%%%%%%%%%%%%%%%%%%%%%%%%%%%%%%%%%%%%%%%%%%%%%%%%%%%
To study the intertwined case we modify  \equa{eqn:free_energy} to shift one of the minima away from one of the axes. One simple choice is to add a term to  $f_c$ so
\begin{align}
f_c \rightarrow  c\psi_1^2 \psi_2^2 + d_1 \psi_1^4 \psi_2^2  \,
\,.
\label{eqn:sectic_free_energy}
\end{align}
and with, now,   $0<c<2$ and $d_1>0$.  We assume $T_{c2}>T_{c1}$ but that the difference in $T_c$ is not too large, and $d_1$ is not too small. In this case (see Appendix~\ref{appendix:phase_diagram} for details),  as temperature is lowered the system first enters a phase with only $\psi_2\neq0$ and then at a lower temperature the phase with $\psi_1\neq 0, ~\psi_2=0$ becomes locally stable although not the global minimum. At a still lower temperature the global minimum is gradually shifted to \RomanNumeralCaps{1}+\RomanNumeralCaps{2} where a nonzero $\psi_1$ component appears.  If we identify $\psi_2$  with density wave order and $\psi_1$  with superconductivity, this scenario may describe stripe ordered cuprates (e.g., La$_{2-x}$Ba$_x$CuO$_4$ around $x=1/8$): the so called pair density wave (PDW) state \cite{Fradkin2015}. The free energy analysis of the $\psi_2/\psi_1$ minimum has previously been discussed \cite{Fradkin2015}; we have generalized the free energy so that it also includes a metastable phase with purely superconducting order and will argue that this generalization is needed to describe recent ultrafast experiments \cite{Cremin2019}. 

The considerations sketched in the previous sections carry over directly to the intertwined order case, as shown by the red line in Fig.~\ref{fig:path}. However, an additional interesting effect may occur if we relax the assumption that the pump heats up the bath appropriate to both order parameters. If the two orders couple to different microscopic degrees of freedom, then one may consider the case when only the free energy landscape for one order is changed.  In particular, in the case of coupled superconducting and charge orders, one may imagine that the charge order couples to phonons much more strongly than do the electrons, so driving the phonons would affect the CDW much more strongly than the superconductivity.  If the system starts in a minimum with both order parameters nonzero (as is the case for intertwined orders) and only $\alpha_2$ is driven to negative, leaving $\alpha_1$ positive, the transient free energy landscape will have only one minimum (\RomanNumeralCaps{1}) and the mean-field dynamics will drive the system into it, as shown by the blue trajectory in Fig.~\ref{fig:path}.  In this process, small fluctuations of the order parameter can be neglected and one can apply deterministic TDGL dynamics to the mean field order parameters. This mechanism does not require faster relaxation for  $\psi_1$; all that is needed is that $\psi_2$ remains suppressed for long enough that the system evolves to the \RomanNumeralCaps{1} minimum. This  timescale is set by the time required for the order parameter to cross the basin boundary, $t_s \sim \frac{1}{\alpha \gamma } \ln \frac{\psi_{2m}}{\psi_{2b}}$ where $\psi_{2m}$ is the value of $\psi_2$ at the original point $\RomanNumeralCaps{1}+\RomanNumeralCaps{2}$ and $\psi_{2b}$ is its value at the intersection between the blue trajectory and the basin boundary.  For shorter pump durations or for pumps that reduce $\alpha_1$ too much, the system would relax back to the global minimum as illustrated schematically by the green trajectory in Fig~\ref{fig:path}. 

\section{Experiments}
\label{sec:experiments}
Competing orders have been reported in many materials, and an increasing number of ultrafast experiments are appearing, including studies of competing charge density waves in tri-tellurides \cite{Kogar2019,Zong2019}, ferromagnetic domain formation in charge ordered manganites \cite{Zhang2016}, and charge and magnetic order in rare earth nickelates. Much attention has focused on reports of  transient  superconductivity appearing in materials that  have low temperature nonsuperconducting density wave states but may reasonably be expected to have competing superconducting states \cite{Fausti2011,Nicoletti2014,Zhang2018,Zhang2018a,Nicoletti2018,Cremin2019,Niwa.2019,Suzuki2019}. The appearance of long lived superconducting-like metastable states has been seen in cuprates \cite{Fausti2011,Nicoletti2014,Zhang2018,Zhang2018a,Nicoletti2018,Cremin2019,Niwa.2019} through optical pump-probe by several independent groups and in FeSe through time resolved APRES \cite{Suzuki2019}. This widespread mystery has heretofore been theoretically addressed via explorations of models in which the nonequilibrium drive changes the microscopic Hamiltonian, creating new physics not existing in equilibrium \cite{Kennes2017,Babadi2017,Sentef2017,Chiriaco2018,Wang.2018} and via TDGL analyses \cite{Kung2013,RossTagaras2019,Dolgirev2019} with deterministic uniform dynamics. 
We argue that in many cases, the physical picture developed here is more relevant.

\subsection{Cuprates}
As an example, we consider the relevant parameters for the cuprate La$_{1.675}$Eu$_{0.2}$Sr$_{0.125}$CuO$_4$ (LESCO$_{1/8}$), the first material found to exhibit such transient phenomenon. At $10 \unit{K}$ where this compound is not superconducting due to competition of charge order, Fausti et al \cite{Fausti2011} pumped the system with a strong infrared pulse. A metastable state appeared within picoseconds and lived for at least nanoseconds. Most strikingly, it exhibits superconducting like terahertz optical response.

In discussing the application of our theory, the first issue is timescales. The timescales associated with gap recovery in cuprate  superconductors are typically of the order of $\tau_{sc} \sim 1 \unit{ps} $ \cite{Smallwood2012}; similar timescales are reported in studies of transient enhancement of the photoresponse in LESCO$_{1/8}$ \cite{Fausti2011} and Y-Bi2212 \cite{Giusti2019}.  Time resolved x-ray and electron diffraction experiments 
reported CDW relaxation timescales in the wide range from 4 to 1000 ps
in Transition Metal Dichalcogenides (TMD) \cite{Eichberger2010, Erasmus2012} where the CDW order is coupled to the lattice. Timescales of only a few $ps$ were reported for the charge order in cuprates \cite{Hinton2013}, but the time scale for the stripe order that strongly competes with the superconductivity is not know, and may be long because the stripe order couples strongly to the lattice \cite{Fausti2011}. Assigning SC to order \RomanNumeralCaps{1} and CDW to order \RomanNumeralCaps{2}, we assume that $\tau_{sc} /\tau_{cdw} =1/3$ for LESCO$_{1/8}$ and $\Delta=\gamma_2\alpha_{2L}/\gamma_1\alpha_{1L} \approx 1/3$ since the $\alpha_{iL} \sim 1$ for both CDW and SC.

The next issue is the Ginsburg parameter $G$. The coherence lengths are of the order of a few $nm$ and Gaussian fluctuations are observed for temperatures within $10\%$ of $T_c$, so $G$ is unlikely to be as small as it is in conventional materials. If there were no competition to CDW, the superconducting $T_c$ of LESCO$_{1/8}$ is about $40 \unit{K}$, indicating a zero temperature superconducting gap $\sim 14 \unit{meV}$ which is perhaps a factor of $\sim 100$ less than the Fermi energy.  We suggest that $\zeta \sim G\sim 10^{-2}$ (lower panel of Fig.~\ref{fig:gaussian}) may be appropriate since the material is effectively two dimensional.  

\begin{figure}
	\includegraphics[width=1 \linewidth]{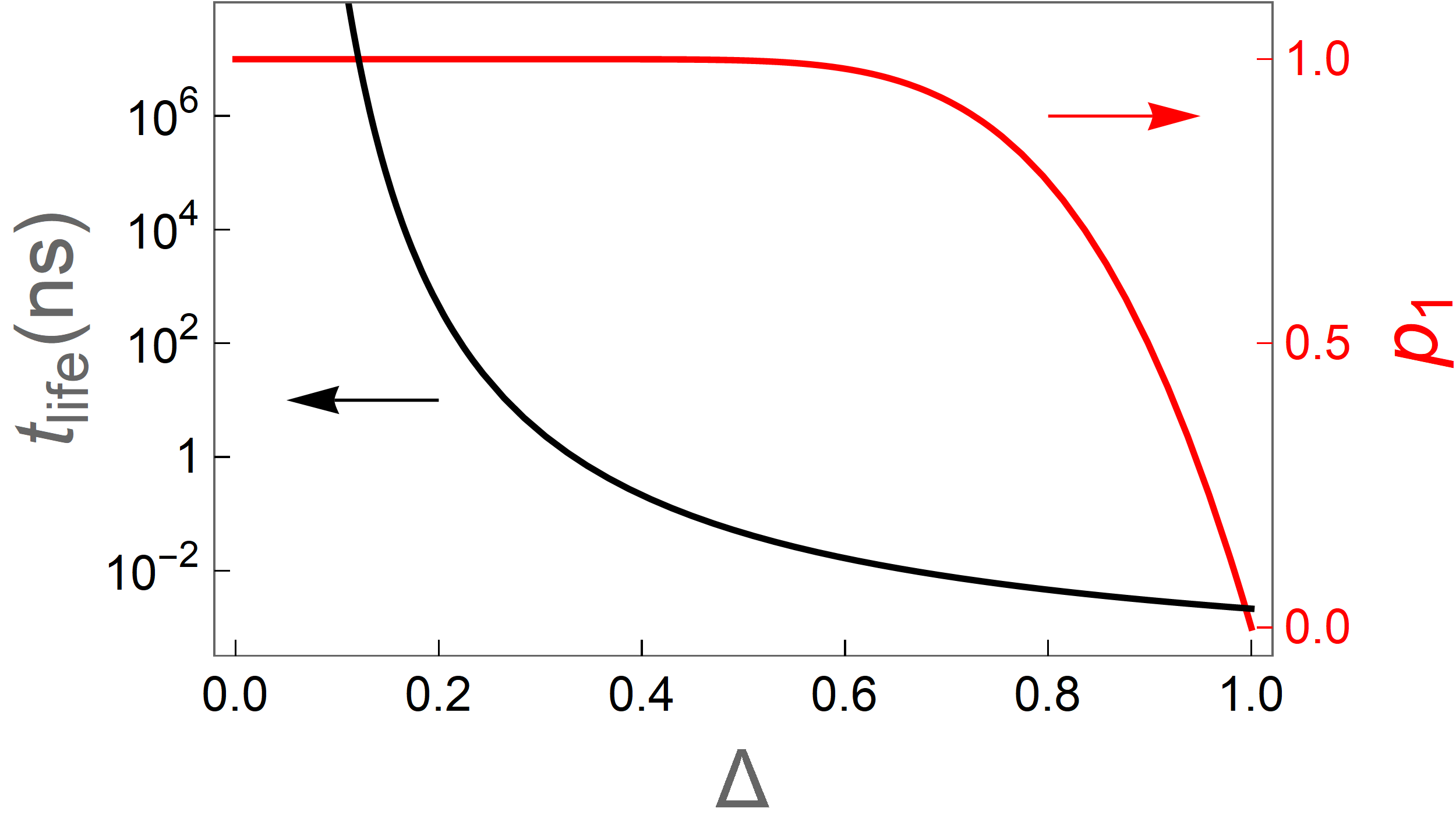}
	\caption{The volume fraction $p_1$ (red curve) of SC domains and the life time $t_{life}$ (black curve) of the metastable state as a function of $\Delta=\gamma_2\alpha_{2L}/\gamma_1\alpha_{1L}$ predicted by Eqs.~\eqref{eqn:p2_2} and \eqref{eqn:td_2}. The Ginzburg parameter is $\zeta=10^{-2}$. }
	\label{fig:p1_tlife}
\end{figure}

The experiment of Fausti et al \cite{Fausti2011} could then be interpreted as destruction of both orders followed by growth of fluctuations described in section~\ref{sec:fast_cooling}.  The pump along a-axis had a fluence of $1 \unit{mJ/cm^2}$, the penetration depth was about $200 \unit{nm}$ and almost all the photon energy was absorbed since the reflectivity at that frequency is nearly zero. Together with the electronic specific heat of $C_{el}/T \approx 3 \unit{mJ \, k^{-2} mol^{-1}}$  \cite{Michon2019} and the lattice constant of $a=3.8 \unit{\AA} ,\, c= 13.2 \unit{\AA}$ \cite{Radaelli.1994}, we estimate that the electronic system was transiently heated up to $T_H=2000 \unit{K}$, much larger than the critical temperatures of both CDW ($80\unit{K}$) and SC. Thus it is reasonable to assume all orders are destroyed by the pump. The cooling time scale \cite{Smallwood2012,Fausti2011} should not be significantly larger than $\tau_{sc}$, thus the equations in the fast cooling limit should give reasonable estimations. Application of \equa{eqn:p2_2} in 2D yields $p_2 \approx 5 \times 10^{-6}$ as the volume fraction of CDW domains in the transient state, meaning most volume is transformed to the SC state. \equa{eqn:td_2} predicts the lifetime of the metastable state to be about one nanosecond. These estimations qualitatively explain the phenomenon seen in LESCO$_{1/8}$ but note that they depend sensitively on the values of $G$ and $\Delta$, see Fig.~\ref{fig:p1_tlife} and Eqs.~\eqref{eqn:p2_2} and \eqref{eqn:td_2}.

%%%%%%%%%%%%%%%%%%%%%%%%%%%%%%%%%%%%%%%%%%%%%%%%%%%%%%
%%%%%%%%%%%%%%%%%%%%%%%%%%%%%%%%%%%%%%%%%%%%%%%%%%%%%%

The cuprate La$_{1.885}$Ba$_{0.115}$CuO$_4$ has a density wave transition at $T_{co}=53 \unit{K}$ followed by a second transition at $T_c=13 \unit{K}$ to a state with both density wave order and weak superconductivity. Cremin et al \cite{Cremin2019} recently reported that upon moderate near infrared ($1.55 \unit{eV}$) pump pulse (fluence of $0.1 \unit{mJ/cm^2}$) along c-axis, the weak superconducting state may be converted to a long-lived strong superconducting like state within picoseconds.  The key observation is that this state can be created only if the static system is in the weakly superconducting state below $T_c$ of bulk superconductivity. If the static temperature is even slightly above $T_c$, the strong superconducting like metastable state is not created. We interpret the result as suggesting that the equilibrium state is an intertwined state with both superconducting and charge order, and the pump  couples more strongly to the charge order while it is not strong enough to kill both orders. The transient phenomenon is due to the second mechanism described in Section~\ref{sec:intertwined}. The transition time is $t_s \sim \frac{1}{\alpha \gamma } \ln \frac{\psi_{2m}}{\psi_{2b}} \sim \unit{ps}$ if one uses $\gamma \sim 1 \unit{ps^{-1}}$. As the static temperature gets close to $T_c$, the transition time diverges logarithmically since $\psi_{2b}$ approaches zero, explaining the key observation.

\subsection{Transition metal tri-tellurides}
Competing phases occur in non-superconducting contexts. In equilibrium, LaTe$_3$ and CeTe$_3$ exhibit long ranged CDW order (denoted as c-CDW). However, when these systems are driven out of equilibrium by a sufficiently strong near infrared pump, a different CDW order (denoted as a-CDW), distinguished from the equilibrium one by the wavevector, appears \cite{Kogar2019,Zhou2019}. This was interpreted as a-CDW states living in topological vortices created in the dominant cCDW state by the Kibble-Zurek mechanism. However, in the experiments by Kogar et al \cite{Kogar2019}, stronger a-CDW signal was observed for a stronger pump while the Kibble-Zurek mechanism says the number of vortices depends only on the cooling time constant through $T_c$, a system parameter that does not quite depend on pump fluence.  Our framework in section \ref{sec:fast_cooling} is an alternative explanation. Assume that the pump destroys the mean field order parameter of the c-CDW but not to zero, its recovery would suppress the growth of fluctuations of a-CDW. Therefore, a stronger pump would suppress c-CDW to a smaller value which gives more space for a-CDW fluctuations to grow. A quantitative application of our theory requires more information on the relaxation rates. Since both  orders are CDWs in this case, their time scales should be comparable and we place LaTe$_3$/CeTe$_3$ close to $\Delta=1$ in Fig.~\ref{fig:criterions}.

\subsection{Manganites}
Zhang and McLeod et al have reported that in charge ordered insulating films made of La$_{2/3}$Ca$_{1/3}$MnO$_3$ \cite{Zhang2016,mcleod.2019}, exposure of pump radiation can create domains of ferromagnetic metallic order, which grow in size with successive pump pulses and at low temperature, do not revert to the ground state on measureable time scales. Analysis of these experiments requires extension of our theory to the case of first order free energy landscapes.

\section{Discussion}
\label{sec:discussion}
We presented a dynamical phase transition theory of a pumped system with competing orders, based on a Landau theory of non conserved order parameters with relaxational dynamics coupled to a quasi thermal bath. We focused on the case where an applied (``pump'') field changes the free energy landscape, thereby
driving any mean field order parameters to vanishingly small values and studied in detail the growth of fluctuations after the pump is removed. We presented a general treatment valid for cooling rates that are fast or slow compared to the basic order parameter time scales, presented a scaling theory valid in the slow cooling limit, and connected our results to the Kibble-Zurek theory of systems quenched through a critical point. 

We computed the probability distribution of order parameter fluctuations, identified the exponential growth of long wave length fluctuations characterized by a universal correlation length growth $\xi\sim \sqrt{8\gamma t}$, leading to a large domain structure. We showed that in physically reasonable cases, a modestly larger relaxation rate of the subdominant order can lead the system to a metastable state of domains, most of which are in this subdominant phase. We derived scaling functions for the volume fraction of different domains and the lifetime of the metastable state. Due to universality of the Landau theory, it applies to solid state systems with competing orders \cite{Fausti2011,Nicoletti2014,Zhang2018,Zhang2018a,Nicoletti2018,Cremin2019,Niwa.2019,Zong2019,Suzuki2019, Kogar2019,Giusti2019,Eichberger2010, Erasmus2012,Hinton2013}, cold atoms \cite{Sanchez.2018} and even the early universe \cite{Kibble1976}. 

The fluctuation theory developed here naturally explains the key features of the observations of metastable states \cite{Fausti2011,Suzuki2019}: (a) the long life time ($>200 \unit{ps}$) of the superconducting like state; (b) the finite frequency conductivity peaks \cite{Nicoletti2014,Hunt.2015} maybe explained by the transiently created domain structure (Fig.~\ref{fig:quartic_free_energy}(b)) which allows coupling of far field radiation to plasmonic modes. 
A further proof of consistency is that  similar transient phenomenon are observed for both infrared and optical pumping, suggesting that the main effect of the pump is incoherent, related to heating of the microscopic degrees of freedom.

Next we discuss the assumptions underlying our approach. (a) We assume the existence of a well-defined, time-dependent temperature T(t) for the high energy electronic degrees of freedom throughout the full time evolution. Indeed, quasiparticle thermalization time ($\sim 10 \unit{fs}$) is usually much shorter than the dynamics of the collective order parameter ($0.1 \sim 1 \unit{ps}$). See, e.g., Ref.~\cite{He.2016}. Therefore, for the slow dynamics of the order parameter we consider,  it is legitimate to assume a well defined temperature for high energy degrees of freedom which acts as the bath for the order parameters. 
(b) Our theory relies on a small Ginzburg parameter $G$, which is the same parameter that controls the validity of mean field theory. Thus our theory applies wherever mean field does, e.g., for SC and CDW where $ \mathrm{gap}/\mathrm{fermi \,\, energy} \ll 1$.  Although the application of mean field arguments to strongly correlated systems such as cuprates maybe questionable, our estimation using $G\sim 10^{-2}$ still renders a trapping probability close to unity. (c) We worked in the context of Ginzburg-Landau theory with relaxational dynamics which holds close to $T_c$ \cite{Lemonik.2017}  or for superconductors rendered gapless by magnetic impurities \cite{Gorkov1968}. However, relaxational dynamics is not essential to our conclusions. If a second order time derivative is added to \equa{eqn:TDGL}, it does not change the picture of exponential growth of the long wave length thermal fluctuations, as long as there is substantial damping to take away the energy. Extension of our approach to the under damped (Hamiltonian dynamics) case is of interest.
(d) For the trapping into the metastable order \RomanNumeralCaps{1}, we need it to relax faster than the equilibrium order \RomanNumeralCaps{2}. This probably happens for competing SC and CDW orders since the latter often couples to the lattice which is heavy.  Since the trapping probability crosses to unity exponentially as $\Delta$ crosses one, order \RomanNumeralCaps{1} just needs to be moderately faster than order \RomanNumeralCaps{2}, as shown by Fig.~\ref{fig:p1_tlife} where the crossing is quick already for a relatively large $G$. 
(e) We assumed the white noise in \equa{etacorrelator} to characterize thermal fluctuations. The underlying assumption is that there is a length scale separation between the long length scale physics of the order parameter dynamics and the presumably short length scale physics of the microscopic degrees of freedom that are integrated out to obtain the order parameter theory. If the microscopic degrees of freedom have a length scale comparable to order parameter length scales, then the partial differential equations we analyse should be replaced by integro-differential equations. Analysis of this situation is beyond the scope of our paper, but we believe that it would not change our main conclusions.

Our work defines directions for future research. On the theoretical side, detailed application of our theory to specific materials, extension to the case of conserved order parameters, to strongly interacting field theories ($G\sim 1$) and to the case of interfaces between domains with different orders \cite{Lorenzo.2016} are all of interest.
Similar conclusions are expected for quenching through a quantum critical point at zero temperature. Instead of thermal fluctuations, quantum fluctuations will be exponentially amplified. Also, generalization of our formalism to the case where the pump couples coherently to the order parameters, as  would happen for Terahertz pumps is of interest. 
On the experimental side, the growing fluctuations and the induced domain structure (Fig.~\ref{fig:quartic_free_energy}(b)) or the topological vortices can be ideally probed by time-space resolved techniques, e.g., ultra fast Terahertz near field microscopy or ultra fast scanning tunneling microscopy. Moreover, the growing SC fluctuation and thus superfluid density indicates increasing Drude weight in the non equilibrium optical conductivity, leading to novel effects on the collective modes \cite{Sun.2016} and THz reflectivity.

\begin{acknowledgements}
We acknowledge support from  the Department of Energy under Grant DE-SC0018218. We thank M. M. Fogler, R. D. Averitt, D. N. Basov, M. Eckstein, W. Yang, D. Golez and Y. He for helpful discussions. 
\end{acknowledgements}

\emph{Note added.---}A very recent paper \cite{dolgirev.2019} uses similar methods to study the post quench growth of order parameter phase fluctuations.
%%%%%%%%%%%%%%%%%%%%%%%%%%%%%%%%%%%%%%%%%%%%%
%\bibliographystyle{apsrev4-2}
\bibliography{Superconductivity}

\begin{thebibliography}{60}%
\makeatletter
\providecommand \@ifxundefined [1]{%
 \@ifx{#1\undefined}
}%
\providecommand \@ifnum [1]{%
 \ifnum #1\expandafter \@firstoftwo
 \else \expandafter \@secondoftwo
 \fi
}%
\providecommand \@ifx [1]{%
 \ifx #1\expandafter \@firstoftwo
 \else \expandafter \@secondoftwo
 \fi
}%
\providecommand \natexlab [1]{#1}%
\providecommand \enquote  [1]{``#1''}%
\providecommand \bibnamefont  [1]{#1}%
\providecommand \bibfnamefont [1]{#1}%
\providecommand \citenamefont [1]{#1}%
\providecommand \href@noop [0]{\@secondoftwo}%
\providecommand \href [0]{\begingroup \@sanitize@url \@href}%
\providecommand \@href[1]{\@@startlink{#1}\@@href}%
\providecommand \@@href[1]{\endgroup#1\@@endlink}%
\providecommand \@sanitize@url [0]{\catcode `\\12\catcode `\$12\catcode
  `\&12\catcode `\#12\catcode `\^12\catcode `\_12\catcode `\%12\relax}%
\providecommand \@@startlink[1]{}%
\providecommand \@@endlink[0]{}%
\providecommand \url  [0]{\begingroup\@sanitize@url \@url }%
\providecommand \@url [1]{\endgroup\@href {#1}{\urlprefix }}%
\providecommand \urlprefix  [0]{URL }%
\providecommand \Eprint [0]{\href }%
\providecommand \doibase [0]{http://dx.doi.org/}%
\providecommand \selectlanguage [0]{\@gobble}%
\providecommand \bibinfo  [0]{\@secondoftwo}%
\providecommand \bibfield  [0]{\@secondoftwo}%
\providecommand \translation [1]{[#1]}%
\providecommand \BibitemOpen [0]{}%
\providecommand \bibitemStop [0]{}%
\providecommand \bibitemNoStop [0]{.\EOS\space}%
\providecommand \EOS [0]{\spacefactor3000\relax}%
\providecommand \BibitemShut  [1]{\csname bibitem#1\endcsname}%
\let\auto@bib@innerbib\@empty
%</preamble>
\bibitem [{\citenamefont {Hohenberg}\ and\ \citenamefont
  {Halperin}(1977)}]{Hohenberg1977}%
  \BibitemOpen
  \bibfield  {author} {\bibinfo {author} {\bibfnamefont {P.~C.}\ \bibnamefont
  {Hohenberg}}\ and\ \bibinfo {author} {\bibfnamefont {B.~I.}\ \bibnamefont
  {Halperin}},\ }\href {\doibase 10.1103/RevModPhys.49.435} {\bibfield
  {journal} {\bibinfo  {journal} {Rev. Mod. Phys.}\ }\textbf {\bibinfo {volume}
  {49}},\ \bibinfo {pages} {435} (\bibinfo {year} {1977})}\BibitemShut
  {NoStop}%
\bibitem [{\citenamefont {Polkovnikov}\ \emph {et~al.}(2011)\citenamefont
  {Polkovnikov}, \citenamefont {Sengupta}, \citenamefont {Silva},\ and\
  \citenamefont {Vengalattore}}]{Polkovnikov.2011}%
  \BibitemOpen
  \bibfield  {author} {\bibinfo {author} {\bibfnamefont {A.}~\bibnamefont
  {Polkovnikov}}, \bibinfo {author} {\bibfnamefont {K.}~\bibnamefont
  {Sengupta}}, \bibinfo {author} {\bibfnamefont {A.}~\bibnamefont {Silva}}, \
  and\ \bibinfo {author} {\bibfnamefont {M.}~\bibnamefont {Vengalattore}},\
  }\href {\doibase 10.1103/RevModPhys.83.863} {\bibfield  {journal} {\bibinfo
  {journal} {Rev. Mod. Phys.}\ }\textbf {\bibinfo {volume} {83}},\ \bibinfo
  {pages} {863} (\bibinfo {year} {2011})}\BibitemShut {NoStop}%
\bibitem [{\citenamefont {Bray}(1994)}]{Bray.1994}%
  \BibitemOpen
  \bibfield  {author} {\bibinfo {author} {\bibfnamefont {A.~J.}\ \bibnamefont
  {Bray}},\ }\href {\doibase 10.1080/00018739400101505} {\bibfield  {journal}
  {\bibinfo  {journal} {Advances in Physics}\ }\textbf {\bibinfo {volume}
  {43}},\ \bibinfo {pages} {357} (\bibinfo {year} {1994})}\BibitemShut
  {NoStop}%
\bibitem [{\citenamefont {Kibble}(1976)}]{Kibble1976}%
  \BibitemOpen
  \bibfield  {author} {\bibinfo {author} {\bibfnamefont {T.~W.~B.}\
  \bibnamefont {Kibble}},\ }\href {http://stacks.iop.org/0305-4470/9/i=8/a=029}
  {\bibfield  {journal} {\bibinfo  {journal} {J. Phys. A}\ }\textbf {\bibinfo
  {volume} {9}},\ \bibinfo {pages} {1387} (\bibinfo {year} {1976})}\BibitemShut
  {NoStop}%
\bibitem [{\citenamefont {Langer}\ \emph {et~al.}(1975)\citenamefont {Langer},
  \citenamefont {Bar-on},\ and\ \citenamefont {Miller}}]{Langer1975}%
  \BibitemOpen
  \bibfield  {author} {\bibinfo {author} {\bibfnamefont {J.~S.}\ \bibnamefont
  {Langer}}, \bibinfo {author} {\bibfnamefont {M.}~\bibnamefont {Bar-on}}, \
  and\ \bibinfo {author} {\bibfnamefont {H.~D.}\ \bibnamefont {Miller}},\
  }\href {\doibase 10.1103/PhysRevA.11.1417} {\bibfield  {journal} {\bibinfo
  {journal} {Phys. Rev. A}\ }\textbf {\bibinfo {volume} {11}},\ \bibinfo
  {pages} {1417} (\bibinfo {year} {1975})}\BibitemShut {NoStop}%
\bibitem [{\citenamefont {Binder}(1987)}]{Binder1987}%
  \BibitemOpen
  \bibfield  {author} {\bibinfo {author} {\bibfnamefont {K.}~\bibnamefont
  {Binder}},\ }\href {\doibase 10.1088/0034-4885/50/7/001} {\bibfield
  {journal} {\bibinfo  {journal} {Reports Prog. Phys.}\ }\textbf {\bibinfo
  {volume} {50}},\ \bibinfo {pages} {783} (\bibinfo {year} {1987})}\BibitemShut
  {NoStop}%
\bibitem [{\citenamefont {Carter}\ and\ \citenamefont
  {Johnson}(1998)}]{Carter1998}%
  \BibitemOpen
  \bibinfo {editor} {\bibfnamefont {W.~C.}\ \bibnamefont {Carter}}\ and\
  \bibinfo {editor} {\bibfnamefont {W.~C.}\ \bibnamefont {Johnson}},\ eds.,\
  \href {\doibase 10.1002/9781118788295} {\emph {\bibinfo {title} {{The
  Selected Works of John W. Cahn}}}}\ (\bibinfo  {publisher} {John Wiley {\&}
  Sons, Inc.},\ \bibinfo {address} {Hoboken, NJ, USA},\ \bibinfo {year}
  {1998})\BibitemShut {NoStop}%
\bibitem [{\citenamefont {Alert}\ \emph {et~al.}(2016)\citenamefont {Alert},
  \citenamefont {Tierno},\ and\ \citenamefont {Casademunt}}]{Tierno2016}%
  \BibitemOpen
  \bibfield  {author} {\bibinfo {author} {\bibfnamefont {R.}~\bibnamefont
  {Alert}}, \bibinfo {author} {\bibfnamefont {P.}~\bibnamefont {Tierno}}, \
  and\ \bibinfo {author} {\bibfnamefont {J.}~\bibnamefont {Casademunt}},\
  }\href {https://doi.org/10.1038/ncomms13067} {\bibfield  {journal} {\bibinfo
  {journal} {Nature Communications}\ }\textbf {\bibinfo {volume} {7}},\
  \bibinfo {pages} {13067} (\bibinfo {year} {2016})}\BibitemShut {NoStop}%
\bibitem [{\citenamefont {Zurek}(1996)}]{Zurek.1996}%
  \BibitemOpen
  \bibfield  {author} {\bibinfo {author} {\bibfnamefont {W.~H.}\ \bibnamefont
  {Zurek}},\ }\href {\doibase 10.1016/S0370-1573(96)00009-9} {\bibfield
  {journal} {\bibinfo  {journal} {Phys. Rep.}\ }\textbf {\bibinfo {volume}
  {276}},\ \bibinfo {pages} {177} (\bibinfo {year} {1996})},\ \Eprint
  {http://arxiv.org/abs/9607135} {9607135} \BibitemShut {NoStop}%
\bibitem [{\citenamefont {Zurek}(1985)}]{Zurek1985}%
  \BibitemOpen
  \bibfield  {author} {\bibinfo {author} {\bibfnamefont {W.~H.}\ \bibnamefont
  {Zurek}},\ }\href {\doibase 10.1038/317505a0} {\bibfield  {journal} {\bibinfo
   {journal} {Nature}\ }\textbf {\bibinfo {volume} {317}},\ \bibinfo {pages}
  {505} (\bibinfo {year} {1985})}\BibitemShut {NoStop}%
\bibitem [{\citenamefont {Biroli}\ \emph {et~al.}(2010)\citenamefont {Biroli},
  \citenamefont {Cugliandolo},\ and\ \citenamefont {Sicilia}}]{Biroli.2010}%
  \BibitemOpen
  \bibfield  {author} {\bibinfo {author} {\bibfnamefont {G.}~\bibnamefont
  {Biroli}}, \bibinfo {author} {\bibfnamefont {L.~F.}\ \bibnamefont
  {Cugliandolo}}, \ and\ \bibinfo {author} {\bibfnamefont {A.}~\bibnamefont
  {Sicilia}},\ }\href {\doibase 10.1103/PhysRevE.81.050101} {\bibfield
  {journal} {\bibinfo  {journal} {Phys. Rev. E}\ }\textbf {\bibinfo {volume}
  {81}},\ \bibinfo {pages} {050101} (\bibinfo {year} {2010})}\BibitemShut
  {NoStop}%
\bibitem [{\citenamefont {Chandran}\ \emph {et~al.}(2012)\citenamefont
  {Chandran}, \citenamefont {Erez}, \citenamefont {Gubser},\ and\ \citenamefont
  {Sondhi}}]{Chandran.2012}%
  \BibitemOpen
  \bibfield  {author} {\bibinfo {author} {\bibfnamefont {A.}~\bibnamefont
  {Chandran}}, \bibinfo {author} {\bibfnamefont {A.}~\bibnamefont {Erez}},
  \bibinfo {author} {\bibfnamefont {S.~S.}\ \bibnamefont {Gubser}}, \ and\
  \bibinfo {author} {\bibfnamefont {S.~L.}\ \bibnamefont {Sondhi}},\ }\href
  {\doibase 10.1103/PhysRevB.86.064304} {\bibfield  {journal} {\bibinfo
  {journal} {Phys. Rev. B}\ }\textbf {\bibinfo {volume} {86}},\ \bibinfo
  {pages} {064304} (\bibinfo {year} {2012})}\BibitemShut {NoStop}%
\bibitem [{\citenamefont {Gr{\"{u}}ner}(1988)}]{Gruner.1988}%
  \BibitemOpen
  \bibfield  {author} {\bibinfo {author} {\bibfnamefont {G.}~\bibnamefont
  {Gr{\"{u}}ner}},\ }\href {\doibase 10.1103/RevModPhys.60.1129} {\bibfield
  {journal} {\bibinfo  {journal} {Rev. Mod. Phys.}\ }\textbf {\bibinfo {volume}
  {60}},\ \bibinfo {pages} {1129} (\bibinfo {year} {1988})}\BibitemShut
  {NoStop}%
\bibitem [{\citenamefont {Fradkin}\ \emph {et~al.}(2015)\citenamefont
  {Fradkin}, \citenamefont {Kivelson},\ and\ \citenamefont
  {Tranquada}}]{Fradkin2015}%
  \BibitemOpen
  \bibfield  {author} {\bibinfo {author} {\bibfnamefont {E.}~\bibnamefont
  {Fradkin}}, \bibinfo {author} {\bibfnamefont {S.~A.}\ \bibnamefont
  {Kivelson}}, \ and\ \bibinfo {author} {\bibfnamefont {J.~M.}\ \bibnamefont
  {Tranquada}},\ }\href {\doibase 10.1103/RevModPhys.87.457} {\bibfield
  {journal} {\bibinfo  {journal} {Rev. Mod. Phys.}\ }\textbf {\bibinfo {volume}
  {87}},\ \bibinfo {pages} {457} (\bibinfo {year} {2015})}\BibitemShut
  {NoStop}%
\bibitem [{\citenamefont {Tokura}(2006)}]{Tokura2006}%
  \BibitemOpen
  \bibfield  {author} {\bibinfo {author} {\bibfnamefont {Y.}~\bibnamefont
  {Tokura}},\ }\href {\doibase 10.1088/0034-4885/69/3/r06} {\bibfield
  {journal} {\bibinfo  {journal} {Reports on Progress in Physics}\ }\textbf
  {\bibinfo {volume} {69}},\ \bibinfo {pages} {797} (\bibinfo {year}
  {2006})}\BibitemShut {NoStop}%
\bibitem [{\citenamefont {Zhang}\ \emph {et~al.}(2016)\citenamefont {Zhang},
  \citenamefont {Tan}, \citenamefont {Liu}, \citenamefont {Teitelbaum},
  \citenamefont {Post}, \citenamefont {Jin}, \citenamefont {Nelson},
  \citenamefont {Basov}, \citenamefont {Wu},\ and\ \citenamefont
  {Averitt}}]{Zhang2016}%
  \BibitemOpen
  \bibfield  {author} {\bibinfo {author} {\bibfnamefont {J.}~\bibnamefont
  {Zhang}}, \bibinfo {author} {\bibfnamefont {X.}~\bibnamefont {Tan}}, \bibinfo
  {author} {\bibfnamefont {M.}~\bibnamefont {Liu}}, \bibinfo {author}
  {\bibfnamefont {S.~W.}\ \bibnamefont {Teitelbaum}}, \bibinfo {author}
  {\bibfnamefont {K.~W.}\ \bibnamefont {Post}}, \bibinfo {author}
  {\bibfnamefont {F.}~\bibnamefont {Jin}}, \bibinfo {author} {\bibfnamefont
  {K.~A.}\ \bibnamefont {Nelson}}, \bibinfo {author} {\bibfnamefont {D.~N.}\
  \bibnamefont {Basov}}, \bibinfo {author} {\bibfnamefont {W.}~\bibnamefont
  {Wu}}, \ and\ \bibinfo {author} {\bibfnamefont {R.~D.}\ \bibnamefont
  {Averitt}},\ }\href {\doibase 10.1038/nmat4695} {\bibfield  {journal}
  {\bibinfo  {journal} {Nat. Mater.}\ }\textbf {\bibinfo {volume} {15}},\
  \bibinfo {pages} {956} (\bibinfo {year} {2016})}\BibitemShut {NoStop}%
\bibitem [{\citenamefont {Basov}\ \emph {et~al.}(2011)\citenamefont {Basov},
  \citenamefont {Averitt}, \citenamefont {van~der Marel}, \citenamefont
  {Dressel},\ and\ \citenamefont {Haule}}]{Basov2011}%
  \BibitemOpen
  \bibfield  {author} {\bibinfo {author} {\bibfnamefont {D.~N.}\ \bibnamefont
  {Basov}}, \bibinfo {author} {\bibfnamefont {R.~D.}\ \bibnamefont {Averitt}},
  \bibinfo {author} {\bibfnamefont {D.}~\bibnamefont {van~der Marel}}, \bibinfo
  {author} {\bibfnamefont {M.}~\bibnamefont {Dressel}}, \ and\ \bibinfo
  {author} {\bibfnamefont {K.}~\bibnamefont {Haule}},\ }\href {\doibase
  10.1103/RevModPhys.83.471} {\bibfield  {journal} {\bibinfo  {journal} {Rev.
  Mod. Phys.}\ }\textbf {\bibinfo {volume} {83}},\ \bibinfo {pages} {471}
  (\bibinfo {year} {2011})}\BibitemShut {NoStop}%
\bibitem [{\citenamefont {Fausti}\ \emph {et~al.}(2011)\citenamefont {Fausti},
  \citenamefont {Tobey}, \citenamefont {Dean}, \citenamefont {Kaiser},
  \citenamefont {Dienst}, \citenamefont {Hoffmann}, \citenamefont {Pyon},
  \citenamefont {Takayama}, \citenamefont {Takagi},\ and\ \citenamefont
  {Cavalleri}}]{Fausti2011}%
  \BibitemOpen
  \bibfield  {author} {\bibinfo {author} {\bibfnamefont {D.}~\bibnamefont
  {Fausti}}, \bibinfo {author} {\bibfnamefont {R.~I.}\ \bibnamefont {Tobey}},
  \bibinfo {author} {\bibfnamefont {N.}~\bibnamefont {Dean}}, \bibinfo {author}
  {\bibfnamefont {S.}~\bibnamefont {Kaiser}}, \bibinfo {author} {\bibfnamefont
  {A.}~\bibnamefont {Dienst}}, \bibinfo {author} {\bibfnamefont {M.~C.}\
  \bibnamefont {Hoffmann}}, \bibinfo {author} {\bibfnamefont {S.}~\bibnamefont
  {Pyon}}, \bibinfo {author} {\bibfnamefont {T.}~\bibnamefont {Takayama}},
  \bibinfo {author} {\bibfnamefont {H.}~\bibnamefont {Takagi}}, \ and\ \bibinfo
  {author} {\bibfnamefont {A.}~\bibnamefont {Cavalleri}},\ }\href {\doibase
  10.1126/science.1197294} {\bibfield  {journal} {\bibinfo  {journal}
  {Science}\ }\textbf {\bibinfo {volume} {331}},\ \bibinfo {pages} {189}
  (\bibinfo {year} {2011})}\BibitemShut {NoStop}%
\bibitem [{\citenamefont {Nicoletti}\ \emph {et~al.}(2014)\citenamefont
  {Nicoletti}, \citenamefont {Casandruc}, \citenamefont {Laplace},
  \citenamefont {Khanna}, \citenamefont {Hunt}, \citenamefont {Kaiser},
  \citenamefont {Dhesi}, \citenamefont {Gu}, \citenamefont {Hill},\ and\
  \citenamefont {Cavalleri}}]{Nicoletti2014}%
  \BibitemOpen
  \bibfield  {author} {\bibinfo {author} {\bibfnamefont {D.}~\bibnamefont
  {Nicoletti}}, \bibinfo {author} {\bibfnamefont {E.}~\bibnamefont
  {Casandruc}}, \bibinfo {author} {\bibfnamefont {Y.}~\bibnamefont {Laplace}},
  \bibinfo {author} {\bibfnamefont {V.}~\bibnamefont {Khanna}}, \bibinfo
  {author} {\bibfnamefont {C.~R.}\ \bibnamefont {Hunt}}, \bibinfo {author}
  {\bibfnamefont {S.}~\bibnamefont {Kaiser}}, \bibinfo {author} {\bibfnamefont
  {S.~S.}\ \bibnamefont {Dhesi}}, \bibinfo {author} {\bibfnamefont {G.~D.}\
  \bibnamefont {Gu}}, \bibinfo {author} {\bibfnamefont {J.~P.}\ \bibnamefont
  {Hill}}, \ and\ \bibinfo {author} {\bibfnamefont {A.}~\bibnamefont
  {Cavalleri}},\ }\href {\doibase 10.1103/PhysRevB.90.100503} {\bibfield
  {journal} {\bibinfo  {journal} {Phys. Rev. B}\ }\textbf {\bibinfo {volume}
  {90}},\ \bibinfo {pages} {100503(R)} (\bibinfo {year} {2014})}\BibitemShut
  {NoStop}%
\bibitem [{\citenamefont {Zhang}\ \emph
  {et~al.}(2018{\natexlab{a}})\citenamefont {Zhang}, \citenamefont {Wang},
  \citenamefont {Wu}, \citenamefont {Liu}, \citenamefont {Shi}, \citenamefont
  {Lin}, \citenamefont {Li}, \citenamefont {Dai}, \citenamefont {Dong},\ and\
  \citenamefont {Wang}}]{Zhang2018}%
  \BibitemOpen
  \bibfield  {author} {\bibinfo {author} {\bibfnamefont {S.~J.}\ \bibnamefont
  {Zhang}}, \bibinfo {author} {\bibfnamefont {Z.~X.}\ \bibnamefont {Wang}},
  \bibinfo {author} {\bibfnamefont {D.}~\bibnamefont {Wu}}, \bibinfo {author}
  {\bibfnamefont {Q.~M.}\ \bibnamefont {Liu}}, \bibinfo {author} {\bibfnamefont
  {L.~Y.}\ \bibnamefont {Shi}}, \bibinfo {author} {\bibfnamefont
  {T.}~\bibnamefont {Lin}}, \bibinfo {author} {\bibfnamefont {S.~L.}\
  \bibnamefont {Li}}, \bibinfo {author} {\bibfnamefont {P.~C.}\ \bibnamefont
  {Dai}}, \bibinfo {author} {\bibfnamefont {T.}~\bibnamefont {Dong}}, \ and\
  \bibinfo {author} {\bibfnamefont {N.~L.}\ \bibnamefont {Wang}},\ }\href
  {\doibase 10.1103/PhysRevB.98.224507} {\bibfield  {journal} {\bibinfo
  {journal} {Phys. Rev. B}\ }\textbf {\bibinfo {volume} {98}},\ \bibinfo
  {pages} {224507} (\bibinfo {year} {2018}{\natexlab{a}})}\BibitemShut
  {NoStop}%
\bibitem [{\citenamefont {Zhang}\ \emph
  {et~al.}(2018{\natexlab{b}})\citenamefont {Zhang}, \citenamefont {Wang},
  \citenamefont {Shi}, \citenamefont {Lin}, \citenamefont {Zhang},
  \citenamefont {Gu}, \citenamefont {Dong},\ and\ \citenamefont
  {Wang}}]{Zhang2018a}%
  \BibitemOpen
  \bibfield  {author} {\bibinfo {author} {\bibfnamefont {S.~J.}\ \bibnamefont
  {Zhang}}, \bibinfo {author} {\bibfnamefont {Z.~X.}\ \bibnamefont {Wang}},
  \bibinfo {author} {\bibfnamefont {L.~Y.}\ \bibnamefont {Shi}}, \bibinfo
  {author} {\bibfnamefont {T.}~\bibnamefont {Lin}}, \bibinfo {author}
  {\bibfnamefont {M.~Y.}\ \bibnamefont {Zhang}}, \bibinfo {author}
  {\bibfnamefont {G.~D.}\ \bibnamefont {Gu}}, \bibinfo {author} {\bibfnamefont
  {T.}~\bibnamefont {Dong}}, \ and\ \bibinfo {author} {\bibfnamefont {N.~L.}\
  \bibnamefont {Wang}},\ }\href {\doibase 10.1103/PhysRevB.98.020506}
  {\bibfield  {journal} {\bibinfo  {journal} {Phys. Rev. B}\ }\textbf {\bibinfo
  {volume} {98}},\ \bibinfo {pages} {020506(R)} (\bibinfo {year}
  {2018}{\natexlab{b}})}\BibitemShut {NoStop}%
\bibitem [{\citenamefont {Nicoletti}\ \emph {et~al.}(2018)\citenamefont
  {Nicoletti}, \citenamefont {Fu}, \citenamefont {Mehio}, \citenamefont
  {Moore}, \citenamefont {Disa}, \citenamefont {Gu},\ and\ \citenamefont
  {Cavalleri}}]{Nicoletti2018}%
  \BibitemOpen
  \bibfield  {author} {\bibinfo {author} {\bibfnamefont {D.}~\bibnamefont
  {Nicoletti}}, \bibinfo {author} {\bibfnamefont {D.}~\bibnamefont {Fu}},
  \bibinfo {author} {\bibfnamefont {O.}~\bibnamefont {Mehio}}, \bibinfo
  {author} {\bibfnamefont {S.}~\bibnamefont {Moore}}, \bibinfo {author}
  {\bibfnamefont {A.~S.}\ \bibnamefont {Disa}}, \bibinfo {author}
  {\bibfnamefont {G.~D.}\ \bibnamefont {Gu}}, \ and\ \bibinfo {author}
  {\bibfnamefont {A.}~\bibnamefont {Cavalleri}},\ }\href {\doibase
  10.1103/PhysRevLett.121.267003} {\bibfield  {journal} {\bibinfo  {journal}
  {Phys. Rev. Lett.}\ }\textbf {\bibinfo {volume} {121}},\ \bibinfo {pages}
  {267003} (\bibinfo {year} {2018})}\BibitemShut {NoStop}%
\bibitem [{\citenamefont {Cremin}\ \emph {et~al.}(2019)\citenamefont {Cremin},
  \citenamefont {Zhang}, \citenamefont {Homes}, \citenamefont {Gu},
  \citenamefont {Sun}, \citenamefont {Fogler}, \citenamefont {Millis},
  \citenamefont {Basov},\ and\ \citenamefont {Averitt}}]{Cremin2019}%
  \BibitemOpen
  \bibfield  {author} {\bibinfo {author} {\bibfnamefont {K.~A.}\ \bibnamefont
  {Cremin}}, \bibinfo {author} {\bibfnamefont {J.}~\bibnamefont {Zhang}},
  \bibinfo {author} {\bibfnamefont {C.~C.}\ \bibnamefont {Homes}}, \bibinfo
  {author} {\bibfnamefont {G.~D.}\ \bibnamefont {Gu}}, \bibinfo {author}
  {\bibfnamefont {Z.}~\bibnamefont {Sun}}, \bibinfo {author} {\bibfnamefont
  {M.~M.}\ \bibnamefont {Fogler}}, \bibinfo {author} {\bibfnamefont {A.~J.}\
  \bibnamefont {Millis}}, \bibinfo {author} {\bibfnamefont {D.~N.}\
  \bibnamefont {Basov}}, \ and\ \bibinfo {author} {\bibfnamefont {R.~D.}\
  \bibnamefont {Averitt}},\ }\href {\doibase 10.1073/pnas.1908368116}
  {\bibfield  {journal} {\bibinfo  {journal} {Proceedings of the National
  Academy of Sciences}\ }\textbf {\bibinfo {volume} {116}},\ \bibinfo {pages}
  {19875} (\bibinfo {year} {2019})}\BibitemShut {NoStop}%
\bibitem [{\citenamefont {Niwa}\ \emph {et~al.}(2019)\citenamefont {Niwa},
  \citenamefont {Yoshikawa}, \citenamefont {Tomari}, \citenamefont {Matsunaga},
  \citenamefont {Song}, \citenamefont {Eisaki},\ and\ \citenamefont
  {Shimano}}]{Niwa.2019}%
  \BibitemOpen
  \bibfield  {author} {\bibinfo {author} {\bibfnamefont {H.}~\bibnamefont
  {Niwa}}, \bibinfo {author} {\bibfnamefont {N.}~\bibnamefont {Yoshikawa}},
  \bibinfo {author} {\bibfnamefont {K.}~\bibnamefont {Tomari}}, \bibinfo
  {author} {\bibfnamefont {R.}~\bibnamefont {Matsunaga}}, \bibinfo {author}
  {\bibfnamefont {D.}~\bibnamefont {Song}}, \bibinfo {author} {\bibfnamefont
  {H.}~\bibnamefont {Eisaki}}, \ and\ \bibinfo {author} {\bibfnamefont
  {R.}~\bibnamefont {Shimano}},\ }\href {\doibase 10.1103/PhysRevB.100.104507}
  {\bibfield  {journal} {\bibinfo  {journal} {Phys. Rev. B}\ }\textbf {\bibinfo
  {volume} {100}},\ \bibinfo {pages} {104507} (\bibinfo {year}
  {2019})}\BibitemShut {NoStop}%
\bibitem [{\citenamefont {Zong}\ \emph {et~al.}(2019)\citenamefont {Zong},
  \citenamefont {Kogar}, \citenamefont {Bie}, \citenamefont {Rohwer},
  \citenamefont {Lee}, \citenamefont {Baldini}, \citenamefont {Erge{\c{c}}en},
  \citenamefont {Yilmaz}, \citenamefont {Freelon}, \citenamefont {Sie},
  \citenamefont {Zhou}, \citenamefont {Straquadine}, \citenamefont {Walmsley},
  \citenamefont {Dolgirev}, \citenamefont {Rozhkov}, \citenamefont {Fisher},
  \citenamefont {Jarillo-Herrero}, \citenamefont {Fine},\ and\ \citenamefont
  {Gedik}}]{Zong2019}%
  \BibitemOpen
  \bibfield  {author} {\bibinfo {author} {\bibfnamefont {A.}~\bibnamefont
  {Zong}}, \bibinfo {author} {\bibfnamefont {A.}~\bibnamefont {Kogar}},
  \bibinfo {author} {\bibfnamefont {Y.~Q.}\ \bibnamefont {Bie}}, \bibinfo
  {author} {\bibfnamefont {T.}~\bibnamefont {Rohwer}}, \bibinfo {author}
  {\bibfnamefont {C.}~\bibnamefont {Lee}}, \bibinfo {author} {\bibfnamefont
  {E.}~\bibnamefont {Baldini}}, \bibinfo {author} {\bibfnamefont
  {E.}~\bibnamefont {Erge{\c{c}}en}}, \bibinfo {author} {\bibfnamefont {M.~B.}\
  \bibnamefont {Yilmaz}}, \bibinfo {author} {\bibfnamefont {B.}~\bibnamefont
  {Freelon}}, \bibinfo {author} {\bibfnamefont {E.~J.}\ \bibnamefont {Sie}},
  \bibinfo {author} {\bibfnamefont {H.}~\bibnamefont {Zhou}}, \bibinfo {author}
  {\bibfnamefont {J.}~\bibnamefont {Straquadine}}, \bibinfo {author}
  {\bibfnamefont {P.}~\bibnamefont {Walmsley}}, \bibinfo {author}
  {\bibfnamefont {P.~E.}\ \bibnamefont {Dolgirev}}, \bibinfo {author}
  {\bibfnamefont {A.~V.}\ \bibnamefont {Rozhkov}}, \bibinfo {author}
  {\bibfnamefont {I.~R.}\ \bibnamefont {Fisher}}, \bibinfo {author}
  {\bibfnamefont {P.}~\bibnamefont {Jarillo-Herrero}}, \bibinfo {author}
  {\bibfnamefont {B.~V.}\ \bibnamefont {Fine}}, \ and\ \bibinfo {author}
  {\bibfnamefont {N.}~\bibnamefont {Gedik}},\ }\href {\doibase
  10.1038/s41567-018-0311-9} {\bibfield  {journal} {\bibinfo  {journal} {Nat.
  Phys.}\ }\textbf {\bibinfo {volume} {15}},\ \bibinfo {pages} {27} (\bibinfo
  {year} {2019})}\BibitemShut {NoStop}%
\bibitem [{\citenamefont {Suzuki}\ \emph {et~al.}(2019)\citenamefont {Suzuki},
  \citenamefont {Someya}, \citenamefont {Hashimoto}, \citenamefont {Michimae},
  \citenamefont {Watanabe}, \citenamefont {Fujisawa}, \citenamefont {Kanai},
  \citenamefont {Ishii}, \citenamefont {Itatani}, \citenamefont {Kasahara},
  \citenamefont {Matsuda}, \citenamefont {Shibauchi}, \citenamefont {Okazaki},\
  and\ \citenamefont {Shin}}]{Suzuki2019}%
  \BibitemOpen
  \bibfield  {author} {\bibinfo {author} {\bibfnamefont {T.}~\bibnamefont
  {Suzuki}}, \bibinfo {author} {\bibfnamefont {T.}~\bibnamefont {Someya}},
  \bibinfo {author} {\bibfnamefont {T.}~\bibnamefont {Hashimoto}}, \bibinfo
  {author} {\bibfnamefont {S.}~\bibnamefont {Michimae}}, \bibinfo {author}
  {\bibfnamefont {M.}~\bibnamefont {Watanabe}}, \bibinfo {author}
  {\bibfnamefont {M.}~\bibnamefont {Fujisawa}}, \bibinfo {author}
  {\bibfnamefont {T.}~\bibnamefont {Kanai}}, \bibinfo {author} {\bibfnamefont
  {N.}~\bibnamefont {Ishii}}, \bibinfo {author} {\bibfnamefont
  {J.}~\bibnamefont {Itatani}}, \bibinfo {author} {\bibfnamefont
  {S.}~\bibnamefont {Kasahara}}, \bibinfo {author} {\bibfnamefont
  {Y.}~\bibnamefont {Matsuda}}, \bibinfo {author} {\bibfnamefont
  {T.}~\bibnamefont {Shibauchi}}, \bibinfo {author} {\bibfnamefont
  {K.}~\bibnamefont {Okazaki}}, \ and\ \bibinfo {author} {\bibfnamefont
  {S.}~\bibnamefont {Shin}},\ }\href {\doibase 10.1038/s42005-019-0219-4}
  {\bibfield  {journal} {\bibinfo  {journal} {Commun. Phys.}\ }\textbf
  {\bibinfo {volume} {2}} (\bibinfo {year} {2019}),\
  10.1038/s42005-019-0219-4}\BibitemShut {NoStop}%
\bibitem [{\citenamefont {Rini}\ \emph {et~al.}(2007)\citenamefont {Rini},
  \citenamefont {Tobey}, \citenamefont {Dean}, \citenamefont {Itatani},
  \citenamefont {Tomioka}, \citenamefont {Tokura}, \citenamefont {Schoenlein},\
  and\ \citenamefont {Cavalleri}}]{Rini2007}%
  \BibitemOpen
  \bibfield  {author} {\bibinfo {author} {\bibfnamefont {M.}~\bibnamefont
  {Rini}}, \bibinfo {author} {\bibfnamefont {R.}~\bibnamefont {Tobey}},
  \bibinfo {author} {\bibfnamefont {N.}~\bibnamefont {Dean}}, \bibinfo {author}
  {\bibfnamefont {J.}~\bibnamefont {Itatani}}, \bibinfo {author} {\bibfnamefont
  {Y.}~\bibnamefont {Tomioka}}, \bibinfo {author} {\bibfnamefont
  {Y.}~\bibnamefont {Tokura}}, \bibinfo {author} {\bibfnamefont {R.~W.}\
  \bibnamefont {Schoenlein}}, \ and\ \bibinfo {author} {\bibfnamefont
  {A.}~\bibnamefont {Cavalleri}},\ }\href {\doibase 10.1038/nature06119}
  {\bibfield  {journal} {\bibinfo  {journal} {Nature}\ }\textbf {\bibinfo
  {volume} {449}},\ \bibinfo {pages} {72} (\bibinfo {year} {2007})}\BibitemShut
  {NoStop}%
\bibitem [{\citenamefont {Binder}\ and\ \citenamefont
  {Stauffer}(1976)}]{Binder1976}%
  \BibitemOpen
  \bibfield  {author} {\bibinfo {author} {\bibfnamefont {K.}~\bibnamefont
  {Binder}}\ and\ \bibinfo {author} {\bibfnamefont {D.}~\bibnamefont
  {Stauffer}},\ }\href {\doibase 10.1080/00018737600101402} {\bibfield
  {journal} {\bibinfo  {journal} {Adv. Phys.}\ }\textbf {\bibinfo {volume}
  {25}},\ \bibinfo {pages} {343} (\bibinfo {year} {1976})}\BibitemShut
  {NoStop}%
\bibitem [{\citenamefont {Lemonik}\ and\ \citenamefont
  {Mitra}(2017)}]{Lemonik.2017}%
  \BibitemOpen
  \bibfield  {author} {\bibinfo {author} {\bibfnamefont {Y.}~\bibnamefont
  {Lemonik}}\ and\ \bibinfo {author} {\bibfnamefont {A.}~\bibnamefont
  {Mitra}},\ }\href {\doibase 10.1103/PhysRevB.96.104506} {\bibfield  {journal}
  {\bibinfo  {journal} {Phys. Rev. B}\ }\textbf {\bibinfo {volume} {96}},\
  \bibinfo {pages} {104506} (\bibinfo {year} {2017})}\BibitemShut {NoStop}%
\bibitem [{\citenamefont {Kung}\ \emph {et~al.}(2013)\citenamefont {Kung},
  \citenamefont {Lee}, \citenamefont {Chen}, \citenamefont {Kemper},
  \citenamefont {Sorini}, \citenamefont {Moritz},\ and\ \citenamefont
  {Devereaux}}]{Kung2013}%
  \BibitemOpen
  \bibfield  {author} {\bibinfo {author} {\bibfnamefont {Y.~F.}\ \bibnamefont
  {Kung}}, \bibinfo {author} {\bibfnamefont {W.-S.}\ \bibnamefont {Lee}},
  \bibinfo {author} {\bibfnamefont {C.-C.}\ \bibnamefont {Chen}}, \bibinfo
  {author} {\bibfnamefont {A.~F.}\ \bibnamefont {Kemper}}, \bibinfo {author}
  {\bibfnamefont {A.~P.}\ \bibnamefont {Sorini}}, \bibinfo {author}
  {\bibfnamefont {B.}~\bibnamefont {Moritz}}, \ and\ \bibinfo {author}
  {\bibfnamefont {T.~P.}\ \bibnamefont {Devereaux}},\ }\href {\doibase
  10.1103/PhysRevB.88.125114} {\bibfield  {journal} {\bibinfo  {journal} {Phys.
  Rev. B}\ }\textbf {\bibinfo {volume} {88}},\ \bibinfo {pages} {125114}
  (\bibinfo {year} {2013})}\BibitemShut {NoStop}%
\bibitem [{\citenamefont {Ross~Tagaras}\ \emph {et~al.}(2019)\citenamefont
  {Ross~Tagaras}, \citenamefont {Weng},\ and\ \citenamefont
  {Allen}}]{RossTagaras2019}%
  \BibitemOpen
  \bibfield  {author} {\bibinfo {author} {\bibfnamefont {M.}~\bibnamefont
  {Ross~Tagaras}}, \bibinfo {author} {\bibfnamefont {J.}~\bibnamefont {Weng}},
  \ and\ \bibinfo {author} {\bibfnamefont {R.~E.}\ \bibnamefont {Allen}},\
  }\href {\doibase 10.1140/epjst/e2018-800102-6} {\bibfield  {journal}
  {\bibinfo  {journal} {The European Physical Journal Special Topics}\ }
  (\bibinfo {year} {2019}),\ 10.1140/epjst/e2018-800102-6}\BibitemShut
  {NoStop}%
\bibitem [{\citenamefont {Dolgirev}\ \emph {et~al.}(2020)\citenamefont
  {Dolgirev}, \citenamefont {Rozhkov}, \citenamefont {Zong}, \citenamefont
  {Kogar}, \citenamefont {Gedik},\ and\ \citenamefont {Fine}}]{Dolgirev2019}%
  \BibitemOpen
  \bibfield  {author} {\bibinfo {author} {\bibfnamefont {P.~E.}\ \bibnamefont
  {Dolgirev}}, \bibinfo {author} {\bibfnamefont {A.~V.}\ \bibnamefont
  {Rozhkov}}, \bibinfo {author} {\bibfnamefont {A.}~\bibnamefont {Zong}},
  \bibinfo {author} {\bibfnamefont {A.}~\bibnamefont {Kogar}}, \bibinfo
  {author} {\bibfnamefont {N.}~\bibnamefont {Gedik}}, \ and\ \bibinfo {author}
  {\bibfnamefont {B.~V.}\ \bibnamefont {Fine}},\ }\href {\doibase
  10.1103/PhysRevB.101.054203} {\bibfield  {journal} {\bibinfo  {journal}
  {Phys. Rev. B}\ }\textbf {\bibinfo {volume} {101}},\ \bibinfo {pages}
  {054203} (\bibinfo {year} {2020})}\BibitemShut {NoStop}%
\bibitem [{\citenamefont {Gor'kov}\ and\ \citenamefont
  {Eliashberg}(1968)}]{Gorkov1968}%
  \BibitemOpen
  \bibfield  {author} {\bibinfo {author} {\bibfnamefont {L.~P.}\ \bibnamefont
  {Gor'kov}}\ and\ \bibinfo {author} {\bibfnamefont {G.~M.}\ \bibnamefont
  {Eliashberg}},\ }\href@noop {} {\bibfield  {journal} {\bibinfo  {journal}
  {Sov. Phys. JETP}\ }\textbf {\bibinfo {volume} {27}},\ \bibinfo {pages} {328}
  (\bibinfo {year} {1968})}\BibitemShut {NoStop}%
\bibitem [{\citenamefont {Cyrot}(1973)}]{Cyrot1973}%
  \BibitemOpen
  \bibfield  {author} {\bibinfo {author} {\bibfnamefont {M.}~\bibnamefont
  {Cyrot}},\ }\href {\doibase 10.1088/0034-4885/36/2/001} {\bibfield  {journal}
  {\bibinfo  {journal} {Reports Prog. Phys.}\ }\textbf {\bibinfo {volume}
  {36}},\ \bibinfo {pages} {103} (\bibinfo {year} {1973})}\BibitemShut
  {NoStop}%
\bibitem [{\citenamefont {Kramers}(1940)}]{Kramers1940}%
  \BibitemOpen
  \bibfield  {author} {\bibinfo {author} {\bibfnamefont {H.}~\bibnamefont
  {Kramers}},\ }\href {\doibase 10.1016/S0031-8914(40)90098-2} {\bibfield
  {journal} {\bibinfo  {journal} {Physica}\ }\textbf {\bibinfo {volume} {7}},\
  \bibinfo {pages} {284} (\bibinfo {year} {1940})}\BibitemShut {NoStop}%
\bibitem [{\citenamefont {Risken}(1996)}]{Risken.1996}%
  \BibitemOpen
  \bibfield  {author} {\bibinfo {author} {\bibfnamefont {H.}~\bibnamefont
  {Risken}},\ }\href@noop {} {\emph {\bibinfo {title} {The Fokker-Planck
  equation: methods of solution and applications}}}\ (\bibinfo  {publisher}
  {Springer-Verlag},\ \bibinfo {address} {Berlin},\ \bibinfo {year}
  {1996})\BibitemShut {NoStop}%
\bibitem [{\citenamefont {He}\ and\ \citenamefont {Millis}(2016)}]{He.2016}%
  \BibitemOpen
  \bibfield  {author} {\bibinfo {author} {\bibfnamefont {Z.}~\bibnamefont
  {He}}\ and\ \bibinfo {author} {\bibfnamefont {A.~J.}\ \bibnamefont
  {Millis}},\ }\href {\doibase 10.1103/PhysRevB.93.115126} {\bibfield
  {journal} {\bibinfo  {journal} {Phys. Rev. B}\ }\textbf {\bibinfo {volume}
  {93}},\ \bibinfo {pages} {115126} (\bibinfo {year} {2016})}\BibitemShut
  {NoStop}%
\bibitem [{\citenamefont {Rothwarf}\ and\ \citenamefont
  {Taylor}(1967)}]{Rothwarf.1967}%
  \BibitemOpen
  \bibfield  {author} {\bibinfo {author} {\bibfnamefont {A.}~\bibnamefont
  {Rothwarf}}\ and\ \bibinfo {author} {\bibfnamefont {B.~N.}\ \bibnamefont
  {Taylor}},\ }\href {\doibase 10.1103/PhysRevLett.19.27} {\bibfield  {journal}
  {\bibinfo  {journal} {Phys. Rev. Lett.}\ }\textbf {\bibinfo {volume} {19}},\
  \bibinfo {pages} {27} (\bibinfo {year} {1967})}\BibitemShut {NoStop}%
\bibitem [{\citenamefont {Lemonik}\ and\ \citenamefont
  {Mitra}(2018)}]{Lemonik.2018}%
  \BibitemOpen
  \bibfield  {author} {\bibinfo {author} {\bibfnamefont {Y.}~\bibnamefont
  {Lemonik}}\ and\ \bibinfo {author} {\bibfnamefont {A.}~\bibnamefont
  {Mitra}},\ }\href {\doibase 10.1103/PhysRevB.98.214514} {\bibfield  {journal}
  {\bibinfo  {journal} {Phys. Rev. B}\ }\textbf {\bibinfo {volume} {98}},\
  \bibinfo {pages} {214514} (\bibinfo {year} {2018})}\BibitemShut {NoStop}%
\bibitem [{\citenamefont {Kogar}\ \emph {et~al.}(2020)\citenamefont {Kogar},
  \citenamefont {Zong}, \citenamefont {Dolgirev}, \citenamefont {Shen},
  \citenamefont {Straquadine}, \citenamefont {Bie}, \citenamefont {Wang},
  \citenamefont {Rohwer}, \citenamefont {Tung}, \citenamefont {Yang},
  \citenamefont {Li}, \citenamefont {Yang}, \citenamefont {Weathersby},
  \citenamefont {Park}, \citenamefont {Kozina}, \citenamefont {Sie},
  \citenamefont {Wen}, \citenamefont {Jarillo-Herrero}, \citenamefont {Fisher},
  \citenamefont {Wang},\ and\ \citenamefont {Gedik}}]{Kogar2019}%
  \BibitemOpen
  \bibfield  {author} {\bibinfo {author} {\bibfnamefont {A.}~\bibnamefont
  {Kogar}}, \bibinfo {author} {\bibfnamefont {A.}~\bibnamefont {Zong}},
  \bibinfo {author} {\bibfnamefont {P.~E.}\ \bibnamefont {Dolgirev}}, \bibinfo
  {author} {\bibfnamefont {X.}~\bibnamefont {Shen}}, \bibinfo {author}
  {\bibfnamefont {J.}~\bibnamefont {Straquadine}}, \bibinfo {author}
  {\bibfnamefont {Y.-Q.}\ \bibnamefont {Bie}}, \bibinfo {author} {\bibfnamefont
  {X.}~\bibnamefont {Wang}}, \bibinfo {author} {\bibfnamefont {T.}~\bibnamefont
  {Rohwer}}, \bibinfo {author} {\bibfnamefont {I.-C.}\ \bibnamefont {Tung}},
  \bibinfo {author} {\bibfnamefont {Y.}~\bibnamefont {Yang}}, \bibinfo {author}
  {\bibfnamefont {R.}~\bibnamefont {Li}}, \bibinfo {author} {\bibfnamefont
  {J.}~\bibnamefont {Yang}}, \bibinfo {author} {\bibfnamefont {S.}~\bibnamefont
  {Weathersby}}, \bibinfo {author} {\bibfnamefont {S.}~\bibnamefont {Park}},
  \bibinfo {author} {\bibfnamefont {M.~E.}\ \bibnamefont {Kozina}}, \bibinfo
  {author} {\bibfnamefont {E.~J.}\ \bibnamefont {Sie}}, \bibinfo {author}
  {\bibfnamefont {H.}~\bibnamefont {Wen}}, \bibinfo {author} {\bibfnamefont
  {P.}~\bibnamefont {Jarillo-Herrero}}, \bibinfo {author} {\bibfnamefont
  {I.~R.}\ \bibnamefont {Fisher}}, \bibinfo {author} {\bibfnamefont
  {X.}~\bibnamefont {Wang}}, \ and\ \bibinfo {author} {\bibfnamefont
  {N.}~\bibnamefont {Gedik}},\ }\href
  {https://doi.org/10.1038/s41567-019-0705-3} {\bibfield  {journal} {\bibinfo
  {journal} {Nature Physics}\ }\textbf {\bibinfo {volume} {16}},\ \bibinfo
  {pages} {159} (\bibinfo {year} {2020})}\BibitemShut {NoStop}%
\bibitem [{\citenamefont {Kennes}\ \emph {et~al.}(2017)\citenamefont {Kennes},
  \citenamefont {Wilner}, \citenamefont {Reichman},\ and\ \citenamefont
  {Millis}}]{Kennes2017}%
  \BibitemOpen
  \bibfield  {author} {\bibinfo {author} {\bibfnamefont {D.~M.}\ \bibnamefont
  {Kennes}}, \bibinfo {author} {\bibfnamefont {E.~Y.}\ \bibnamefont {Wilner}},
  \bibinfo {author} {\bibfnamefont {D.~R.}\ \bibnamefont {Reichman}}, \ and\
  \bibinfo {author} {\bibfnamefont {A.~J.}\ \bibnamefont {Millis}},\ }\href
  {\doibase 10.1038/nphys4024} {\bibfield  {journal} {\bibinfo  {journal} {Nat.
  Phys.}\ }\textbf {\bibinfo {volume} {13}},\ \bibinfo {pages} {479} (\bibinfo
  {year} {2017})}\BibitemShut {NoStop}%
\bibitem [{\citenamefont {Babadi}\ \emph {et~al.}(2017)\citenamefont {Babadi},
  \citenamefont {Knap}, \citenamefont {Martin}, \citenamefont {Refael},\ and\
  \citenamefont {Demler}}]{Babadi2017}%
  \BibitemOpen
  \bibfield  {author} {\bibinfo {author} {\bibfnamefont {M.}~\bibnamefont
  {Babadi}}, \bibinfo {author} {\bibfnamefont {M.}~\bibnamefont {Knap}},
  \bibinfo {author} {\bibfnamefont {I.}~\bibnamefont {Martin}}, \bibinfo
  {author} {\bibfnamefont {G.}~\bibnamefont {Refael}}, \ and\ \bibinfo {author}
  {\bibfnamefont {E.}~\bibnamefont {Demler}},\ }\href {\doibase
  10.1103/PhysRevB.96.014512} {\bibfield  {journal} {\bibinfo  {journal} {Phys.
  Rev. B}\ }\textbf {\bibinfo {volume} {96}},\ \bibinfo {pages} {014512}
  (\bibinfo {year} {2017})}\BibitemShut {NoStop}%
\bibitem [{\citenamefont {Sentef}\ \emph {et~al.}(2017)\citenamefont {Sentef},
  \citenamefont {Tokuno}, \citenamefont {Georges},\ and\ \citenamefont
  {Kollath}}]{Sentef2017}%
  \BibitemOpen
  \bibfield  {author} {\bibinfo {author} {\bibfnamefont {M.~A.}\ \bibnamefont
  {Sentef}}, \bibinfo {author} {\bibfnamefont {A.}~\bibnamefont {Tokuno}},
  \bibinfo {author} {\bibfnamefont {A.}~\bibnamefont {Georges}}, \ and\
  \bibinfo {author} {\bibfnamefont {C.}~\bibnamefont {Kollath}},\ }\href
  {\doibase 10.1103/PhysRevLett.118.087002} {\bibfield  {journal} {\bibinfo
  {journal} {Phys. Rev. Lett.}\ }\textbf {\bibinfo {volume} {118}},\ \bibinfo
  {pages} {087002} (\bibinfo {year} {2017})}\BibitemShut {NoStop}%
\bibitem [{\citenamefont {Chiriac{\`{o}}}\ \emph {et~al.}(2018)\citenamefont
  {Chiriac{\`{o}}}, \citenamefont {Millis},\ and\ \citenamefont
  {Aleiner}}]{Chiriaco2018}%
  \BibitemOpen
  \bibfield  {author} {\bibinfo {author} {\bibfnamefont {G.}~\bibnamefont
  {Chiriac{\`{o}}}}, \bibinfo {author} {\bibfnamefont {A.~J.}\ \bibnamefont
  {Millis}}, \ and\ \bibinfo {author} {\bibfnamefont {I.~L.}\ \bibnamefont
  {Aleiner}},\ }\href {\doibase 10.1103/PhysRevB.98.220510} {\bibfield
  {journal} {\bibinfo  {journal} {Phys. Rev. B}\ }\textbf {\bibinfo {volume}
  {98}},\ \bibinfo {pages} {220510(R)} (\bibinfo {year} {2018})}\BibitemShut
  {NoStop}%
\bibitem [{\citenamefont {Wang}\ \emph {et~al.}(2018)\citenamefont {Wang},
  \citenamefont {Chen}, \citenamefont {Moritz},\ and\ \citenamefont
  {Devereaux}}]{Wang.2018}%
  \BibitemOpen
  \bibfield  {author} {\bibinfo {author} {\bibfnamefont {Y.}~\bibnamefont
  {Wang}}, \bibinfo {author} {\bibfnamefont {C.-C.}\ \bibnamefont {Chen}},
  \bibinfo {author} {\bibfnamefont {B.}~\bibnamefont {Moritz}}, \ and\ \bibinfo
  {author} {\bibfnamefont {T.~P.}\ \bibnamefont {Devereaux}},\ }\href {\doibase
  10.1103/PhysRevLett.120.246402} {\bibfield  {journal} {\bibinfo  {journal}
  {Phys. Rev. Lett.}\ }\textbf {\bibinfo {volume} {120}},\ \bibinfo {pages}
  {246402} (\bibinfo {year} {2018})}\BibitemShut {NoStop}%
\bibitem [{\citenamefont {Smallwood}\ \emph {et~al.}(2012)\citenamefont
  {Smallwood}, \citenamefont {Hinton}, \citenamefont {Jozwiak}, \citenamefont
  {Zhang}, \citenamefont {Koralek}, \citenamefont {Eisaki}, \citenamefont
  {Lee}, \citenamefont {Orenstein},\ and\ \citenamefont
  {Lanzara}}]{Smallwood2012}%
  \BibitemOpen
  \bibfield  {author} {\bibinfo {author} {\bibfnamefont {C.~L.}\ \bibnamefont
  {Smallwood}}, \bibinfo {author} {\bibfnamefont {J.~P.}\ \bibnamefont
  {Hinton}}, \bibinfo {author} {\bibfnamefont {C.}~\bibnamefont {Jozwiak}},
  \bibinfo {author} {\bibfnamefont {W.}~\bibnamefont {Zhang}}, \bibinfo
  {author} {\bibfnamefont {J.~D.}\ \bibnamefont {Koralek}}, \bibinfo {author}
  {\bibfnamefont {H.}~\bibnamefont {Eisaki}}, \bibinfo {author} {\bibfnamefont
  {D.-H.}\ \bibnamefont {Lee}}, \bibinfo {author} {\bibfnamefont
  {J.}~\bibnamefont {Orenstein}}, \ and\ \bibinfo {author} {\bibfnamefont
  {A.}~\bibnamefont {Lanzara}},\ }\href {\doibase 10.1126/science.1217423}
  {\bibfield  {journal} {\bibinfo  {journal} {Science}\ }\textbf {\bibinfo
  {volume} {336}},\ \bibinfo {pages} {1137} (\bibinfo {year}
  {2012})}\BibitemShut {NoStop}%
\bibitem [{\citenamefont {Giusti}\ \emph {et~al.}(2019)\citenamefont {Giusti},
  \citenamefont {Marciniak}, \citenamefont {Randi}, \citenamefont {Sparapassi},
  \citenamefont {Boschini}, \citenamefont {Eisaki}, \citenamefont {Greven},
  \citenamefont {Damascelli}, \citenamefont {Avella},\ and\ \citenamefont
  {Fausti}}]{Giusti2019}%
  \BibitemOpen
  \bibfield  {author} {\bibinfo {author} {\bibfnamefont {F.}~\bibnamefont
  {Giusti}}, \bibinfo {author} {\bibfnamefont {A.}~\bibnamefont {Marciniak}},
  \bibinfo {author} {\bibfnamefont {F.}~\bibnamefont {Randi}}, \bibinfo
  {author} {\bibfnamefont {G.}~\bibnamefont {Sparapassi}}, \bibinfo {author}
  {\bibfnamefont {F.}~\bibnamefont {Boschini}}, \bibinfo {author}
  {\bibfnamefont {H.}~\bibnamefont {Eisaki}}, \bibinfo {author} {\bibfnamefont
  {M.}~\bibnamefont {Greven}}, \bibinfo {author} {\bibfnamefont
  {A.}~\bibnamefont {Damascelli}}, \bibinfo {author} {\bibfnamefont
  {A.}~\bibnamefont {Avella}}, \ and\ \bibinfo {author} {\bibfnamefont
  {D.}~\bibnamefont {Fausti}},\ }\href {\doibase
  10.1103/PhysRevLett.122.067002} {\bibfield  {journal} {\bibinfo  {journal}
  {Phys. Rev. Lett.}\ }\textbf {\bibinfo {volume} {122}},\ \bibinfo {pages}
  {067002} (\bibinfo {year} {2019})}\BibitemShut {NoStop}%
\bibitem [{\citenamefont {Eichberger}\ \emph {et~al.}(2010)\citenamefont
  {Eichberger}, \citenamefont {Sch{\"{a}}fer}, \citenamefont {Krumova},
  \citenamefont {Beyer}, \citenamefont {Demsar}, \citenamefont {Berger},
  \citenamefont {Moriena}, \citenamefont {Sciaini},\ and\ \citenamefont
  {Miller}}]{Eichberger2010}%
  \BibitemOpen
  \bibfield  {author} {\bibinfo {author} {\bibfnamefont {M.}~\bibnamefont
  {Eichberger}}, \bibinfo {author} {\bibfnamefont {H.}~\bibnamefont
  {Sch{\"{a}}fer}}, \bibinfo {author} {\bibfnamefont {M.}~\bibnamefont
  {Krumova}}, \bibinfo {author} {\bibfnamefont {M.}~\bibnamefont {Beyer}},
  \bibinfo {author} {\bibfnamefont {J.}~\bibnamefont {Demsar}}, \bibinfo
  {author} {\bibfnamefont {H.}~\bibnamefont {Berger}}, \bibinfo {author}
  {\bibfnamefont {G.}~\bibnamefont {Moriena}}, \bibinfo {author} {\bibfnamefont
  {G.}~\bibnamefont {Sciaini}}, \ and\ \bibinfo {author} {\bibfnamefont
  {R.~J.~D.}\ \bibnamefont {Miller}},\ }\href {\doibase 10.1038/nature09539}
  {\bibfield  {journal} {\bibinfo  {journal} {Nature}\ }\textbf {\bibinfo
  {volume} {468}},\ \bibinfo {pages} {799} (\bibinfo {year}
  {2010})}\BibitemShut {NoStop}%
\bibitem [{\citenamefont {Erasmus}\ \emph {et~al.}(2012)\citenamefont
  {Erasmus}, \citenamefont {Eichberger}, \citenamefont {Haupt}, \citenamefont
  {Boshoff}, \citenamefont {Kassier}, \citenamefont {Birmurske}, \citenamefont
  {Berger}, \citenamefont {Demsar},\ and\ \citenamefont
  {Schwoerer}}]{Erasmus2012}%
  \BibitemOpen
  \bibfield  {author} {\bibinfo {author} {\bibfnamefont {N.}~\bibnamefont
  {Erasmus}}, \bibinfo {author} {\bibfnamefont {M.}~\bibnamefont {Eichberger}},
  \bibinfo {author} {\bibfnamefont {K.}~\bibnamefont {Haupt}}, \bibinfo
  {author} {\bibfnamefont {I.}~\bibnamefont {Boshoff}}, \bibinfo {author}
  {\bibfnamefont {G.}~\bibnamefont {Kassier}}, \bibinfo {author} {\bibfnamefont
  {R.}~\bibnamefont {Birmurske}}, \bibinfo {author} {\bibfnamefont
  {H.}~\bibnamefont {Berger}}, \bibinfo {author} {\bibfnamefont
  {J.}~\bibnamefont {Demsar}}, \ and\ \bibinfo {author} {\bibfnamefont
  {H.}~\bibnamefont {Schwoerer}},\ }\href {\doibase
  10.1103/PhysRevLett.109.167402} {\bibfield  {journal} {\bibinfo  {journal}
  {Phys. Rev. Lett.}\ }\textbf {\bibinfo {volume} {109}},\ \bibinfo {pages}
  {167402} (\bibinfo {year} {2012})}\BibitemShut {NoStop}%
\bibitem [{\citenamefont {Hinton}\ \emph {et~al.}(2013)\citenamefont {Hinton},
  \citenamefont {Koralek}, \citenamefont {Lu}, \citenamefont {Vishwanath},
  \citenamefont {Orenstein}, \citenamefont {Bonn}, \citenamefont {Hardy},\ and\
  \citenamefont {Liang}}]{Hinton2013}%
  \BibitemOpen
  \bibfield  {author} {\bibinfo {author} {\bibfnamefont {J.~P.}\ \bibnamefont
  {Hinton}}, \bibinfo {author} {\bibfnamefont {J.~D.}\ \bibnamefont {Koralek}},
  \bibinfo {author} {\bibfnamefont {Y.~M.}\ \bibnamefont {Lu}}, \bibinfo
  {author} {\bibfnamefont {A.}~\bibnamefont {Vishwanath}}, \bibinfo {author}
  {\bibfnamefont {J.}~\bibnamefont {Orenstein}}, \bibinfo {author}
  {\bibfnamefont {D.~A.}\ \bibnamefont {Bonn}}, \bibinfo {author}
  {\bibfnamefont {W.~N.}\ \bibnamefont {Hardy}}, \ and\ \bibinfo {author}
  {\bibfnamefont {R.}~\bibnamefont {Liang}},\ }\href {\doibase
  10.1103/PhysRevB.88.060508} {\bibfield  {journal} {\bibinfo  {journal} {Phys.
  Rev. B}\ }\textbf {\bibinfo {volume} {88}},\ \bibinfo {pages} {060508(R)}
  (\bibinfo {year} {2013})}\BibitemShut {NoStop}%
\bibitem [{\citenamefont {Michon}\ \emph {et~al.}(2019)\citenamefont {Michon},
  \citenamefont {Girod}, \citenamefont {Badoux}, \citenamefont
  {Ka{\v{c}}mar{\v{c}}{\'{i}}k}, \citenamefont {Ma}, \citenamefont {Dragomir},
  \citenamefont {Dabkowska}, \citenamefont {Gaulin}, \citenamefont {Zhou},
  \citenamefont {Pyon}, \citenamefont {Takayama}, \citenamefont {Takagi},
  \citenamefont {Verret}, \citenamefont {Doiron-Leyraud}, \citenamefont
  {Marcenat}, \citenamefont {Taillefer},\ and\ \citenamefont
  {Klein}}]{Michon2019}%
  \BibitemOpen
  \bibfield  {author} {\bibinfo {author} {\bibfnamefont {B.}~\bibnamefont
  {Michon}}, \bibinfo {author} {\bibfnamefont {C.}~\bibnamefont {Girod}},
  \bibinfo {author} {\bibfnamefont {S.}~\bibnamefont {Badoux}}, \bibinfo
  {author} {\bibfnamefont {J.}~\bibnamefont {Ka{\v{c}}mar{\v{c}}{\'{i}}k}},
  \bibinfo {author} {\bibfnamefont {Q.}~\bibnamefont {Ma}}, \bibinfo {author}
  {\bibfnamefont {M.}~\bibnamefont {Dragomir}}, \bibinfo {author}
  {\bibfnamefont {H.~A.}\ \bibnamefont {Dabkowska}}, \bibinfo {author}
  {\bibfnamefont {B.~D.}\ \bibnamefont {Gaulin}}, \bibinfo {author}
  {\bibfnamefont {J.~S.}\ \bibnamefont {Zhou}}, \bibinfo {author}
  {\bibfnamefont {S.}~\bibnamefont {Pyon}}, \bibinfo {author} {\bibfnamefont
  {T.}~\bibnamefont {Takayama}}, \bibinfo {author} {\bibfnamefont
  {H.}~\bibnamefont {Takagi}}, \bibinfo {author} {\bibfnamefont
  {S.}~\bibnamefont {Verret}}, \bibinfo {author} {\bibfnamefont
  {N.}~\bibnamefont {Doiron-Leyraud}}, \bibinfo {author} {\bibfnamefont
  {C.}~\bibnamefont {Marcenat}}, \bibinfo {author} {\bibfnamefont
  {L.}~\bibnamefont {Taillefer}}, \ and\ \bibinfo {author} {\bibfnamefont
  {T.}~\bibnamefont {Klein}},\ }\href {\doibase 10.1038/s41586-019-0932-x}
  {\bibfield  {journal} {\bibinfo  {journal} {Nature}\ }\textbf {\bibinfo
  {volume} {567}},\ \bibinfo {pages} {218} (\bibinfo {year}
  {2019})}\BibitemShut {NoStop}%
\bibitem [{\citenamefont {Radaelli}\ \emph {et~al.}(1994)\citenamefont
  {Radaelli}, \citenamefont {Hinks}, \citenamefont {Mitchell}, \citenamefont
  {Hunter}, \citenamefont {Wagner}, \citenamefont {Dabrowski}, \citenamefont
  {Vandervoort}, \citenamefont {Viswanathan},\ and\ \citenamefont
  {Jorgensen}}]{Radaelli.1994}%
  \BibitemOpen
  \bibfield  {author} {\bibinfo {author} {\bibfnamefont {P.~G.}\ \bibnamefont
  {Radaelli}}, \bibinfo {author} {\bibfnamefont {D.~G.}\ \bibnamefont {Hinks}},
  \bibinfo {author} {\bibfnamefont {A.~W.}\ \bibnamefont {Mitchell}}, \bibinfo
  {author} {\bibfnamefont {B.~A.}\ \bibnamefont {Hunter}}, \bibinfo {author}
  {\bibfnamefont {J.~L.}\ \bibnamefont {Wagner}}, \bibinfo {author}
  {\bibfnamefont {B.}~\bibnamefont {Dabrowski}}, \bibinfo {author}
  {\bibfnamefont {K.~G.}\ \bibnamefont {Vandervoort}}, \bibinfo {author}
  {\bibfnamefont {H.~K.}\ \bibnamefont {Viswanathan}}, \ and\ \bibinfo {author}
  {\bibfnamefont {J.~D.}\ \bibnamefont {Jorgensen}},\ }\href {\doibase
  10.1103/PhysRevB.49.4163} {\bibfield  {journal} {\bibinfo  {journal} {Phys.
  Rev. B}\ }\textbf {\bibinfo {volume} {49}},\ \bibinfo {pages} {4163}
  (\bibinfo {year} {1994})}\BibitemShut {NoStop}%
\bibitem [{\citenamefont {{Zhou}}\ \emph {et~al.}(2019)\citenamefont {{Zhou}},
  \citenamefont {{Williams}}, \citenamefont {{Malliakas}}, \citenamefont
  {{Kanatzidis}}, \citenamefont {{Kemper}},\ and\ \citenamefont
  {{Ruan}}}]{Zhou2019}%
  \BibitemOpen
  \bibfield  {author} {\bibinfo {author} {\bibfnamefont {F.}~\bibnamefont
  {{Zhou}}}, \bibinfo {author} {\bibfnamefont {J.}~\bibnamefont {{Williams}}},
  \bibinfo {author} {\bibfnamefont {C.~D.}\ \bibnamefont {{Malliakas}}},
  \bibinfo {author} {\bibfnamefont {M.~G.}\ \bibnamefont {{Kanatzidis}}},
  \bibinfo {author} {\bibfnamefont {A.~F.}\ \bibnamefont {{Kemper}}}, \ and\
  \bibinfo {author} {\bibfnamefont {C.-Y.}\ \bibnamefont {{Ruan}}},\
  }\href@noop {} {\bibfield  {journal} {\bibinfo  {journal} {arXiv e-prints}\
  ,\ \bibinfo {eid} {arXiv:1904.07120}} (\bibinfo {year} {2019})},\ \Eprint
  {http://arxiv.org/abs/1904.07120} {arXiv:1904.07120 [cond-mat.mes-hall]}
  \BibitemShut {NoStop}%
\bibitem [{\citenamefont {McLeod}\ \emph {et~al.}(2019)\citenamefont {McLeod},
  \citenamefont {Zhang}, \citenamefont {Gu}, \citenamefont {Jin}, \citenamefont
  {Zhang}, \citenamefont {Post}, \citenamefont {Zhao}, \citenamefont {Millis},
  \citenamefont {Wu}, \citenamefont {Rondinelli}, \citenamefont {Averitt},\
  and\ \citenamefont {Basov}}]{mcleod.2019}%
  \BibitemOpen
  \bibfield  {author} {\bibinfo {author} {\bibfnamefont {A.~S.}\ \bibnamefont
  {McLeod}}, \bibinfo {author} {\bibfnamefont {J.}~\bibnamefont {Zhang}},
  \bibinfo {author} {\bibfnamefont {M.~Q.}\ \bibnamefont {Gu}}, \bibinfo
  {author} {\bibfnamefont {F.}~\bibnamefont {Jin}}, \bibinfo {author}
  {\bibfnamefont {G.}~\bibnamefont {Zhang}}, \bibinfo {author} {\bibfnamefont
  {K.~W.}\ \bibnamefont {Post}}, \bibinfo {author} {\bibfnamefont {X.~G.}\
  \bibnamefont {Zhao}}, \bibinfo {author} {\bibfnamefont {A.~J.}\ \bibnamefont
  {Millis}}, \bibinfo {author} {\bibfnamefont {W.}~\bibnamefont {Wu}}, \bibinfo
  {author} {\bibfnamefont {J.~M.}\ \bibnamefont {Rondinelli}}, \bibinfo
  {author} {\bibfnamefont {R.~D.}\ \bibnamefont {Averitt}}, \ and\ \bibinfo
  {author} {\bibfnamefont {D.~N.}\ \bibnamefont {Basov}},\ }\href@noop {}
  {\enquote {\bibinfo {title} {Multi-messenger nano-probes of hidden magnetism
  in a strained manganite},}\ } (\bibinfo {year} {2019}),\ \Eprint
  {http://arxiv.org/abs/1910.10361} {arXiv:1910.10361 [cond-mat.mes-hall]}
  \BibitemShut {NoStop}%
\bibitem [{\citenamefont {Guardado-Sanchez}\ \emph {et~al.}(2018)\citenamefont
  {Guardado-Sanchez}, \citenamefont {Brown}, \citenamefont {Mitra},
  \citenamefont {Devakul}, \citenamefont {Huse}, \citenamefont {Schau\ss{}},\
  and\ \citenamefont {Bakr}}]{Sanchez.2018}%
  \BibitemOpen
  \bibfield  {author} {\bibinfo {author} {\bibfnamefont {E.}~\bibnamefont
  {Guardado-Sanchez}}, \bibinfo {author} {\bibfnamefont {P.~T.}\ \bibnamefont
  {Brown}}, \bibinfo {author} {\bibfnamefont {D.}~\bibnamefont {Mitra}},
  \bibinfo {author} {\bibfnamefont {T.}~\bibnamefont {Devakul}}, \bibinfo
  {author} {\bibfnamefont {D.~A.}\ \bibnamefont {Huse}}, \bibinfo {author}
  {\bibfnamefont {P.}~\bibnamefont {Schau\ss{}}}, \ and\ \bibinfo {author}
  {\bibfnamefont {W.~S.}\ \bibnamefont {Bakr}},\ }\href {\doibase
  10.1103/PhysRevX.8.021069} {\bibfield  {journal} {\bibinfo  {journal} {Phys.
  Rev. X}\ }\textbf {\bibinfo {volume} {8}},\ \bibinfo {pages} {021069}
  (\bibinfo {year} {2018})}\BibitemShut {NoStop}%
\bibitem [{\citenamefont {Hunt}\ \emph {et~al.}(2015)\citenamefont {Hunt},
  \citenamefont {Nicoletti}, \citenamefont {Kaiser}, \citenamefont {Takayama},
  \citenamefont {Takagi},\ and\ \citenamefont {Cavalleri}}]{Hunt.2015}%
  \BibitemOpen
  \bibfield  {author} {\bibinfo {author} {\bibfnamefont {C.~R.}\ \bibnamefont
  {Hunt}}, \bibinfo {author} {\bibfnamefont {D.}~\bibnamefont {Nicoletti}},
  \bibinfo {author} {\bibfnamefont {S.}~\bibnamefont {Kaiser}}, \bibinfo
  {author} {\bibfnamefont {T.}~\bibnamefont {Takayama}}, \bibinfo {author}
  {\bibfnamefont {H.}~\bibnamefont {Takagi}}, \ and\ \bibinfo {author}
  {\bibfnamefont {A.}~\bibnamefont {Cavalleri}},\ }\href {\doibase
  10.1103/PhysRevB.91.020505} {\bibfield  {journal} {\bibinfo  {journal} {Phys.
  Rev. B}\ }\textbf {\bibinfo {volume} {91}},\ \bibinfo {pages} {020505}
  (\bibinfo {year} {2015})}\BibitemShut {NoStop}%
\bibitem [{\citenamefont {Del~Re}\ \emph {et~al.}(2016)\citenamefont {Del~Re},
  \citenamefont {Fabrizio},\ and\ \citenamefont {Tosatti}}]{Lorenzo.2016}%
  \BibitemOpen
  \bibfield  {author} {\bibinfo {author} {\bibfnamefont {L.}~\bibnamefont
  {Del~Re}}, \bibinfo {author} {\bibfnamefont {M.}~\bibnamefont {Fabrizio}}, \
  and\ \bibinfo {author} {\bibfnamefont {E.}~\bibnamefont {Tosatti}},\ }\href
  {\doibase 10.1103/PhysRevB.93.125131} {\bibfield  {journal} {\bibinfo
  {journal} {Phys. Rev. B}\ }\textbf {\bibinfo {volume} {93}},\ \bibinfo
  {pages} {125131} (\bibinfo {year} {2016})}\BibitemShut {NoStop}%
\bibitem [{\citenamefont {Sun}\ \emph {et~al.}(2016)\citenamefont {Sun},
  \citenamefont {Basov},\ and\ \citenamefont {Fogler}}]{Sun.2016}%
  \BibitemOpen
  \bibfield  {author} {\bibinfo {author} {\bibfnamefont {Z.}~\bibnamefont
  {Sun}}, \bibinfo {author} {\bibfnamefont {D.~N.}\ \bibnamefont {Basov}}, \
  and\ \bibinfo {author} {\bibfnamefont {M.~M.}\ \bibnamefont {Fogler}},\
  }\href {\doibase 10.1103/PhysRevLett.117.076805} {\bibfield  {journal}
  {\bibinfo  {journal} {Phys. Rev. Lett.}\ }\textbf {\bibinfo {volume} {117}},\
  \bibinfo {pages} {076805} (\bibinfo {year} {2016})}\BibitemShut {NoStop}%
\bibitem [{\citenamefont {Dolgirev}\ \emph {et~al.}(2019)\citenamefont
  {Dolgirev}, \citenamefont {Michael}, \citenamefont {Zong}, \citenamefont
  {Gedik},\ and\ \citenamefont {Demler}}]{dolgirev.2019}%
  \BibitemOpen
  \bibfield  {author} {\bibinfo {author} {\bibfnamefont {P.~E.}\ \bibnamefont
  {Dolgirev}}, \bibinfo {author} {\bibfnamefont {M.~H.}\ \bibnamefont
  {Michael}}, \bibinfo {author} {\bibfnamefont {A.}~\bibnamefont {Zong}},
  \bibinfo {author} {\bibfnamefont {N.}~\bibnamefont {Gedik}}, \ and\ \bibinfo
  {author} {\bibfnamefont {E.}~\bibnamefont {Demler}},\ }\href@noop {}
  {\enquote {\bibinfo {title} {Universal dynamics of order parameter
  fluctuations in pump-probe experiments},}\ } (\bibinfo {year} {2019}),\
  \Eprint {http://arxiv.org/abs/1910.02518} {arXiv:1910.02518 [cond-mat.other]}
  \BibitemShut {NoStop}%
\bibitem [{\citenamefont {Landauer}\ and\ \citenamefont
  {Swanson}(1961)}]{Landauer1961}%
  \BibitemOpen
  \bibfield  {author} {\bibinfo {author} {\bibfnamefont {R.}~\bibnamefont
  {Landauer}}\ and\ \bibinfo {author} {\bibfnamefont {J.~A.}\ \bibnamefont
  {Swanson}},\ }\href {\doibase 10.1103/PhysRev.121.1668} {\bibfield  {journal}
  {\bibinfo  {journal} {Phys. Rev.}\ }\textbf {\bibinfo {volume} {121}},\
  \bibinfo {pages} {1668} (\bibinfo {year} {1961})}\BibitemShut {NoStop}%
\end{thebibliography}%

\appendix
\begin{widetext}
\section{Equilibrium Free Energy and Phase Diagram}
\label{appendix:phase_diagram}
\subsection{Competing orders}
The phase diagram is shown in Fig.~\ref{fig:static_phase_diagram}.
We write Eq.~\eqref{eqn:free_energy} for the spatially uniform case using Eq.~\eqref{eqn:fc} and writing $\psi_1^2=R^2cos^2\theta$, $\psi_2=R^2sin^2\theta$. We obtain
\begin{equation}
	f=-\frac{\alpha_1+\alpha_2}{2}R^2-\frac{\alpha_1-\alpha_2}{2}R^2cos2\theta+R^4\frac{1+\frac{c}{2}}{2}+R^4\frac{1-\frac{c}{2}}{2}cos^22\theta
\end{equation}
Minimizing with respect to $cos2\theta$ gives
\begin{equation}
	cos2\theta=\frac{\alpha_1-\alpha_2}{2R^2\left(2-c\right)}
\end{equation}
so
\begin{equation}
	f=-\frac{\alpha_1+\alpha_2}{2}R^2-\frac{(\alpha_1-\alpha_2)^2}{8\left(1-\frac{c}{2}\right)}+R^4\frac{1+\frac{c}{2}}{2}
\end{equation}
and minimizing over $R$ gives
\begin{equation}
	R^2=\frac{\alpha_1+\alpha_2}{2+c}
\end{equation}
so
\begin{equation}
	f=-\frac{\left(\alpha_1+\alpha_2\right)^2}{4(2+c)}-\frac{(\alpha_1-\alpha_2)^2}{8\left(1-\frac{c}{2}\right)}
\end{equation}
and
\begin{equation}
	cos2\theta=\frac{\alpha_1-\alpha_2}{2(\alpha_1+\alpha_2)}\frac{2+c}{2-c}
\end{equation}
The alternative solution is to set one of the $\psi=0$, obtaining
\begin{equation}
	f=-\frac{\alpha_i^2}{2}
\end{equation}
Thus we see that if $c>2$ then the mixed solution costs energy and lower energy solutions are $2\theta=0$ and $\pi$. Expanding around the $\theta=0$ solution we obtain
\begin{equation}
	f(\theta)-f(\theta=0)=\frac{\alpha_1^2-\alpha_1\alpha_2}{4}\theta^2-\frac{\alpha_1^2}{8}\left(1-\frac{c}{2}\right)2\theta^2+\mathcal{O}\theta^4
\end{equation}
or
\begin{equation}
	f(\theta)-f(\theta=0)=\frac{\alpha_1^2}{4}\left(\frac{c}{2}-\frac{\alpha_1}{\alpha_2}\right)\theta^2+\mathcal{O}\theta^4
\end{equation}
so we see that the minimum at $\psi_2=0$ is only stable if $\frac{c}{2}>\frac{\alpha_1}{\alpha_2}$; expanding around the other minimum changes the sign of the $\theta^2$ term and interchanges $\alpha_2$ and $\alpha_1$, justifying the inequalities presented in the main text. 

\subsection{Intertwined orders}
We write Eq.~\eqref{eqn:free_energy} for the spatially uniform case using Eq.~\eqref{eqn:fc}  and Eq.~\eqref{eqn:sectic_free_energy}, now in their original form
\begin{equation}
	f=-\alpha_1\psi_1^2-\alpha_2\psi_2^2+\psi_1^4+\psi_2^4+c\psi_1^2\psi_2^2+d_1\psi_1^4\psi_2^2
\end{equation}
We suppose $0<c<2$ and $d_1>0$, assume $T_{c2}>T>T_{c1}$ and consider the physics as $T$ is decreased below $T_{c2}$. Initially we have a solution with $\psi_2^2=\frac{\alpha_2}{2}$ and $\psi_1=0$. As the temperature is decreased, $\alpha_1-c\alpha_2/2$ may become positive; if this occurs,  a $\psi_1$ component is added to the solution with $\psi_2\neq 0$.  This instability takes place at a temperature lower than $T_{c1}$ if $c>0$ and at a temperature higher than $T_{c1}$ if $c<0$.  We interpret this mixed state as having intertwined order, since both order parameters are non-zero. As $T$ is decreased below $T_{c1}$ a second extremum (saddle point) appears at $\psi_1^2=\frac{\alpha_1}{2},~\psi_2=0$. After $\alpha_1$ becomes large enough such that $\frac{c\left|\alpha_1\right|}{2}+d\frac{\alpha_1^2}{4}>\alpha_2$, this saddle point becomes stable to variations in $\psi_2$ and thus a local minimum. 
\begin{figure}
	\includegraphics[width=0.8 \linewidth]{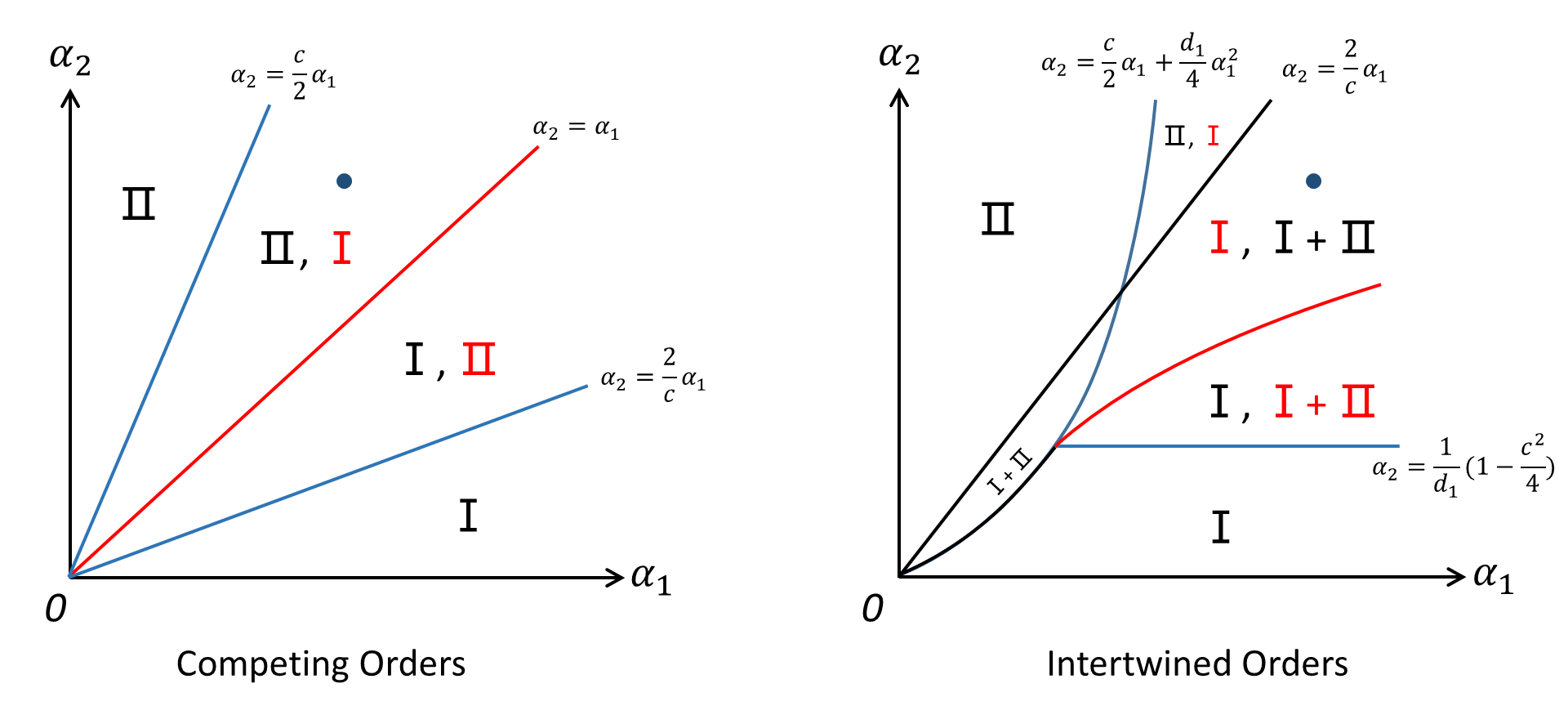}
	\caption{Equilibrium phase diagrams on the $\alpha_1$ v.s. $\alpha_2$ plane. Black roman numeral indicates the corresponding phase is a global minimum. Red one indicates local minimum. $\RomanNumeralCaps{1}+\RomanNumeralCaps{2}$ means a minimum with both orders nonzero. The competing orders case corresponds to $c>2$ and $d_1=0$. The intertwined orders case corresponds to $0<c<2$ and $d_1>0$. The systems are assumed to be at the blue dots in equilibrium.}
	\label{fig:static_phase_diagram}
\end{figure}

\section{The Fokker-Planck Equation and its Approximations}
\label{FP_equation}
If one defines the probability functional $\rho[\psi]$, the stochastic TDGL equation \eqref{eqn:TDGL} is equivalent to the Fokker-Plank equation \cite{Hohenberg1977} for the probability functional $\rho[\psi(\mathbf{r},t)]$:
\begin{align}
	\partial_t \rho = \frac{1}{E_c} \sum_i \int d^D\mathbf{r} \, 
	\gamma_i
	\partial_{\psi_i} \left(  \rho  \partial_{\psi_i} F + T \partial_{\psi_i} \rho \right)
	\,.
	\label{eqn:Fokker_Planck}
\end{align}
where $\partial_{\psi_i}$ should be understood as functional derivative: $\partial_{\psi_i} \equiv \frac{\partial}{\partial \psi_i(\mathbf{r},t)}$. The averages $\langle \psi^2 \rangle$ taken throughout the paper is over the probability functional $\rho[\psi(\mathbf{r},t)]$. The order parameter can be written as a uniform field plus small fluctuations:
\begin{align}
	\psi_i(\mathbf{r},t) = \bar{\psi}_{i}(t) + \delta\psi_i(\mathbf{r},t) = \bar{\psi}_{i}(t) + \sum_{k \neq 0} \psi_{i}(k) e^{i \mathbf{k} \mathbf{r} }
	\,.
	\label{eqn:psi_decompose}
\end{align}
Our mean field plus fluctuation theory can be viewed as an expansion in terms of the Ginzburg parameter $G$, or equivalently, the small noise term  $\eta(\mathbf{r},t)$. 
The uniform background is the zeroth order term in the random noise $\eta(\mathbf{r},t)$. The TDGL equation thus leads to the coupled equations of the uniform background and the fluctuations
\begin{align}
	\frac{1}{\gamma_i} \partial_t \bar{\psi}_{i}(t) &= \left( -\frac{\delta F}{\delta \psi_i (\mathbf{r},t)} + \eta_{i}(\mathbf{r},t) \right)_{0} 
	= -\partial_{\bar{\psi}_{i}}F + \eta_{i0}(t) + O(\eta^2) \,,
	\notag \\
	\frac{1}{\gamma_i} \partial_t \psi_{i}(k,t) &=
	\left( -\frac{\delta F}{\delta \psi_i (\mathbf{r},t)} + \eta_i(\mathbf{r},t) \right)_{k} 
	= 2\alpha_{ik} \psi_{i}(k) - \left(4 \psi_{i}^3 + 2c \psi_i \psi_{j}^2 + O(\psi^5) \right)_k + \eta_{ik}(t)
	\,
	\label{eqn:psi_coupled_equation}
\end{align}
where $()_k$ mean the Fourier component with momentum $k$, $\eta_{ik}(t)$ means the $k$ momentum component of the noise, $j\neq i$ represents the other order different from order $i$.
Multiplying the second equation by $\psi_{-k}(t)$ and taking the average over the probability functional $\rho$, one obtains the equation of motion for the second moment:
\begin{align}
	\frac{1}{\gamma_i} \partial_t 
	\langle \psi_{i}(k)\psi_{i}(-k) \rangle 
	= 4\alpha_{ik} \langle \psi_{i}(k)\psi_{i}(-k) \rangle  - 
	2\left\langle
	\psi_{i,-k}
	\left(4 \psi_{i}^3 + 2c \psi_i \psi_{j}^2 + O(\psi^5) \right)_k 
	\right\rangle
	+ 2  \left\langle \psi_{i}(-k) \eta_{ik}(t) \right\rangle
	\,.
	\label{eqn:psi_second_moment_eom}
\end{align}
If one keeps only $O(\eta^2)$ terms, the equation for the second moment $\langle \psi_{i}(k) \psi_{j}(-k) \rangle$  simplifies to
\begin{align}
	\partial_t  \langle  \psi_{i}(k) \psi_{j}(-k)  \rangle 
	= -  \left( \partial_{\bar{\psi}_{\mu}} \partial_{\bar{\psi}_{\nu}}  F + 2\xi_{\mu\nu}^2 k^2 \right)
	\left( 
	\gamma_{i\nu}  \langle \psi_{\mu}(k) \psi_{j}(-k) \rangle 
	+
	\gamma_{j\nu}  \langle \psi_{\mu}(k) \psi_{i}(-k) \rangle 
	\right)
	+ 2  T_v \gamma_{ij}
	\,
	\label{eqn:psi_second_moment_eom_simple}
\end{align}
where 
\begin{align}
	\xi_{\mu\nu} =  
	\begin{pmatrix}
		\xi_{10} & 0 \\
		0 & \xi_{20}
	\end{pmatrix}
	\,,\quad
	\gamma_{\mu\nu} =  
	\begin{pmatrix}
		\gamma_{1} & 0 \\
		0 & \gamma_{2}
	\end{pmatrix}
	\,
\end{align}
and repeated indices should be summed over. Note that  $T_v = T/(E_c V)$ is the temperature normalized to the condensation energy of the whole volume.
Therefore, the fluctuation just evolves in a time dependent quadratic potential determined by the curvature of the local free energy landscape taken at $\bar{\psi}(t)$.   In principle, one could make a `mean field' approximation by assuming the probability function $\rho$ is always a Gaussian function and easily take into account the fluctuation correction to \equa{eqn:psi_second_moment_eom_simple}. But this is out of the scope of this paper.

The linearized Fokker-Plank equation close to point $O$ reads
\begin{align}
\partial_t \rho = \partial_{\psi(k)} \left( -2 \gamma \alpha_{k}(t) \psi(k) \rho + \gamma T_v(t) \partial_{\psi(k)} \rho \right)
\,.
\label{eqn:diffusion}
\end{align}
where $\alpha$ and $T_v$ are time dependent and $\alpha_{k}(t) = \alpha(t) - \xi_0^2 k^2$. This is a diffusion equation for the probability $\rho(\psi_k)$ in the quadratic potential $-\alpha_k \psi(k)^2$ with diffusion constant $\gamma T_v$.
The exact solution to \equa{eqn:diffusion} is the Gaussian function \equa{eqn:gausssian_solution_diffusion} with the variance satisfying (here and below we denote $\psi(k)$ as $\psi_{k}$ for simplicity)
\begin{align}
\partial_t \langle \psi_{k}^2 \rangle =  4\gamma \alpha_{k}(t) \langle \psi_{k}^2 \rangle + 2 T_v(t) \gamma
\,.
\label{eqn:time_dependent_equation}
\end{align}
The solution to \equa{eqn:time_dependent_equation} is 
\begin{align}
\langle \psi_k^2 \rangle_t =  \langle \psi_k^2 \rangle_{t^\prime} e^{2 S_k(t,t^\prime)} + 2\gamma \int_{{t^{\prime}}}^{t} dt^{\prime\prime} T_v(t^{\prime\prime}) e^{2 S_k(t,t^{\prime\prime})}
\,
\label{eqn:time_dependent_solution}
\end{align}
which could be obtained by squaring
\begin{align}
\psi_k(t) =  \int_{0}^{t} dt^\prime G_k(t,t^\prime) \eta_k(t^\prime) + \psi_k(0) e^{S_k(t)}
\label{eqn:linear_response}
\end{align}
where $G_k(t,t^\prime) = \Theta(t-t^\prime) e^{S_k(t,t^\prime)}$ is the Green's function for \equa{eqn:fourier_mode_linear_evolution} of the main text and $S_k(t,t^\prime) = 2 \gamma \int_{t^\prime}^{t} dx \alpha_k(x) $ is the accumulated exponent.

\section{Dynamics}
\label{sec:dynamics}
We evaluate Eq.~\eqref{variance1} of the main text, which together with the initial condition term is
\begin{equation}
	D_k(t)\equiv\left<\psi_k(t)\psi_{-k}(t)\right>=
    D_k(-t_{pump}) e^{2S_k(t,-t_{pump})} +
	2\gamma \int_{-t_{pump}}^{t}dt^\prime e^{2S_k(t,t^\prime)}T(t^\prime) 
	\label{variance11_appendix}
\end{equation}
where the accumulated exponent is
\begin{equation}
	S_k(t,t^\prime)=2 \gamma \int_{t^\prime}^tdt^{\prime\prime}\left(\alpha(t^{\prime\prime})-\xi_0^2k^2\right)
	\label{Sdef1}
\end{equation}
and $T$ should be understood as the dimensionless quantity $T/(E_c V)$ here and in the following.
The integral in Eq.~\eqref{variance11} describes the contributions to the variance $D_k(t)$ from noise fluctuations that are created at time $t^\prime$ and then propagated forward by the equation of motion. The pump and cooling profile determines the time dependence of $D_k$ and the needed expressions  may be straightforwardly evaluated for any pump and cooling profile. Within the linear cooling profile approximation of Figure~\ref{fig:quench_profile}, $S_k$ and $T$ are combinations of quadratic, linear and constant functions of time and analytic results can be written down in terms of error functions, Gaussians and exponentials. We present here further analytical work based on the linear cooling profile that brings insight. 

At times $t<t_0$, $S_k<0$ for all $k$ so fluctuations created at a time $t<t_0$ decay as time increases to $t_0$. At times $t>t_0$ long wavelength fluctuations ($k^2\xi_0^2<\alpha(t)$)  increase exponentially with time. Thus $t_0$ is a convenient reference point and we are interested in times $t$ greater than  $t_0$. Separating the integral into times greater and less than $t_0$ and noting that $S(t,t^\prime)=S(t,t_0)+S(t_0,t^\prime)$ and that $S(t,t^\prime)=-S(t^\prime,t)$ we have
\begin{equation}
	D_k(t)= e^{2S_k(t,t_0)}\left(D_k^{(1)}+D_k^{(2)}(t)\right)
	\label{variance111}
\end{equation}
with
\begin{equation}
	D_k^{(1)}=
	 D_k(-t_{pump}) e^{2S_k(t_0,-t_{pump})}
	 +
	2\gamma \int_{-t_{pump}}^{t_0}dt^\prime e^{2S_k(t_0,t^\prime)}T(t^\prime)
	\label{D1def}
\end{equation}
and
\begin{equation}
	D_k^{(2)}(t)=2\gamma\int_{t_0}^{t}dt^\prime e^{-2S_k(t^\prime,t_0)}T(t^\prime)
\,.
	\label{D2def}
\end{equation}
The first term ($D^{(1)}$) describes the  contribution to the variance of  fluctuations created before $t_0$ and propagated forward to $t$ and the second term ($D^{(2)}$), which we have rearranged for later convenience, describes the additional contributions of fluctuations occurring after $t_0$.  We are interested in the case in which the fluctuations at $t=t_0$ are very small, and we wish to focus on long times such that the growing modes have increased to an amplitude of the order of unity. In this circumstance a general asymptotic analysis is possible but for ease of writing we will focus on the linear cooling profile for which
\begin{align}
T(t)=\left\{
\begin{array}{lc}
    T_H & (-t_{pump}<t<0)
\\
    T_H\left(1-\frac{t}{t_0}\right)+T_C\frac{t}{t_0} & (0<t<t_0)
\\
    T_C\frac{t_m-t}{t_m-t_0}+T_L\frac{t-t_0}{t_m-t_0} & (t_0<t<t_m)
\\
	T_L & (t>t_m)
\end{array}
\right.
\end{align}
and
\begin{align}
\alpha(t)=\left\{
\begin{array}{lc}
	\alpha_H\hspace{.7in} & (-t_{pump}<t<0)
    \\
    \alpha_H\left(1-\frac{t}{t_0}\right)\hspace{.2in} & (0<t<t_0)
	\\
    \alpha_L\frac{t-t_0}{t_m-t_0}\hspace{0.3in} & (t_0<t<t_m)
    \\
    \alpha_L\hspace{.75in} & (t>t_m)
\end{array}
\right.
\,.
\end{align}
We begin with $D^{(1)}$ which we rewrite as
\begin{equation}
	D^{(1)}_k=D_H+D_{KZ}
	\label{D1split}
\end{equation}
where $D_H$ describes the propagation forward in time of the fluctuations existing before the pump was turned on and created by the pump.  Using the linear cooling profile formulas and the definition of $S$
\begin{equation}
	D_H=e^{-2\gamma t_0\left(|\alpha_H|+2\xi_0^2k^2\right)}
	\left(\frac{T_H\left(1-e^{-4\gamma t_{pump}\left(|\alpha_H|+\xi_0^2k^2\right)}\right)}
	{2\left(|\alpha_H|+\xi_0^2k^2\right)}+\frac{T_L
e^{-4\gamma t_{pump}\left(|\alpha_H|+\xi_0^2k^2\right)}	
}{2\left(2\alpha_L+\xi_0^2k^2\right)}\right)
\label{eqn:DH}
\end{equation}
The requirement that the mean field order parameter be completely suppressed means that $e^{-2|\alpha_H|(2t_{pump}+t_0)}\ll G/\alpha_L$ so
\begin{equation}
	D_H \approx \frac{T_H}{2\left(|\alpha_H|+\xi_0^2k^2\right)}
	e^{-2\gamma t_0\left(|\alpha_H|+2\xi_0^2k^2\right)}
\end{equation}
which represents the hot thermal fluctuations created by the pump propagated to $t=t_0$.

We now turn to $D_{KZ}$ which represents the fluctuations created as the system cools from $t=0$ to $t=t_0$ after the pump is turned off:
\begin{eqnarray}
	D_{KZ}&=&2\gamma\int_{0}^{t_0}dt^\prime e^{2S_k(t_0,t^\prime)}T(t^\prime)
	\\
	&=&2\gamma t_0\int_0^1du~e^{-2\gamma t_0 |\alpha_H|u^2-4\gamma t_0 k^2\xi_0^2 u}\left(T_Hu+T_c(1-u)\right)
\end{eqnarray}
where in the second equality we have defined $u=(t_0-t)/t_0$.
In the rapid cooling limit $|\alpha_H|\gamma t_0\ll1$ $D_{KZ}$ is $\mathcal{O}(\gamma t_0)$ and is much smaller than $D_H$. In the slow cooling limit the integral is dominated by small $u$, and we may extend the upper limit to infinity and rescale $u=v/(\sqrt{2|\alpha_H|\gamma t_0})$ obtaining
\begin{equation}
	D_{KZ}=\frac{T_c}{\alpha_{KZ}}\int_0^\infty dv e^{-v^2-2 k^2\xi_{KZ}^2v}
\end{equation}
where we have defined the important length and time scales
\begin{equation}
	\alpha_{KZ}=\sqrt{\frac{|\alpha_H|}{2\gamma t_0}}=\sqrt{\frac{|\alpha_H|+\alpha_L}{2\gamma t_m}};
	\hspace{0.1in}\xi_{KZ}^2=\xi_0^2/\alpha_{KZ};
	\hspace{0.1in}
	t_{KZ} = \sqrt{\frac{t_m}{2\gamma (|\alpha_H|+\alpha_L)}}
	\,.
\end{equation}
We see that for $k^2\xi_{KZ}^2\ll1$, $D_{KZ}\approx \frac{T_c}{2\alpha_{KZ}}\sqrt{\pi}$ and for $k^2\xi_{KZ}^2\gg 1$, $D_{KZ}\sim \frac{T_c}{2k^2\xi_0^2}$. This is the expected behavior of a critical theory with mean field exponents and an effective distance from criticality determined by the cooling rate, consistent with the general analysis of Kibble and Zurek \cite{Kibble1976,Zurek.1996,Zurek1985}. 

We now present a qualitative evaluation of  $D^{(2)}(t)$. We have $S_k(t,t_0)=S_0(t,t_0)-2\xi_0^2k^2(t-t_0)$ with
\begin{equation}
	S_0(t,t_0)=\alpha_L\frac{(t-t_0)^2}{t_m-t_0}\Theta(t_m-t)+2\alpha_L\left(t-\frac{t_m+t_0}{2}\right)\Theta(t-t_m)
	\,.
\end{equation}
We are interested in growing modes, for which $\frac{S_0(t,t_0)}{t-t_0}>k^2\xi_0^2$; roughly these are those for which $\alpha_L>k^2\xi_0^2$.
In the slow cooling case $\alpha_L\gamma(t_m-t_0)=\frac{\alpha_L^2}{|\alpha_H|+\alpha_L}\gamma t_m >1$. In the fast cooling limit, we may set $t_m-t_0=0$ and write
\begin{equation}
	D_k^{(2)}(t)=2\gamma T_L\int_{t_0}^tdt^\prime e^{-4\gamma(t^\prime-t_0)\left(\alpha_L-\xi_0^2k^2\right)}=\frac{T_L}{2 (\alpha_L -\xi_0^2 k^2)}\left(1-e^{-4\gamma(t-t_0)\left(\alpha_L-\xi_0^2k^2\right)}\right)\approx \frac{T_L}{2\gamma \alpha_L}
\end{equation}
Here we have neglected $k$-dependence, which is on a scale that is not relevant at large enough $t$. In the ultra slow cooling case we have for $t<t_m$ and defining $u=\frac{t^\prime-t_0}{t_m-t_0}$ 
\begin{equation}
	D^{(2)}(t)=2\gamma(t_m-t_0)\int_{0}^{u_{max}}du e^{-2\alpha_L\gamma(t_m-t_0)u^2+4\gamma (t_m-t_0)uk^2\xi_0^2}\left(T_c(1-u)+T_Lu\right)
\end{equation}
with $u_{max}=\frac{t-t_0}{t_m-t_0}$. The argument of the exponential is maximized at $u=u^\star=\frac{\xi_0^2k^2}{\alpha_L}$ (for growing modes $u_{max}>2\xi_0^2k^2/\alpha_L$ so $u^\star$ is within  integration range). Defining $u=u^\star+\frac{v}{2\sqrt{\gamma(t_m-t_0)}}$ and noting that $t_m-t_0=t_m\frac{\alpha_L}{|\alpha_H|+\alpha_L}$ we have (after integrating over $v$ using saddle point approximation)
\begin{equation}
	D_k^{(2)}(t)=\frac{\sqrt{\pi}}{2}e^{\xi_{KZ}^4k^4} \frac{1}{\alpha_{KZ}} \left(T_c(1-u^\star)+T_Lu^\star\right)
\end{equation}
Note that we have kept only half of the Gaussian integral. This is valid for the modes $\xi_0^2 k^2 \lesssim \frac{1}{4\gamma (t-t_0)} \ll \frac{1}{\sqrt{8\gamma (t_m-t_0)/\alpha_L}} $ which are the only relevant ones at long time $t-t_0$. The exponent is also small in this limit so to an adequate approximation we have
\begin{equation}
	D^{(2)}(t)=\sqrt{\pi} \frac{ T_c}{2\alpha_{KZ}}
\end{equation}
which is the same as the $k\rightarrow 0$ limit of $D^{(1)}_k \approx D_{KZ}$. This is expected since $D^{(2)}_k$ can be interpreted as the fluctuations created after $t_0$ and back propagated to $t_0$. This is symmetric to $D^{(1)}_k$ for $k \rightarrow 0$ in the slow cooling limit.

\section{Exact solution in terms of error functions}
\label{appendix:exact_solution}
The $D_k^{(1)}=D_H + D_{KZ}$ in \equa{variance111} has the interpretation of the fluctuation prepared at time $t_0$. The exact form of $D_H$ is \equa{eqn:DH}. In the linear cooling profile approximation used in this paper, the exact form of $D_{KZ}$ is
\begin{align}
D_{KZ}= \frac{T_c}{2 \alpha_{KZ}}
\Bigg[&
\sqrt{\pi}
\left( 1- \left( \frac{T_H}{T_c} -1 \right) \frac{\xi_{KZ}^2 k^2}{\alpha_H}
\right) e^{\xi_{KZ}^4 k^4} 
\mathrm{Erf}\left[\xi_{KZ}^2 k^2,\, \xi_{KZ}^2 k^2 + \sqrt{2\gamma |\alpha_{H}| t_0} \right]
\notag\\
&-
\left( \frac{T_H}{T_c} -1 \right) \frac{t_{KZ}}{t_0} \left(e^{2 S_k(t_0,0)}-1 \right)
\Bigg]
\label{DKZ_exact}
\end{align}
where $\mathrm{Erf}[x_1,x_2] = \frac{2}{\sqrt{\pi}} \int_{x_1}^{x_2} e^{-x^2} dx $ is the error function. 

For the $D_k^{(2)}$ term, it is simpler to neglect the time dependence of the noise, i.e., take $T_c = T_L$, after which one obtains  
\begin{equation}
D_k^{(2)}(t) =
\left\{
\begin{array}{lr} 
g_k(t_m,t) \,, & t< t_m  \\
g_k(t_m,t_m) + 
\frac{T_L}{2(\alpha_L - \xi_0^2k^2)}
e^{-2S_k(t_m,t_0)}
\left( 1-e^{-2S_k(t,t_m)} \right)
\,,&  t \geq t_m
\end{array}
\right .
\label{D2_exact}
\end{equation}
where we have defined  
\begin{align}
	g_k(t_m,t) = \sqrt{\pi} \frac{ T_L}{2\alpha_{KZ}} e^{\xi_{KZ}^4 k^4}   
	\mathrm{Erf} 
	\left[-\xi_{KZ}^2 k^2,-\xi_{KZ}^2 k^2 + t/t_{KZ} \right]
	\,.
	\label{eqn:g}
\end{align}
One can take various limits of \equa{DKZ_exact} and \eqref{D2_exact} to get the results in the previous section.

\section{The fast cooling limit: $t_m =0$}
%%%%%%%%%%%%%%%%%%%%%%%%%%%%%%%%%%%%%%%%%%%%%%%%%%%%%%
\begin{figure}
	\includegraphics[width= 0.8\linewidth]{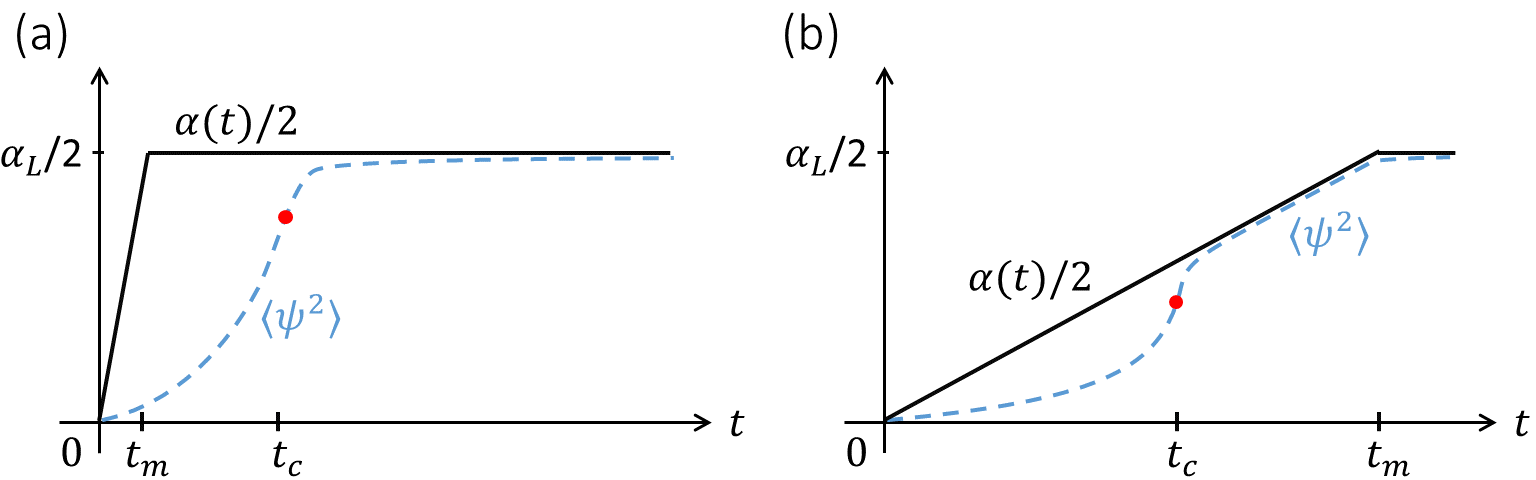}
	\caption{Illustration of the evolution of order parameter fluctuation in (a) the fasting cooling case and (b) the slow cooling case. }
	\label{fig:Slow_quench_dynamics}
\end{figure}
%%%%%%%%%%%%%%%%%%%%%%%%%%%%%%%%%%%%%%%%%%%%%%%%%%%%%%
For $t_m =0$ and neglecting the initial condition at $t=t_{pump}$, \equa{variance11} reduces to
\begin{equation}
	D_k =\frac{1}{2}
	\left(\frac{T_L}{\alpha_L-\xi_{0}^2 k^2} +
	 \frac{T_H}{|\alpha_H|+\xi_{0}^2 k^2} \right)
e^{4\gamma t(\alpha_L -\xi_0^2 k^2)} 
- \frac{1}{2}\frac{T_L}{\alpha_L-\xi_{0}^2 k^2}
	\,.
	\label{eqn:fast_quench}
\end{equation}
Summing up contribution from all the Fourier modes, one obtains the real space correlation function
\begin{align}
	\langle \psi(0) \psi(r) \rangle 
	&= V (2\pi)^{-D} \int d^D \mathbf{k} e^{i\mathbf{k} \mathbf{r}} \langle \psi_{k}^2 \rangle 
	\,.
	\label{eqn:correlation function1}
\end{align}
At long time $4 \gamma \alpha_l t \gg 1 $, $D_k$ is approximately a gaussian function in $k$ and the Fourier transform becomes
\begin{align}
	\langle \psi(0) \psi(r) \rangle  
	&\approx  
	\left(\frac{G(T_L)}{2\alpha_L} +\frac{G(T_H)}{2|\alpha_H|}
	\right) 
	\left( \frac{1}{16\pi \gamma t} \right)^{D/2} e^{4 \alpha_L \gamma t} e^{-r^2/\left( 2 \xi(t)^2 \right)} 
	\notag\\
	&= \frac{\alpha_L}{2} 
	\left(	\left(1 +\frac{G(T_H) \alpha_L}{G(T_L) |\alpha_H|}
	\right)  (4\pi)^{-D/2} \alpha_L^{D/2-2} G(T_L) \right)
	\left( \frac{1}{4 \alpha_L \gamma t} \right)^{D/2} e^{4 \alpha_L \gamma t} e^{-r^{ 2}/\left( 2 \xi(t)^2 \right)}
	\notag\\
	&= \frac{\alpha_l}{2} \zeta
	\left( \frac{1}{4 \alpha_L \gamma t} \right)^{D/2} e^{4 \alpha_L \gamma t} e^{-r^{ 2}/\left( 2 \xi(t)^2 \right)}
	\,
	\label{eqn:correlation function2}
\end{align}
where $\xi(t) = \xi_0 \sqrt{8\gamma t}$ is the universal correlation growth law and
\begin{align}
\zeta =\left(1 +\frac{G(T_H) \alpha_L}{G(T_L) |\alpha_H|}
\right)  (4\pi)^{-D/2} \alpha_L^{D/2-2} G(T_L) \sim 2 (4\pi)^{-D/2} \alpha_L^{D/2-2} G(T_L)
\label{eqn:zeta}
\end{align}
is the Ginzburg parameter for critical phenomenon at equilibrium.
Thus the fluctuation $\langle \psi_i(0)^2 \rangle$ grows exponentially with time and $\psi_1$ grows faster due to a larger $\alpha_{1L} \gamma_1$. Setting $\langle \psi(0)^2 \rangle = \alpha_L/2 $ gives the crossover time
\begin{align}
	4 \alpha_L \gamma t_c = \ln \frac{1}{\zeta} + \frac{D}{2} \ln (4 \alpha \gamma t_c)
	\,.
	\label{eqn:t0}
\end{align}
At this time, keeping the first two terms in the $\ln$ expansion of $t_c$, the ratio between the fluctuations in the two directions is
\begin{align}
	\frac{\langle \psi_2^2 \rangle}{\langle \psi_1^2 \rangle} \approx
	\frac{\alpha_{1L}}{\alpha_{2L}} \frac{G_2}{G_1} 
	\left( \frac{\gamma_1}{\gamma_2} \right)^{D/2}
	\left(
	\frac{1}{\zeta_1} \left(\ln \frac{1}{\zeta_1} \right)^{D/2}
	\right)^{\frac{\alpha_{2}\gamma_2}{\alpha_{1}\gamma_1}-1}
	\,.
	\label{eqn:psi_ratio}
\end{align}

\section{The slow cooling case in competing order systems}
\label{appendix:slow_cooling}
The slow cooling case is characterized by a large $t_m$.
After $t_1$, $\psi_1$ starts to grow exponentially while $\psi_2$ has been growing for a time of $t_1-t_2$. Assume both order parameters are in the exponential growing stage and nonlinearity is not yet on set, they obey the equation 
\begin{align}
\left<\psi_i(0)\psi_i(r)\right>_t
&= \frac{\sqrt{\pi} G_i/\alpha_{iKZ}}{\left(16\pi\gamma_i\left(t-t_i\right)\right)^\frac{D}{2}}
e^{2\gamma_i (\alpha_{iL}+|\alpha_{iH}|) (t-t_i)^2/t_m} 
e^{-\frac{r^2}{2\xi_i(t)^2}}
\,.
\label{local_i}
\end{align}
The crossover of $\psi_i$ to nonlinearity happens at $\left<\psi_i(0)^2\right>_{t} = \alpha_i(t)$ which yields
\begin{align}
1=\zeta_i x_i^{-\frac{D}{4}-1/2}
e^{x_i} 
\,,\quad
x_i = \ln\frac{1}{\zeta_{mi}} + \frac{D}{4} \ln x_i
\,,
\label{eqn:t_10}
\end{align}
where $x_i = 2\gamma_i (\alpha_{iL}+|\alpha_{iH}|) (t-t_i)^2/t_m$ and 
\begin{align}
\zeta_{mi} = 2^{-7D/4+1} \pi^{1/2-D/2} \left( \frac{ \gamma_i t_m}{\alpha_{iL}+|\alpha_{iH}|} \right)^{-D/4+1} 
G_i
= 2^{-3D/4} \pi^{1/2} \left( \frac{  \alpha_L^{2} \gamma_i t_m}{\alpha_{iL}+|\alpha_{iH}|} \right)^{-D/4+1}  
\zeta_i
\sim \zeta_i
\,.
\label{eqn:t_10}
\end{align}
To leading order in the $\ln$ expansion we have for order \RomanNumeralCaps{1}:
\begin{align}
	t_c -t_1 \approx 
	\left(
	\frac{t_m}{2(\alpha_{1L}+|\alpha_{1H}|) \gamma_1}
	\ln \frac{1}{\zeta_{m1}}
	\right)^{1/2}
\,.
\label{eqn:t_10}
\end{align}
At time $t_c$, the correlation length of the fluctuations is $\xi \approx 2\sqrt{2} \left(\frac{\gamma t_m}{2(\alpha_{1L}+|\alpha_{1H}|)}  \ln \frac{1}{\zeta} \right)^{1/4} \xi_0 \sim \left(\gamma t_m  \ln \frac{1}{\zeta} \right)^{1/4} \xi_0$, which predicts a logarithmic correction to the  $\xi \sim t_m^{1/4}$ Kibble-Zurek scaling \cite{Zurek1985}. Note that the logarithmic correction could be numerically large for very small $\zeta$. In 2D, the number density of topological vortices created is thus $n\sim 1/\xi(t_c)^2 \sim (\gamma t_m \ln \frac{1}{\zeta})^{-1/2} \xi_0^{-2}$. For the mean field plus fluctuation theory to be valid, we also require $G\alpha(t)^{D/2-2} \ll 1$ at the predicted $t_c$ for $D<4$, in other words:
\begin{align}
\left(
\frac{t_m}{2(\alpha_{1L}+|\alpha_{1H}|) \gamma_1}
\ln \frac{1}{\zeta_{m1}}
\right)^{1/2}
\frac{\alpha_{1L}+|\alpha_{1H}|}{t_m} \gg \left( \frac{1}{G} \right)^{1/(D/2-2)}
\,
\label{eqn:t_10}
\end{align}
which yields
\begin{align}
 t_m \gamma_1 \ll  \left( \frac{1}{G} \right)^{-1/(D/4-1)}
\frac{\alpha_{1L}+|\alpha_{1H}|}{2 }
\ln \frac{1}{\zeta_{m1}} 
\sim 
\left( \frac{1}{\zeta} \right)^{\frac{1}{1-D/4}}
\,.
\label{eqn:t_10}
\end{align}

Trapping into the metastable minimum requires that $\left\langle \psi_2(0)^2 \right\rangle \ll \left\langle \psi_1(0)^2 \right\rangle $ at $t_m$. Simply comparing the exponents yields
\begin{align}
	t_m
	\ll
	\frac{
		\left(
		\left(	\frac{1}{ \gamma_2} 
		\frac{1}{a_{2L}-a_{2H}}
		\right)^{1/2}
		-
		\left(	\frac{1}{ \gamma_1} 
		\frac{1}{a_{1L}-a_{1H}}
		\right)^{1/2}
		\right)^2
	}
	{
		2 \lambda_d^2
	}
	\ln \frac{1}{\zeta_{m1}}
	\,.
	\label{eq:criterion2_simple}
\end{align}
where $\lambda_d= (t_1-t_2)/t_m$.
This imposes the criterion
\begin{align}
\Delta 
\ll
\frac{1}{r}
\left(
\sqrt{\frac{2\lambda_d^2 \left( \alpha_{1L}+|\alpha_{1H}| \right)}{\ln \frac{1}{\zeta_{m1}}}  \gamma_1 t_m } +1
\right)^{-2}
\approx 
\frac{1}{r}
\left(
\sqrt{\frac{2\lambda_d^2 \left( \alpha_{1L}+|\alpha_{1H}| \right)}{\ln \frac{1}{\zeta_{1}}}  \gamma_1 t_m } +1
\right)^{-2}
\equiv f_2(\gamma_1 t_m)
\,.
\label{eq:criterion2_simple}
\end{align}

By assuming $\zeta = 10^{-4}$, $\gamma_1 = 2 \unit{ps}^{-1}$, $(\alpha_{1l},\alpha_{2l}) = (1, 1.1)$ and $(\alpha_{1h},\alpha_{2h}) = (-1, -0.9)$, one obtains $t_{mu} \approx 4.6 \unit{ps}$ and $t_{ms} \approx 460 \unit{ps}$. Since the typical cooling time due to electron phonon thermalization ranges from $1 \unit{ps}$ to $100 \unit{ps}$, most ultrafast experiments are in the regime analyzed in this paper (Fig.~\ref{fig:criterions}). 

\section{The pumping process}
The pump brings $\alpha_{iL}$ to $\alpha_{iH} <0$ as shown in Fig.~\ref{fig:quench_profile} which induces the order parameter dynamics from point \RomanNumeralCaps{2} to $O$ in Fig.~\ref{fig:quartic_free_energy}(a). At mean field level, this nonlinear dynamics is described by \equa{eqn:psi_coupled_equation} and the uniform component obeys the exact solution
\begin{align}
	\bar{\psi}_2^2 (t) = \frac{-\alpha_{2H}/2}{(1-\alpha_{2H}/\alpha_{2L})e^{-4\alpha_{2H} \gamma_2 (t+t_{pump})} -1}
	\,.
	\label{eqn:pump_dynamics}
\end{align}
The long time asymptotic form is $\bar{\psi}_2^2= \frac{-\alpha_{2H}/2}{(1-\alpha_{2H}/\alpha_{2L})} e^{4\alpha_{2H} \gamma_2 (t+t_{pump})} $ which means $\bar{\psi}_2$ approaches zero exponentially but never reaches it during finite amount of time. At time zero, the pump is removed and $\bar{\psi}_2$ reaches a small value 
\begin{align}
	\bar{\psi}_{20}^2= \frac{-\alpha_{2H}/2}{(1-\alpha_{2H}/\alpha_{2L})} e^{4\alpha_{2H} \gamma_2 t_{pump}}
	\,.
	\label{eq:pump_initial_condition}
\end{align}
We first consider the fast cooling limit $t_m=0$. After time zero, $\bar{\psi}_2$ goes back towards minimum \RomanNumeralCaps{2} following the dynamics
\begin{align}
	\bar{\psi}_{2}^2 (t) =
	\frac{\alpha_{2L}/2}
	{\left( \frac{\alpha_{2L}}{2\psi_{20}^2} -1 \right)
		e^{-4\alpha_{2L} \gamma_2 t} + 1}
	\approx 
	\bar{\psi}_{20}^2 e^{4\alpha_{2L} \gamma_2 t}
	\quad \text{for} \,\, 
	4 \alpha_{2L} \gamma_2 t \ll 
	\ln \left(\frac{\alpha_{2L}}{2\bar{\psi}_{20}^2} \right)
	\,\, \text{and} \,\,
	\bar{\psi}_{20}^2 \ll \alpha_{2L}/2
	\,.
	\label{eqn:recover_dynamics}
\end{align}
Thus at time $t_c$ when $\psi_1$ fluctuation crossovers to nonlinearity, $\bar{\psi}_2$ has recovered by an exponential factor. The more accurate picture for the probability distribution is that of Fig.~\ref{fig:gaussian}(a) but shifted in $\psi_2$ direction by the amount of $\bar{\psi}_{2}^2 (t_c)$. For the probability of trapping into phase \RomanNumeralCaps{1} to be still close to one, we require that  
\begin{align}
	\bar{\psi}_{2}^{2/\Delta} (t_c) \ll \langle \psi_{1}^2 \rangle_{t_c}
	\,
	\label{eqn:pump_criterion1}
\end{align}
which yields
\begin{align}
	\psi_{20}^{2/\Delta} 
	&\ll 
	\frac{\alpha_{1L}}{2} \zeta_1 \left(4\alpha_{1L} \gamma_1 t_c \right)^{-D/2}
	\,
	\label{eqn:pump_criterion2}
\end{align}
and further leads to the criterion
\begin{align}
	t_{pump}
	&\gg 
	\frac{\Delta }{4 |\alpha_{2H}| \gamma_2}
	\ln \frac{1}{\zeta_1}
	\equiv t_d
	\,
	\label{eqn:pump_criterion3}
\end{align}
for the pump pulse in the leading order.
Despite the logarithmic factor, this time scale can be made small with a larger $|\alpha_{2H}|$, i.e., a stronger pump will prepare the $\bar{\psi}_{20}$ with a smaller value at time zero.

If the cooling rate is finite, it is simpler to consider the case $t_m > t_{mu}$ such that the crossover happens at $t_c$ before $t_m$. The pumping time is effectively longer than $t_{pump}$ in this case since the suppression process of $\psi_2$ lasts until $t_2$, when $\alpha_2(t)$ crosses zero. 
Applying \equa{eqn:pump_criterion1} to this case yields
\begin{align}
	t_{pump}
	\gg & 
	\frac{(\alpha_{2L}+|\alpha_{2H}|) t_m}{2 |\alpha_{2H}|}
	\left(
	\sqrt{\frac{1}{2\gamma_1 (\alpha_{1L}+|\alpha_{1H}|) t_m} \ln\frac{1}{\zeta_1} }
	+ \frac{|\alpha_{1H}|}{\alpha_{1L}+|\alpha_{1H}|}
	\right)
	\notag\\
	&\left(
	\sqrt{\frac{1}{2\gamma_1 (\alpha_{1L}+|\alpha_{1H}|) t_m} \ln\frac{1}{\zeta_1} }
	+ \frac{|\alpha_{1H}|}{\alpha_{1L}+|\alpha_{1H}|}
	- \frac{2|\alpha_{2H}|}{\alpha_{2L}+|\alpha_{2H}|}
	\right)
	\,
	\label{eqn:pump_criterion_linear_cooling}
\end{align}
to leading order. For sufficiently large $|\alpha_{iH}|$, the right hand side of \equa{eqn:pump_criterion_linear_cooling} becomes negative and thus the criterion is satisfied by any $t_{pump}>0$. This is because in the cooling process before $t_2$, the high temperature stage already suppresses $\psi_2$ well enough. 

In realistic situation, the pump might not be strong enough and the proportion of phase \RomanNumeralCaps{1} domains created, $p_1$, can be calculated as a function of pump fluence/duration. It should crossover sharply from $0$ to $1$ at the boundary of \equa{eqn:pump_criterion3} or \equa{eqn:pump_criterion_linear_cooling} depending on which regime the cooling rate is in. 

\section{Joint probability function}
\label{appendix:joint_probability}
The probability that $\psi(0)=A$ and $\psi(r)=B$ is
\begin{equation}
	P(A,B)=\mathcal{N}\int\mathcal{D}\psi Exp\left[-\sum_kD^{-1}_k\psi_k\psi_{-k}\right]\delta\left(\psi(0)-A\right)\delta\left(\psi(r)-B\right)
	\label{Pdef}
\end{equation}
where
\begin{equation}
	D_k =e^{4\alpha_k \gamma t} \frac{1}{\alpha_k} \frac{T}{E_c V} = 2 \left\langle \psi^2_k \right\rangle_t 
	\label{Ddef}
\end{equation}
and $\mathcal{N}$ is the normalization of the functional integral and $\delta$ is a functional delta function. 
Representing the delta functions by  integrals gives
\begin{equation}
	P(A,B)=\mathcal{N}\int d\lambda_1 d\lambda_2 \int\mathcal{D}\psi Exp\left[-\sum_kD^{-1}_k\psi_k\psi_{-k}+i\lambda_1\left(\psi(0)-A\right)+i\lambda_2\left(\psi(r)-B\right)\right]
	\label{P1}
\end{equation}
or, Fourier transforming the real-space $\psi$
\begin{equation}
	P(A,B)=\mathcal{N}\int d\lambda_1 d\lambda_2 \int\mathcal{D}\psi Exp\left[-\sum_kD^{-1}_k\psi_k\psi_{-k}+i\left(\lambda_1+e^{ik\cdot r}\lambda_2\right)\psi_k-i\lambda_1A-i\lambda_2B\right]
	\label{P2} \,.
\end{equation}
We can now perform the integral over the $\psi_k$ and arrive at
\begin{equation}
	P(A,B)= \mathcal{N}^\prime \int d\lambda_1 d\lambda_2 Exp\left[-\sum_k\frac{D_k}{4}\left(\lambda_1^2+\lambda_2^2+2cos(k\cdot r)\lambda_1\lambda_2\right)-i\lambda_1A-i\lambda_2B\right]
	\label{P3}
\end{equation}
The sum over $k$ results in
\begin{equation}
	P(A,B)=  \mathcal{N}^\prime \int d\lambda_1 d\lambda_2 Exp
	\left[
	- \left(\lambda_1, \lambda_2 \right) \hat{M} (\lambda_1, \lambda_2)^T
	-i(\lambda_1A+\lambda_2B)
	\right]
	\label{P3}
\end{equation}
where 
\begin{align}
	\hat{M} &= \frac{1}{ 2} 
	\begin{pmatrix}
		\langle \psi(0)^2 \rangle   & \langle \psi(0) \psi(r) \rangle   \\
		\langle \psi(0) \psi(r) \rangle   & \langle \psi(0)^2 \rangle  
	\end{pmatrix} 
	= \frac{1}{ 2} \frac{G}{2\alpha} \left( \frac{1}{16\pi \gamma t} \right)^{D/2} e^{4 \alpha \gamma t}  
	\begin{pmatrix}
		1   & e^{-r^2/2\xi(t)^2}   \\
		e^{-r^2/2\xi(t)^2}   & 1
	\end{pmatrix} 
	\notag\\
	&=\frac{\alpha_1}{4}
	\begin{pmatrix}
		1   & e^{-r^2/2\xi(t_c)^2}   \\
		e^{-r^2/2\xi(t_c)^2}   & 1
	\end{pmatrix} 
	\,
	\label{eqn:green_function}
\end{align}
at $t=t_c$. We finally perform the $\lambda$ integrals, getting
\begin{align}
	P(A,B)&=  \mathcal{N}^\prime \frac{\pi}{\sqrt{\mathrm{Det}[M]}} Exp
	\left[
	- \frac{1}{4} (A,B) \hat{M}^{-1} (A,B)^T
	\right]
	\notag\\
	&=
	\frac{1}{\pi}
	\frac{1/\alpha}{\sqrt{1- e^{-r^2/\xi(t_c)^2} }}	
	Exp\left[
	-\frac{1/\alpha}{1- e^{-r^2/\xi(t_c)^2} }
	\begin{pmatrix}
		A  & B
	\end{pmatrix}
	\begin{pmatrix}
		1   & -e^{-r^2/2\xi(t_c)^2}   \\
		-e^{-r^2/2\xi(t_c)^2}   & 1
	\end{pmatrix} 
	\begin{pmatrix}
		A \\
		B
	\end{pmatrix}
	\right]
	\,.
	\label{P3}
\end{align}

\section{The coefficient $\lambda$ and $\vartheta$} 
\label{appendix:coefficients}
The coefficients are:
\begin{align}
	\lambda = (c^2 -4)^{(1/\Delta-1)/2} 
	\left(\frac{-2\alpha_1 + c\alpha_2}{\left(-2\alpha_2 + c\alpha_1\right)^{1/\Delta}}
	\right)^{1/2}
	\,,
\end{align}
%\begin{align}
%\vartheta = 2 \lambda \Gamma\left( \frac{1}{2} (1+\Delta) \right)
%(2\pi)^{(1-\Delta)D/4} \Delta^{\Delta D/4} \alpha_1^{-1/2} 
%\left( \frac{\alpha_2}{\alpha_1} \right)^{-\Delta(1-D/4)} 
%\left( \frac{\xi_{20}}{\xi_{10}} \right)^{-D\Delta/2} 
%\,.
%\end{align}
and
\begin{align}
	\vartheta = \frac{2}{\pi} \lambda   \Gamma\left[\frac{1}{2}(1+1/\Delta) \right] 
	\alpha_1^{(1/\Delta-1)/2} \Delta^{-D/(4\Delta)} 
	\left( \frac{\alpha_1}{\alpha_2}\right)^{\frac{1}{2\Delta} (1-D/2) } 
	\left(\frac{\xi_{10}}{\xi_{20}}\right)^{D/(2\Delta)}
	\,.
\end{align}
Note that $\alpha_{i}$ should be interpreted as $\alpha_{iL}$ in the fast cooling limit.

\section{Non-equilibrium phase diagrams}
%%%%%%%%%%%%%%%%%%%%%%%%%%%%%%%%%%%%%%%%%%%%%%%%%%%%%%
\begin{figure}
	\includegraphics[width= 0.9\linewidth]{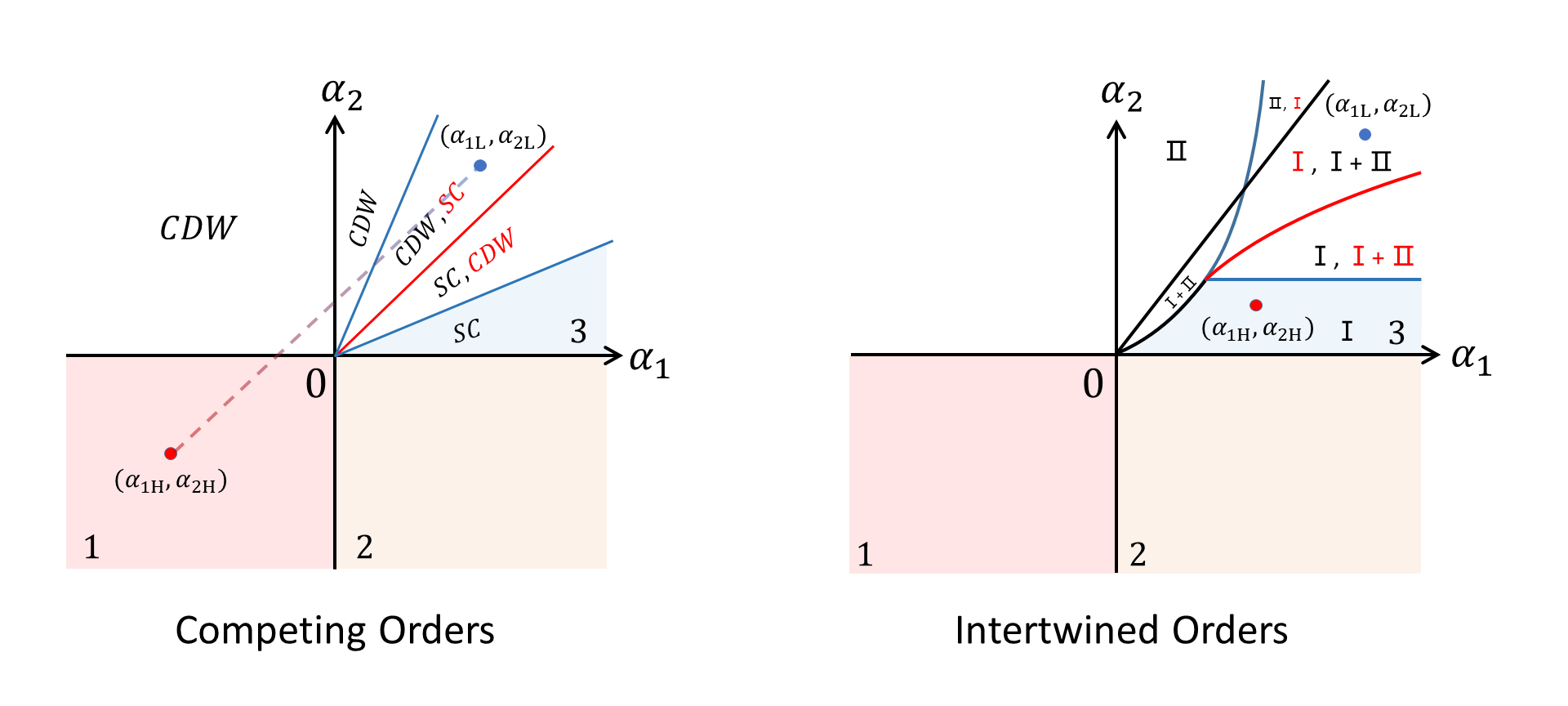}
	\caption{Nonequilibrium phase diagram for the pumped parameter $\alpha_{iH}$. If  $\alpha_{iH}$ lies in the colored regions, as illustrated by the red dots, the system can be trapped into the metastable SC sate. Dashed line is the trajectory of $\alpha_i(t)$ in the cooling process. }
	\label{fig:quartic_phase_diagram_nonequilibrium}
\end{figure}
%%%%%%%%%%%%%%%%%%%%%%%%%%%%%%%%%%%%%%%%%%%%%%%%%%%%%%

`Nonequilibrium' phase diagrams  Fig.~\ref{fig:quartic_phase_diagram_nonequilibrium} can be drawn for $\alpha_{iH}$. The system is originally at the blue dot $\alpha_{iL}$. If the pump brings the $\alpha_{iH}$ to any of the colored regions, metastable trapping into the SC (\RomanNumeralCaps{1}) state could happen. However, different regions have different stories as described in the main text. For example, it $\alpha_{iH}$ is in region  $1$ and $t_{pump} + t_0/2$ is much larger than $t_d=\frac{\Delta}{4 |\alpha_{2H}| \gamma_2} \ln \frac{1}{\zeta_1} $, the system can be brought to disordered state by the pump. The subsequent dynamics of fluctuation will lead the system into the metastable SC state if the relaxation in the SC direction is substantially faster. 

\section{Nano Granules}
If the system is a nano granule whose size is smaller than a coherence length, one can neglect the spatial fluctuation and treat the order parameter as uniform, i.e., one could keep the $k=0$ mode only. 
The initial dynamics is an expansion of the Gaussian probability due to thermal noise, regardless of the flow direction due to the potential. After the time $4 \alpha \gamma t \sim 1$, the dynamics starts to be dominated by the flow. At this time, $\psi_{i}^2 \sim T_v/\alpha_i$ and is much smaller than $\alpha_i$ if the system volume $V$ is not too small. Thus nonlinearity is not yet onset. After passing the crossover point $\psi_{i}^2 \sim T_v/\alpha_i$ to flow dynamics, the order parameter in each basin will be finally attracted to the corresponding minima, as shown in Fig.~\ref{fig:basin}. One immediately observes that if $\psi_{i}$ distribution is still tiny at the crossover point, most of the $\psi$ lies inside basin \RomanNumeralCaps{1} due to the nearly vertical shape of the basin boundary close to the origin.

To estimate of probability of trapping into phase \RomanNumeralCaps{1}, one can draw a rectangle centered at $O$ with half lengths of $L_i=\sqrt{\langle\psi_{i}^2\rangle}$. The length of its edge embedded in basin \RomanNumeralCaps{2} is $l_2 = 4 \lambda L_2^{\gamma_1\alpha_1/(\gamma_2 \alpha_2)}$ while the total length is $l=4L_1+4L_2$. Therefore, the probability of trapping into phase \RomanNumeralCaps{1} is roughly
\begin{align}
p_1 \sim 1- l_2/l = 1 - \kappa T_v^\delta
\,.
\label{eqn:probability}
\end{align}
where $\delta = \frac{1}{2} \left( \gamma_1\alpha_1/(\gamma_2 \alpha_2)-1 \right) >0 $ and $\kappa$ is order one. Since $T_v$ is a very small number, this probability is almost unity. After trapped into it, the life time of this metastable state is is exponentially large: $T_{life} \sim \frac{1}{\gamma} e^{U/T_v}$ where $U$ is the dimensionless energy barrier between the global and metastable minima. The detailed calculation for the lifetime is described by Kramer's theory \cite{Kramers1940,Landauer1961}.

%%%%%%%%%%%%%%%%%%%%%%%%%%%%%%%%%%%%%%%%%%%%%%%%%%%%%%%
%\begin{figure}[b]
%	\includegraphics[width=2.8 in]{pump_probe_schematic.png}
%	\caption{Schematic of the stages of order parameter evolution in pump-probe experiment.}
%	\label{fig:pump_probe_schematic}
%\end{figure}
%%%%%%%%%%%%%%%%%%%%%%%%%%%%%%%%%%%%%%%%%%%%%%%%%%%%%%%

%\emph{Granular case---}For a nano granule whose size is smaller than the coherence length $\xi_i$, spatial fluctuations can be neglected and $\psi_i$ is uniform throughout the volume $V$ of the sample. The TDGL equation with the stochastic force is equivalent to the diffusion equation for the probability density $\rho(\psi_i)$ in the potential landscape
%\begin{align}
%\partial_t \rho = \sum_i \left( \partial_i (\rho \gamma_i \partial_i F ) + D_i \partial_i^2 \rho
%\right)  \,
%\label{eqn:diffusion}
%\end{align}
%where $D_i = \gamma_i T^\prime = \gamma_i k_B T/(E_c V)$ is the diffusion constant. Note that $\partial_i$ in this case means $\partial_{\psi_i}$. 

\end{widetext}

\end{document}